\documentclass[bibyear]{aa}
\usepackage[varg]{txfonts}
\usepackage{graphics}
\usepackage{float}
\def\cc{\,{\rm cm^{-3}}} 
\def\cm2{\,{\rm cm^{-2}}}

\def\kms{\,{\rm {km\,s^{-1}}}} 
\def\kkms{\,{\rm {K\,km\,s^{-1}}}}

\def\co{\,{\rm ^{12}CO}} 
\def\thirco{\,{\rm ^{13}CO}} 
\def\h2{\,{\rm H_2}} 
\def\hco{\,{\rm HCO^{+}}} 
\def\C34S{\,{\rm C^{34}S}} 
\def\hi{\,{\rm H I}}
\def\ci{\,{\rm [C I]}}

\def\cii{\,{\rm [C II]}}

\def\etal{\rm et\,al.\ }
\def\aua{{\it A\&A} } 
 
\def\apj{{\it ApJ} } 
\def\aj{{\it AJ} } 
\def\apjs{{\it ApJS} } 
\def\apjl{{\it ApJL} } 
\def\araa{{\it ARAA} } 
\def\mnras{{\it MNRAS} }

\def\pasj{{\it PASJ} } 
\def\pasp{{\it PASP} } 
\begin{document} 

\title{Central molecular zones in galaxies: \\
Multitransition survey of dense gas tracers HCN, HNC, and HCO$^{+}$ }

\author{F.P. Israel 
        \inst{1} 
           } 

\institute{Leiden Observatory, Leiden University, P.O. Box 9513, 2300 RA Leiden, the Netherlands}
 
\titlerunning{HCN, HNC, and HCO$^+$ in galaxy centers}

\offprints{F.P. Israel} 
 
\date{Received ????; accepted ????} 
 
\abstract{New measurements of 46 nearby galaxy centers in up to three
  transitions of HCN, HNC, and $\hco$ combined with literature surveys
  establish a database of 130 galaxies measured in both HCN and
  HCO$^+$, and 94 galaxies in HNC as well, allowing a systematic
  exploration of the relations between normalized luminosities and
  line ratios. The almost linear relations between luminosities are
  predominantly caused by distance effects and do not reflect galaxy
  physical properties. Individual galaxies show significant dispersion
  in both their luminosity and line ratio, which will be analyzed in
  more detail in a later paper. Very few line ratios correlate either
  with luminosities or with other line ratios. Only the normalized
  transition ladders of HCN and HCO$^+$ and the $J$=1-0
  $\co$/$\thirco$ isotopologue ratio are positively correlated with CO
  and far infrared (FIR) luminosity. On average, HCN and HCO$^+$ have
  very similar intensities and trace the same gas. In galaxies
  dominated by an active nucleus, HCO$^+$ intensities appear to be
  depressed relative to HCN intensities. Only a small fraction of CO
  emission is associated with gas emitting in HCN and $\hco$, yet a
  significant fraction of even that gas appears to be translucent
  molecular gas. In the observed galaxy centers, the HCN/CO line
  intensity ratio is not a proxy for the dense gas fraction, and the
  FIR/HCN and FIR/CO ratios are not proxies for the star formation
  efficiency.  A proper understandig of star formation requires a more
  appropriate determination of gas mass than provided by the
  intensities of individual HCN or CO transitions. The observed
  molecular line emission is fully consistent with UV-photon heating
  boosted by significant mechanical heating.  The molecular gas
  sampled by HCN and $\hco$ has low kinetic temperatures
  $T_{\rm kin}\,=\,10-50$ K, low densities
  $n_{rm H}\,=\,10^4-10^5\,\cc$, and low optical depths in the
  ground-state lines. Most of the gas sampled by CO has densities
  lower by one to two orders of magnitude. For a mechanical heating
  fraction of 0.5, a modest energy input of only $G\,=\,300$ G$_{0}$
  is required. } 

\maketitle

\section{Introduction}

This paper presents multitransition measurements of the HCN, HNC and
$\hco$ molecular species tracing the dense molecular gas in the
centers of nearby galaxies. Many late-type galaxies contain massive
concentrations of molecular hydrogen gas close to the nucleus. These
concentrations form the reservoirs that feed mass inflow into
supermassive black holes (SMBH), mass outflow from the nucleus into
space, and bursts of circumnuclear star formation. In a previous paper
(Israel 2020), we have explored the physical characteristics of the
inner-disk gas from measurements of the lower $J$ transitions of $\co$
and $\thirco$, augmented by literature measurements of neutral ($\ci$)
and ionized ($\cii$) atomic carbon.

\begin{table}
\begin{center}
{\small %
\caption[]{\label{crit}Critical densities}
\begin{tabular}{lrrrrr}
\noalign{\smallskip}     
\hline
\noalign{\smallskip} 
Molecule&Tran-     &Frequency&E$_{u}$/k$^{a}$   &$n^{a,b}_{crit}$&$n^{c}_{eff}$\\
        &sition    & (GHz)   & (K)        & ($\cc$)    & ($\cc$)\\
(1)     & (2)      &     (3) & (4)        & (5)        & (6) \\
\noalign{\smallskip}      
\hline
\noalign{\smallskip} 
CO     & 1-0 & 115.271 &  5.5 & 4 (2) & ...   \\
       & 2-1 & 230.538 & 16.6 & 5 (3) & ...   \\
       & 3-2 & 345.796 & 33.2 & 2 (4) & ...   \\
       & 4-3 & 461.041 & 55.3 & 2 (5) & ...   \\
HCN    & 1-0 &  88.632 &  4.3 & 2 (5) & 3 (3) \\
       & 3-2 & 265.886 & 25.5 & 8 (6) & 1 (4) \\
       & 4-3 & 354.506 & 42.5 & 2 (7) & 2 (5) \\
HCO$^+$& 1-0 &  89.189 &  4.3 & 3 (4) & 4 (2) \\
       & 3-2 & 267.558 & 25.7 & 1 (6) & 4 (3) \\ 
       & 4-3 & 356.734 & 42.8 & 3 (6) & 2 (4) \\ 
HNC    & 1-0 &  90.664 &  4.4 & 1 (5) & 2 (3) \\ 
       & 3-2 & 271.981 & 26.1 & 4 (6) & 3 (4) \\  
       & 4-3 & 362.630 & 43.5 & 9 (6) & 1 (5) \\ 
C$_2$H & 1-0 &  87.317 &  4.3 & 2 (5) & 2 (4) \\
CS     & 2-1 &  97.981 &  7.1 & 9 (4) & 9 (3) \\ 
       & 3-2 & 146.969 & 14.1 & 3 (5) & 3 (4) \\ 
  \noalign{\smallskip}     
\hline
\end{tabular}
}%
\end{center} 
Notes: a. Jansen (1995) and Shirley (2015); b. Calculated
for $T_{\rm kin}$ = 30 K in the optically thin limit; c.
Shirley (2015), with log $N_{ref}$ = 14.0 (HCN, $\hco$, and HNC)
and log $N_{ref}$ = 13.5 (CS, C$_{2}$H)
\end{table}   

Gas inflow and outflow involve both dense clouds and diffuse
intercloud gas together responsible for the carbon and carbon monoxide
emission studied in that paper.  Star formation, in contrast, is
exclusively associated with the dense gas. The determination of the
properties of that gas requires observation of molecules that, unlike
CO, need relatively high densities for excitation at modest
temperatures.  Molecules such as HCN, $\hco$, HNC, and CS are suitable
for this purpose even though they have abundances much lower than CO
and much weaker emission lines. In Table\,\ref{crit} we list, for
these molecules in the relevant transitions, the line frequency, the
minimum temperature for excitation $T_{min}\,=\,E_{upper}$/k, and the
temperature-dependent critical density $n_{crit}$ for an assumed
kinetic temperature $T_{kin}$ = 30 K (cf. Paper I). Critical densities
are lower for warmer gas and higher for cooler gas, roughly by factors
of up to two going from 30 K to either 10 K or 100 K and HCN traces
the highest densities. As molecular emission remains detectable,
however, at densities well below $n_{crit}$, we also list the
effective de-excitation density $n_{eff}$. Somewhat arbitrarily, this
is the density that produces a line of intensity 1 $\kkms$, for a
given temperature and column density (see Evans 1999; Shirley
2015). The molecules and transitions in Table\,\ref{crit} cover a wide
range of excitation conditions, including densities 10$^2$ - 10$^7$
$\cc$ and temperatures 4-55 K. In their study of molecular cloud
ensembles in the inner Galaxy, Evans $\etal$ (2020) found that the
integrated HCN and $\hco$ luminosity can be dominated by emission from
even more tenuous molecular gas with densities as low as 10$^2$ $\cc$.

After the brightest individual galaxies had been observed in the
relatively accessible $J$=1-0 lines of HCN, $\hco$, and HNC, the first
surveys with the $IRAM$ 30m telescope, encompassing some fifteen
galaxies, were published by Nguyen-Q-Rieu $\etal$ (1992) and
H\"uttemeister $\etal$ (1995).  Among early observations carried out
between 1993 and 1997, the large HCN and CO survey of 53 galaxies by
Gao $\&$ Solomon (2004a) and the survey of 37 galaxies in HCN, HNC,
$\hco$, CO, CN, and CS by Baan $\etal$ (2008) stand out. The latter
0combined all observations available at the time into a heterogeneous
database of 117 galaxies with 23 galaxies detected in all three lines,
HCN(1-0), HNC(1-0), and $\hco$(1-0). They concluded that the emission
from these lines and the far-infrared (FIR) continuum are tightly
correlated, and suggested that the ratio of the HCN-to-CO luminosity
traces the fraction of dense molecular gas. Kohno $\etal$ (2001, 2008)
and Imanishi $\etal$ (2007) then suggested that the ground-state
CO-HCN-$\hco$ line intensity ratios from galaxy centers dominated by a
burst of star-formation (SB) differ from those of galaxy centers
dominated by nuclear activity (AGN). In their multi-transition $IRAM$
30m survey of the HCN and $\hco$ emission from a dozen galaxies, Krips
$\etal$ (2008) found higher average molecular gas densities in SB
galaxies and higher average temperatures and HCN/$\hco$ line ratios in
AGN galaxies. An HCN/$\hco$ overabundance was, however, also found in
luminous (LIRG) and ultra-luminous (ULIRG) galaxies representing
extreme SBs (Grac\'ia-Carpo $\etal$ 2008). Bussman $\etal$ (2008)
conducted a survey of HCN(3-2) emission and found it to behave
differently from HCN(1-0). The major study of dense gas in luminous
galaxies by Baan $\etal$ (2008) referred to earlier took these results
a step further by constructing diagnostic diagrams to investigate
source differentiation as a function of initial conditions and
radiative environment.

More recently, the installation of very sensitive multi-mixer
receivers (EMIR) with very wide back-end coverage (FTS, WILMA) at the
$IRAM$ 30m telescope has enormously expanded the possibilities for
extra-galactic molecular line measurements as illustrated, for
instance, by the spectral scans published by Costagliola $\etal$
(2011) and Jiang $\etal$ (2011). Later in this paper we will refer to
their findings as well as those of others obtained since then.  We
will do this in the context of analyzing and discussing an extensive
and homogeneous database of HCN, HNC, HCO$^+$, and CO line intensities
in various transitions encompassing the newly obtained observations as
well as directly comparable data from the published literature. The
database will be used to investigate the overall relations between the
various lines and transitions, including verification of previous
claims.  Detailed modelling of the results is deferred to a later
paper.

This paper is structured as follows. Section 2 presents the new
observations. Section 3 describes the expansion of the database with
line intensities from surveys published by others and the
normalization applied to data observed at different resolutions in
order to obtain a homogeneous sample of intercomparable intensities.
Section 4 systematically analyzes the trends and correlations, either
present or absent, between the various observed line intensities and
line ratios. Section 5 discusses the meaning of these findings.
Section 6 summarizes the most important points.

\section{Observations}

\begin{table}
\begin{center}
{\small %
\caption[]{\label{sample} List of observed galaxy centers} 
\begin{tabular}{lrrrrcc}
\noalign{\smallskip}     
\hline
\noalign{\smallskip} 
NGC & Dist.&lg$I_{FIR}$&lg$L_{FIR}$& $I_{\rm CO}$(1-0)& $D_{25}$ & Type$^a$ \\
IC  &(Mpc)&(Wm$^{-2}$)&($L_{\odot}$)& ($K km/s$)& (') & \\
(1)      & (2)  &    (3) & (4)   & (5)  & (6)      &  (7) \\
\noalign{\smallskip}      
\hline
\noalign{\smallskip} 
N 253    &  3.4 & -10.42 & 10.13 & 1030 & 25x7.4   & S   \\
N 470    & 31.7 & -12.44 & 10.04 &   28 & 2.8x1.7  & S   \\ 
N 520    & 30.5 & -11.79 & 10.66 &  113 & 4.5x1.8  & S   \\
N 660    & 12.2 & -11.46 & 10.20 &  154 & 9.1      & S   \\
N 891    &  9.4 & -11.53 &  9.90 &  137 & 14x2.5   & (S) \\
N 972    & 21.4 & -11.75 & 10.39 &   67 & 3.4x1.7  & S   \\
Maff2    &  3.1 & -11.23 &  9.23 &  220 & 5.8x1.6  & S   \\
N1055    & 13.4 & -11.84 &  9.89 &   77 & 7.6x2.7  & S   \\
N1068$^b$& 15.2 & -11.04 & 10.80 &  168 & 7.1x6.0  & A   \\ 
N1084    & 18.6 & -12.33 &  9.69 &   30 & 3.3x1.2  & (S) \\ 
N1097$^b$& 16.5 & -11.81 & 10.10 &  136 & 9.3x6.6  & A+S \\ 
N1365$^b$& 21.5 & -11.36 & 10.78 &  260 &  11x6.2  & A+S \\  
I342     &  3.8 & -11.36 &  9.28 &  161 &  21x21   & S   \\ 
N1808    & 12.3 & -11.31 & 10.35 &  135 & 6.5x3.9  & A   \\ 
N2146    & 16.7 & -11.16 & 10.77 &  187 & 6.0x3.4  & S   \\ 
N2559    & 21.4 & -11.78 & 10.36 &   78 & 4.1x2.1  & ... \\ 
N2623    & 79.4 & -11.94 & 11.34 &   18 & 2.4x0.7  & L   \\ 
N2903    &  7.3 & -11.65 &  9.56 &   80 &  13x6.0  & S   \\ 
N3034    &  5.9 & -10.28 & 10.74 &  680 &  11x4.3  & S   \\
N3079$^b$& 20.7 & -11.60 & 10.51 &  235 & 7.9x1.4  & A   \\
N3310    & 19.2 & -11.79 & 10.26 &    7 & 3.1X2.4  & S   \\
N3627    &  6.5 & -11.61 &  9.50 &   74 & 9.1x4.2  & A+S \\
N3628    &  8.5 & -11.54 &  9.80 &  203 &  15x3.0  & S   \\
N3690    & 48.5 & -11.32 & 11.53 &   69 & 2.9x2.1  & L   \\
N4030    & 26.4 & -11.95 & 10.37 &   42 & 4.3      & ... \\
N4038    & 23.3 & -11.65 & 10.56 &   47 & 5.2x3.1  & S   \\
N4102    & 17.3 & -11.62 & 10.34 &   75 & 2.8x1.2  & A   \\  
N4321    & 14.1 & -11.88 &  9.90 &   82 & 7.4x6.3  & S   \\
N4414    &  9.0 & -11.77 &  9.62 &   51 & 3.6x2.0  & A?  \\
N4527    & 13.5 & -11.79 &  9.95 &   88 & 6.2x2.1  & S?  \\
N4569    & 12.3 & -12.19 &  9.47 &   89 & 9.5x4.4  & A+S \\
N4666    & 27.5 & -11.74 & 10.62 &   74 & 4.6x1.3  & S   \\       
N4826$^b$&  3.8 & -11.66 &  9.34 &   91 &  10x5.4  & A+S \\      
N5033$^b$& 17.2 & -12.00 &  9.95 &   53 &  11x5.0  & A   \\      
N5055    &  8.3 & -11.66 &  9.66 &   70 &  13x7.2  & (A) \\      
N5194$^b$&  9.1 & -11.59 &  9.81 &   48 &  11x6.9  & A   \\      
N5236    &  4.0 & -11.22 &  9.46 &  195 &  13x12   & S   \\      
N5775    & 28.9 & -11.97 & 10.43 &   48 & 4.2x1.0  & ... \\    
N6240    &  109 & -11.96 & 11.59 &   70 & 2.1x1.1  & L   \\    
N6701    & 59.1 & -12.23 & 10.80 &   45 & 1.5x1.3  & A+S \\    
N6946    &  5.5 & -11.06 &  9.90 &  228 & 11.5x9.8 & S   \\
N6951$^b$& 24.3 & -12.04 & 10.21 &   50 & 3.9x3.2  & A+S \\    
N7469$^*$& 67.0 & -11.88 & 11.25 &   55 & 1.5x1.1  & L   \\    
N7714    & 38.5 & -12.30 & 10.35 &    4 & 1.9x1.4  & S   \\    
N7771    & 58.0 & -11.97 & 11.04 &  100 & 2.5x1.0  & L   \\    
Arp220   & 82.9 & -11.31 & 12.04 &  110 & 1.5x1.2  & U   \\
\noalign{\smallskip}     
\hline
\end{tabular}
}%
\end{center} 
Notes: a. S = Starburst, A = AGN, L = LIRG, U = ULIRG;
parentheses indicate marginal case. b. Seyfert nucleus.
\end{table}   

\begin{figure*}
\resizebox{18cm}{!}{\rotatebox{0}{\includegraphics*{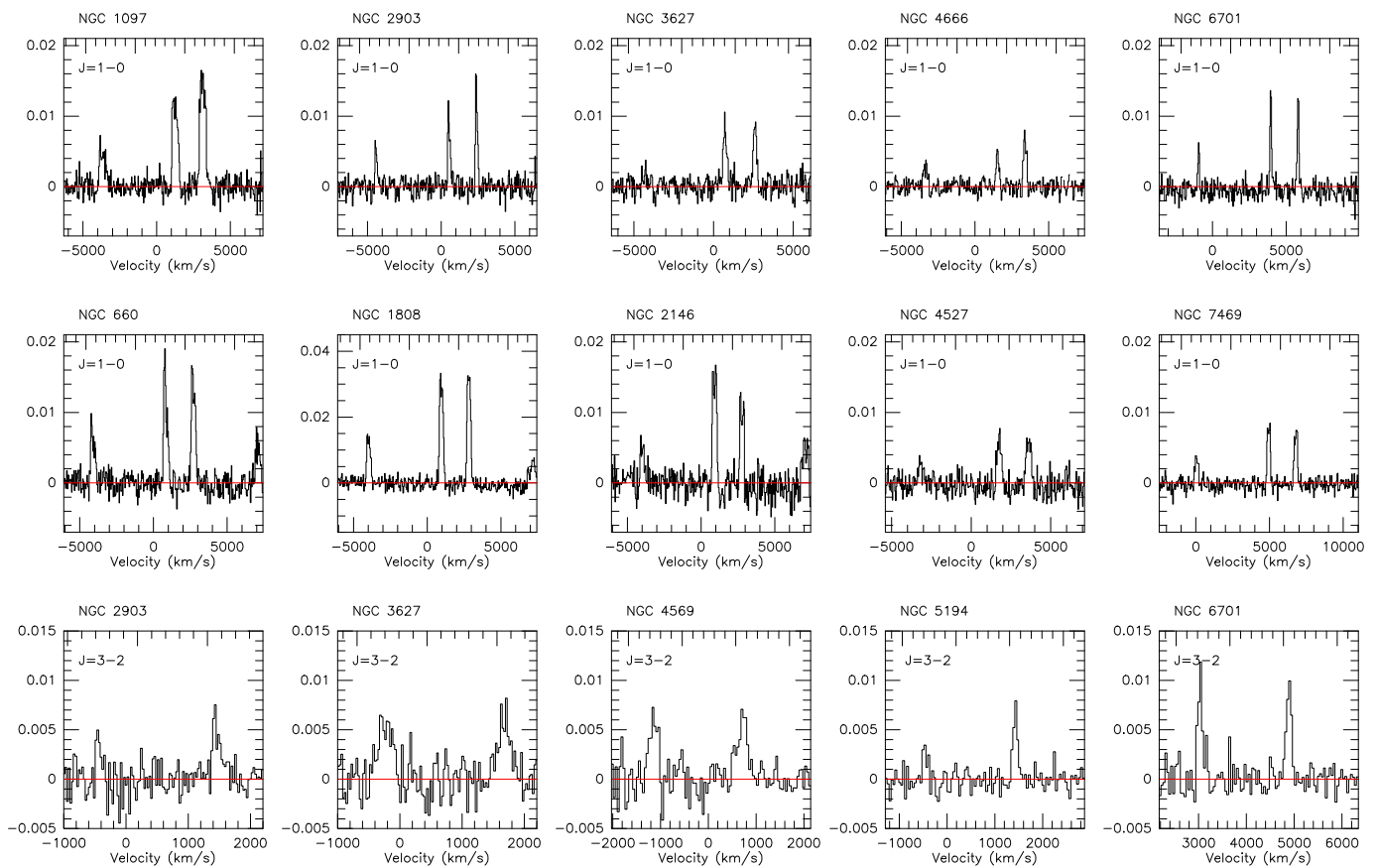}}}
\caption[] {Sample of molecular line profiles of galaxy centers
  observed with the $IRAM$ 30m telescope. Galaxy and transition depicted
  are identified at the top of each panel.  The top row shows $J$ 1-0
  observations of (in each panel from left to right) HNC, HCO$^{+}$,
  and HCN observed in 2010. The middle row shows the same lines and
  includes the C$_2$H at the very right, observed in 2011. The bottom
  row shows $J$=3-2 observations of HCO$^{+}$ (left) and HCN
  (right). In all panels the vertical scale is intensity
  $T_{\rm A}^{*}$ (K) and the horizontal scale is velocity V(LSR) in
  $\kms$.  }
\label{iramprofiles}
\end{figure*}

\begin{figure*}
\resizebox{18cm}{!}{\rotatebox{0}{\includegraphics*{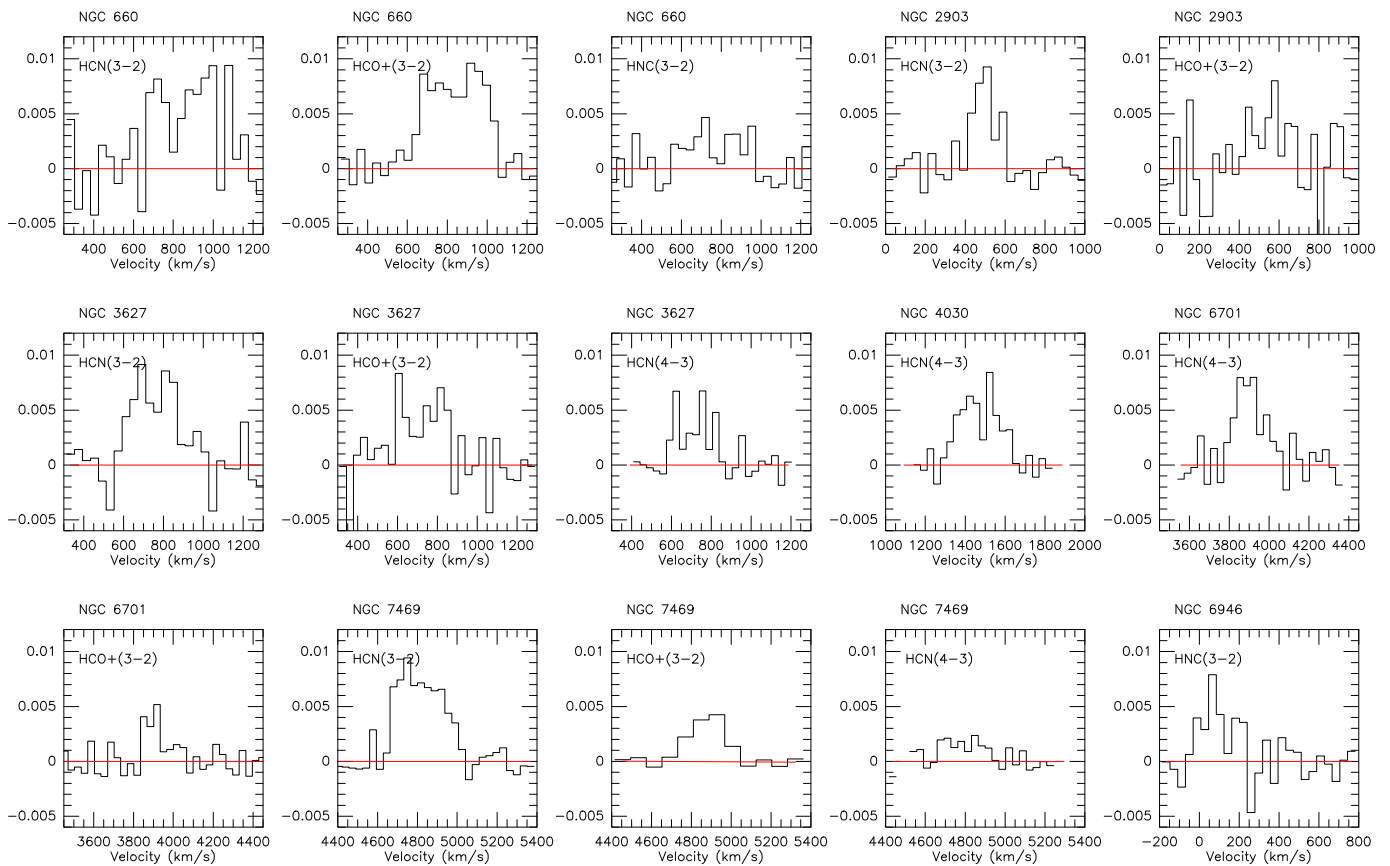}}}
\caption[] {Sample of $J$=3-2 and $J$=4-3 molecular line profiles of
  galaxy centers observed with the 15m $JCMT$. Galaxy and transition are
  depicted at the top of each panel. In all panels the vertical scale
  is intensity $T_{\rm A}^{*}$ (K) and the horizontal scale is velocity
  V(LSR) in $\kms$.
 }
\label{jcmtprofiles}
\end{figure*}

\subsection{Observing list}

We based our selection on the multi-transition $\co$ and $\thirco$
survey of 126 galaxies by Israel (2020), which originally consisted of
the brightest galaxies in the $IRAS$ infrared sky survey but was later
expanded by adding other galaxies bright in $\co$. Brightness is an
important criterion because the $\thirco$ line, with its low optical
depth, is relatively weak. Because the HCN, HCO$^+$ and HNC lines are
even weaker, the survey presented in this paper is limited to a subset
of the brightest galaxies from the CO survey. Because of their greater
distance, few infrared-luminous galaxies are bright enough to be
included in this limited sample.

Table\,\ref{sample} provides the names of the 46 galaxies observed,
their distances, FIR intensities and luminosities, $\co$(1-0)
intensities in the $IRAM$ $22"$ beam, and overall angular size (Israel
2020 and references therein). The list contains a single ULIRG and
five LIRGs. The remainder are lower-luminosity galaxies with a central
starburst an AGN, or both.

We made observations in the $J$=1-0, $J$=3-2 and $J$=4-3 transitions
of HCN, $\hco$ and HNC with the $IRAM$ 30m and the $JCMT$ 15m
telescopes. The $J$=2-1 transitions are at frequencies with poor
atmospheric transmission and observations were not attempted. The
$J$=3-2 transitions were observed with both $IRAM$ and $JCMT$,
allowing determination of beam-dilution effects. As a byproduct of
the early (2008) observations with the $IRAM$ ABCD receivers we also
obtained CS(3-2) observations of thirteen galaxies, whereas the later
$IRAM$ EMIR observations also covered C$_{2}$H(1-0) emission at the
band-edge in eleven galaxies.
 
\begin{table*}
  \caption[]{\label{fluxes} Galaxy centers: line intensities
    $\int(T_{mb}dV)$ ($\kkms$) for HCN, HCO$^{+}$, HNC}
\begin{center} 
{\small %
\begin{tabular}{lrrrrrrrrrr} 
\noalign{\smallskip}     
\hline
\noalign{\smallskip} 
NGC  & HCN            &              &              &            & HCO$^+$      &               &              &              & HNC          & \\  
IC   & (1-0)          & (3-2)        & (3-2)        & (4-3)      & (1-0)        & (3-2)         & (3-2)        & (4-3)        & (1-0)        & (3-2)   \\
     & $IRAM$         &$IRAM$        &$JCMT$        &$JCMT$      &$IRAM$        &$IRAM$         &$JCMT$        &$JCMT$        &$IRAM$        &$JCMT$   \\
     & 27.6"          &  9.7"        &  19.0"       &  13.7"     & 27.6"        &  9.7"         & 19.0"        & 13.7"        & 27.6"        & 19.0"   \\
(1)  &    (2)         &     (3)      &     (4)      &     (5)    &     (6)      &     (7)       &     (8)      &     (9)      &    (10)      &  (11)   \\  
\noalign{\smallskip}     
\hline
  \noalign{\smallskip}
253  & 68.4$\pm1.4^a$ & 120$\pm$12   & 77.2$\pm$14.2&61.1$\pm$1.0 &57.9$\pm1.2^a$&112$\pm$25    &62.4$\pm$1.1  &66.0$\pm$1.0 &35.5$\pm3.2^a$ &34.3$\pm$0.7 \\
470  &  0.75$\pm$0.21 & ...          & ...          &...          &  ...         & ...          & ...          & ...          & ...          &...          \\
520  &  2.14$\pm$0.14 & ...          & 1.85$\pm$0.41&...          & 3.12$\pm$0.13& ...          & 2.34$\pm$0.27& ...          & 1.31$\pm$0.12&...          \\
660  &  6.54$\pm$0.65 & 5.34$\pm$0.68& 4.20$\pm$0.90&2.55$\pm$0.50& 5.13$\pm$0.75& 3.35$\pm$0.90& 4.33$\pm$0.29& ...          & 2.97$\pm$0.31&1.74$\pm$0.32\\
891  &  1.30$\pm$0.12 & 1.90$\pm$0.55& 1.66$\pm$0.38&...          & 1.33$\pm$0.19& 1.29$\pm$0.28& 1.36$\pm$0.35& ...          & 0.44$\pm$0.08&...          \\
972  &  1.64$\pm$0.18 & ...          & 1.60$\pm$0.55&...          & 1.91$\pm$0.17& ...          & 0.71$\pm$0.35& ...          & 0.76$\pm$0.17&...          \\
1055 &  2.35$\pm$0.15 & ...          & 1.55$\pm$0.49&...          & 1.40$\pm$0.14& ...          & 0.44$\pm$0.15& ...          & 0.79$\pm$0.14&...          \\
Maf2 & 15.4$\pm$0.5   & 16.6$\pm$0.7 & 8.68$\pm$0.50&3.79$\pm$0.79& 9.48$\pm$0.30&12.7$\pm$1.4  & ...          & 3.48$\pm$0.70& 7.3$\pm0.7^d$&4.04$\pm$0.46\\
1068 & 25.0$\pm1.2^b$ &19.0$\pm1.2^c$& 14.3$\pm$0.6 &10.9$\pm$0.6 &14.8$\pm0.8^b$& 7.6$\pm0.8^c$ & 5.43$\pm$0.64& 4.85$\pm$0.50& 9.3$\pm1.2^b$&4.08$\pm$0.55\\
1084 &  0.71$\pm$0.10 & ...          & 1.85$\pm$0.40&...          & 0.62$\pm$0.10& ...          & 1.26$\pm$0.30& ...          & 0.28$\pm$0.09&0.37$\pm$0.17\\
1097 &  9.31$\pm$0.18 & ...          & 5.99$\pm$0.81&...          & 6.67$\pm$0.20& ...          & 4.52$\pm$0.66& ...          & 3.06$\pm$0.19&...          \\
I342 & 10.8$\pm$0.3   & 8.3$\pm$0.9  & 3.53$\pm$0.35&2.7$\pm0.3^i$& 8.41$\pm$0.24& 7.49$\pm$0.94& 5.05$\pm$0.75&3.8$\pm0.3^i$ & 4.75$\pm0.11^j$&1.47$\pm$0.24\\
1365 & 14.2$\pm$0.7   &12.0$\pm$0.4  & 9.20$\pm$0.57&...          &11.7$\pm$1.4  & 9.80$\pm$0.36& ...          & ...          & 7.24$\pm$0.52&3.59$\pm$0.74\\
1808 & 12.5$\pm$0.35  & ...          & ...          &...          &11.7$\pm$0.4  & ...          & ...          & ...          & 5.12$\pm$0.35&...          \\
2146 &  4.67$\pm$0.36 & 4.4$\pm0.3^c$& 4.30$\pm$0.49&4.49$\pm$0.89& 6.70$\pm$0.36& 5.2$\pm0.6^c$& ...          & ...          & 1.86$\pm$0.35&2.33$\pm$0.83\\
2559 &  2.96$\pm$0.13 & ...          & 2.52$\pm$0.33&2.40$\pm$0.48& 3.17$\pm$0.18& ...          & ...          & ...          & 1.03$\pm$0.13&0.14$\pm$0.13\\
2623 &  1.77$\pm$0.38 & 2.07$\pm$0.76& 1.82$\pm$0.34&1.29$\pm$0.39& 1.87$\pm$0.25& 3.18$\pm$0.48& ...          & ...          & ...         &1.83$\pm0.2^f$\\
2903 &  3.35$\pm$0.12 & ...          & 1.88$\pm$0.19&1.36$\pm$0.23& 1.81$\pm$0.32&...           & 1.62$\pm$0.32& ...          & 1.23$\pm$0.12&0.61$\pm$0.48\\
3034 & 29$\pm$3$^c$   & 27$\pm3^c$   & 13.4$\pm$0.7 &7.51$\pm$0.55&40.2$\pm2.5^c$&34.0$\pm2.0^c$ &  27.2$\pm$1.1& 19.3$\pm$1.9 &12$\pm1.5^e$ &3.58$\pm$1.52\\
3079 &  7.57$\pm$0.58 & 15.8$\pm$1.5 & 6.91$\pm$0.46&3.44$\pm$0.47& 6.54$\pm$0.23&16.2$\pm$1.8  & 6.14$\pm$1.05& 3.78$\pm$0.49& 2.2$\pm0.14^j$&3.1$\pm$1.2\\
3310 &  0.30$\pm$0.08 & ...          & 1.41$\pm$0.72&...          & 0.67$\pm$0.40&...           & 0.74$\pm$0.28& ...          & 0.32$\pm$0.12&...          \\
3627 &  2.79$\pm$0.14 & ...          & 2.23$\pm$0.40&1.64$\pm$0.16& 2.93$\pm$0.15& 2.1$\pm0.2^j$& ...          & ...          & 0.78$\pm$0.19&0.58$\pm$0.20\\
3628 &  5.35$\pm$0.36 & 3.25$\pm$0.73& 3.42$\pm$0.20&2.01$\pm$0.17& 5.60$\pm$0.23& 4.67$\pm$0.93& ...          & ...          & 3.4$\pm0.3^d$&...          \\
3690 &  2.51$\pm$0.30 &1.58$\pm0.38^g$& 0.99$\pm$0.21&0.44$\pm$0.19& 3.62$\pm$0.18&4.72$\pm0.44^g$& ...        & ...          & 0.90$\pm$0.19&...          \\
4030 &  3.45$\pm$0.14 & 2.00$\pm$0.50& 1.37$\pm$0.45&1.57$\pm$0.17& 1.76$\pm$0.17& 0.90$\pm$0.26& ...          & ...          & 1.80$\pm$0.17&...          \\
4038 &  2.57$\pm$0.10 & 0.91$\pm$0.19& 0.95$\pm$0.21&1.10$\pm$0.71& 3.53$\pm$0.10& 1.53$\pm$0.33& 1.44$\pm$0.28& ...          & 1.13$\pm$0.07&...          \\
4102 &  3.75$\pm$0.32 & 3.64$\pm$0.68& 2.01$\pm$0.62&...          & 2.28$\pm$0.51& 2.02$\pm$0.61& ...          & 2.74$\pm$0.17& 2.81$\pm$0.23&   ...      \\
4321 &  4.93$\pm$0.28 & 3.09$\pm$0.26& 2.23$\pm$0.60&1.75$\pm$0.46& 3.19$\pm$0.12& 2.12$\pm$0.39& ...          & ...          & 1.64$\pm$0.08&...          \\
4414 &  3.76$\pm$0.39 & ...          & 1.05$\pm$0.39&0.92$\pm$0.17& 2.68$\pm$0.30& 1.34$\pm$0.41& ...          & ...          & 1.03$\pm$0.15&...          \\
4527 &  3.41$\pm$0.24 & 1.90$\pm$0.21& 1.33$\pm$0.44&...          & 3.07$\pm$0.23& 1.87$\pm$0.20& ...          & ...          & 1.54$\pm$0.28&...          \\
4569 &  3.21$\pm$0.15 & 2.35$\pm$0.38& 0.76$\pm$0.23&0.69$\pm$0.25& 2.50$\pm$0.14& 2.08$\pm$0.33& ...          & ...          & 0.73$\pm$0.15&...          \\
4666 &  2.19$\pm$0.07 & 1.82$\pm$0.46& 0.60$\pm$0.38&1.36$\pm$0.11& 1.42$\pm$0.10& 0.63$\pm$0.24& ...          & ...          & 1.36$\pm$0.14&...          \\
4826 &  7.19$\pm$0.39 & 1.77$\pm$0.36& 1.94$\pm$0.70&1.22$\pm$0.32& 4.46$\pm$0.10& 1.33$\pm$0.21& ...          & ...          & 3.03$\pm$0.15&1.32$\pm$0.31\\
5033 &  1.28$\pm$0.13 &...           & 1.26$\pm$0.51&...          & 0.95$\pm$0.15&...           & 0.62$\pm$0.26& ...          & 1.07$\pm$0.17&...          \\
5055 &  2.59$\pm$0.44 & 1.85$\pm$0.48& 0.90$\pm$0.60&0.61$\pm$0.22& 1.43$\pm$0.28&...           & ...          & ...          & 0.71$\pm$0.15&...          \\
5194 &  7.15$\pm$0.29 & 2.31$\pm$0.24& 1.38$\pm$0.70&1.36$\pm$0.43& 3.36$\pm$0.15& 1.43$\pm$0.29& ...          & ...          & 2.74$\pm$0.14&0.48$\pm$0.29\\
5236 &  8.54$\pm$0.26 &...           & 3.86$\pm$0.30&1.76$\pm$0.20&10.4$\pm$0.4  & ...          & ...          & 2.52$\pm$0.36& 4.3$\pm0.3^d$ &1.30$\pm$0.28\\
5775 &  1.39$\pm$0.36 & 0.85$\pm$0.19&...           &*            & 1.30$\pm$0.39& 0.46$\pm$0.31& ...          & ...          & 0.27$\pm$0.17&...          \\
6240 &  3.80$\pm$0.20 & 7.8$\pm0.31^h$& 4.70$\pm$0.29&4.61$\pm$1.42& 6.37$\pm$0.20& 7.1$\pm1.9^h$& 1.79$\pm$1.15 & ...          & 0.78$\pm$0.15&...          \\
6701 &  2.32$\pm$0.06 & 2.55$\pm$0.25& 1.09$\pm$0.20&1.85$\pm$0.46& 2.37$\pm$0.07& 2.41$\pm$0.24& 3.43$\pm$0.75& ...          & 0.96$\pm$0.09&0.50$\pm$0.50\\
6946 &  9.9$\pm0.1^c$ & 9.2$\pm0.6^c$& 3.85$\pm$0.36&1.53$\pm$0.22& 8.5$\pm0.1^c$& 8.6$\pm0.6^c$ & 3.19$\pm$0.38& ...          & 4.0$\pm0.4^d$&1.98$\pm$0.54\\
6951 &  2.68$\pm$0.13 & 1.99$\pm$0.38& 1.13$\pm$0.39&...          & 1.95$\pm$0.08& 2.24$\pm$0.41& 1.30$\pm$0.44& ...          & 1.50$\pm$0.09&0.67$\pm$0.20\\
7469 &  2.56$\pm$0.13 & 2.76$\pm0.34^g$&2.01$\pm$0.51&0.73$\pm$0.25&2.68$\pm$0.12&2.08$\pm0.42^g$&1.51$\pm$0.19& ...          & 1.32$\pm$0.13&1.09$\pm$0.44\\
7714 &  0.56$\pm$0.18 & ...          & ...          &...          & 0.18$\pm$0.17& ...          & ...          & ...          &    ...       &...          \\
7771 &  5.35$\pm$0.60 &3.33$\pm0.58^g$&1.19$\pm$0.23&...          & 5.59$\pm$0.26&1.98$\pm0.37^g$& 2.48$\pm$0.56& ...          & 2.31$\pm$0.15&1.21$\pm$0.43\\
A220 &8.16$\pm$0.17$^g$&18.0$\pm0.5^g$&12.5$\pm$0.6 &3.89$\pm$0.59&3.77$\pm0.21^g$&4.57$\pm0.23^g$& ...         & 2.96$\pm$0.49 & 7.8$\pm0.8^g$&...         \\
\noalign{\smallskip}     
\hline
\end{tabular}
}%
\end{center} 
References to intensities taken from the literature: a. Aladro $\etal$
(2015); b. Average from Krips $\etal$ (2008), Costagliola $\etal$
(2011), and Aladro $\etal$ (2015); c. Krips $\etal$ (2008);
d. H\"uttemeister $\etal$ (1995); e. Average from H\"uttemeister
$\etal$ (1995) and Aladro $\etal$ (2015); f. P\'erez-Beaupuits $\etal$
(2007); g. Graci\'a-Carpio $\etal$ 2008: h. Li $\etal$ (2018);
i. Tan $\etal$ (2018); j. Li et al. (2021)
\end{table*}  

\begin{table}
\begin{center}
{\small %
\caption[]{\label{morefluxes}  Galaxy centers: CS, C$_2$H line intensities
  $\int(T_{mb}dV)$}
\begin{tabular}{lrr}
\noalign{\smallskip}     
 \hline
\noalign{\smallskip} 
NGC   & CS(3-2)          & C$_2$H(1-0)         \\
IC    & $16.8"$          & $27.6"$             \\
      & $\kkms$          & $\kkms$             \\
(1)   &    (2)           &   (3)               \\
\noalign{\smallskip}      
\hline
\noalign{\smallskip} 
N 253 &     ...          & 34.64$\pm$0.45$^a$  \\
N 660 & 1.70$\pm$0.22    & 2.77$\pm$0.28       \\
N 891 & 0.41$\pm$0.07    & 0.57$\pm$0.24       \\
Maff2 & 4.41$\pm$0.41    &     ...             \\
N1055 & 0.40$\pm$0.14    &     ...             \\
N1068 &     ...          & 8.84$\pm$2.20$^{ab}$ \\  
N1365 & 2.80$\pm$0.35    &     ...             \\  
..I342  & 4.03$\pm$0.20    &     ...             \\
N1808 &     ...          & 5.43$\pm$1.09       \\  
N2146 &1.22$\pm$0.23$^c$ & 3.85$\pm$0.79       \\
N2623 & 0.90$\pm$0.11    &     ...             \\ 
N2903 & 0.58$\pm$0.08    &     ...             \\ 
N3034 &12.1$\pm$0.5$^c$  & 20.61$\pm$0.23$^a$   \\
N3079 & 4.98$\pm$0.29    & 2.90$\pm$0.30$^{bd}$ \\
N3628 & 3.77$\pm$0.23    &     ...             \\
N3690 &0.68$\pm$0.09$^c$ & 1.20$\pm$0.18$^e$    \\
N4102 & 1.09$\pm$0.20    &     ...             \\  
N5055 & 2.89$\pm$0.42    &     ...             \\      
N5194 &1.27$\pm$0.07$^c$ & 1.26$\pm$0.07$^a$   \\
N5236 & 3.36$\pm$0.22    & 4.01$\pm$0.08$^a$   \\      
N5775 &0.27$\pm$0.13$^c$ & 1.90$\pm$0.70       \\
N6240 &0.60$\pm$0.04$^c$ & 1.19$\pm$0.20$^{be}$ \\
N7469 &    ...           & 1.71$\pm$0.29$^{ab}$ \\
N7771 &    ...           & 1.10$\pm$0.18$^b$    \\
\noalign{\smallskip}     
\hline
\end{tabular}
}%
\end{center} 
References to measurements from the literature: a. Aladro
$\etal$ (2015); b. Costagliola $\etal$ (2011); d. Li $\etal$ (2019);
e. Jiang $\etal$ (2011).
\end{table}

\subsection{$IRAM$ 30m observations}

We used the $IRAM$ 30 m telescope on Pico de Veleta (Granada, Spain)
\footnote{$IRAM$ is supported by INSU/CNRS (France), MPG (Germany),
  and IGN (Spain). The observations enabling the research in this
  paper have received funding from the European Commission Seventh
  Framework Programme (FP/2007-2013) under grant agreement No 283393
  (RadioNet3)} in three observing runs (April 2008, November 2010, and
February 2011) in beam-switching mode with a throw of 4$\arcmin$. In
the 2008 run, we observed 13 galaxies with the IRAM 3-mm and 1.3-mm
SIS (ABCD) receivers coupled to 4-MHz back-ends, one line per
spectrum. In the 2010 and 2011 runs we used the EMIR receiver to
observe both the J=1-0 transitions of HCN, HCO$^{+}$, and HNC, and the
$J$=3-2 transitions of HCN and HCO$^{+}$ in single spectra and also
reobserved the weaker galaxies from the 2008 run. A sample of the
$IRAM$ observations is shown in Fig.\,\ref{iramprofiles}. Individual
spectra were reduced with the CLASS package; line parameters were
determined by fitting gaussians after baseline subtraction.
Intensities were converted to main-beam brightness temperatures using
main beam efficiencies $\eta_{\rm mb}$ of 0.82 at 89 GHz, 0.74 at 146
GHz, and 0.52 at 255 GHz.

\subsection{$JCMT$ 15 m observations}

The observations with the 15 m James Clerk Maxwell Telescope ($JCMT$)
on Mauna Kea (Hawaii) \footnote{Between 1987 and 2015, the James Clerk
  Maxwell Telescope (JCMT) was operated by the Joint Astronomy Centre
  on behalf of the Particle Physics and Astronomy Research Council of
  the United Kingdom, the Netherlands Organization for Scientific
  Research (until 2013), and the National Research Council of Canada.}
were obtained at various periods between 2010 and 2013, with a
beam-switch throw of 3 $\arcmin$, and mostly in queue or backup
service mode. The $J$=3-2 measurements were done with the A3 receiver,
and the $J$=4-3 measurements with the central pixel of the HARP array
receiver.  We used the FITS protocol to transport the data from SPECX
into CLASS. The effectively available band-pass was limited to 1050
$\kms$ in the $J$=3-2 transition and 800 $\kms$ in the $J$=4-3
transition. Where weak lines covered much of these ranges, we used the
high S/N CO profiles from Israel (2020) as a template to separate the
emission from the baseline. We determined integrated intensities by
fitting gaussians to the observed profiles after linear baseline
subtraction. Because of the limited free baseline, several of the
integrated $JCMT$ fluxes have larger uncertainties than those
determined with the $IRAM$ telescope.  A sample of the $JCMT$
observations is shown in Fig.\,\ref{jcmtprofiles}.  Antenna
temperatures were converted to main-beam brightness temperatures with
efficiencies $\eta_{\rm mb}(265)$ = 0.69 and $\eta_{\rm mb}(355)$ =
0.63.

\begin{table*}
  \begin{center}
  \caption[]{\label{norm}Galaxy centers: molecular line intensities
    $\int(T_{mb}dV)$ ($\kkms$) normalized to 22'' beams (see text)}
{\small %
\begin{tabular}{l|rrrr|rr|rrr} 
\noalign{\smallskip}     
\hline
\noalign{\smallskip}
NGC   &\multicolumn{4}{c}{HCN}&\multicolumn{2}{c}{HNC}&\multicolumn{3}{c}{HCO$^+$} \\
      &(1-0)&(2-1)$^a$& (3-2) & (4-3) & (1-0) & (3-2) & (1-0) & (3-2) & (4-3) \\
(1)   & (2)    & (3)  & (4)   & (5)   & (6    & (7)   & (8)   & (9)   & (10)) \\
\noalign{\smallskip}     
\hline
\noalign{\smallskip}
N0253 &  87.60 &  ... & 70.12 & 44.78 & 45.44 & 31.15 & 74.11 & 54.93 & 43.71 \\
N0520 &   2.85 &  ... &  1.7  &    ...&  1.74 &    ...&  4.15 &  2.20 &    ...\\
N0660 &   9.29 &  6.4 &  3.99 &  2.15 &  4.22 &  1.35 &  7.29 &  4.58 &    ...\\
N0891 &   1.96 &  ... &  1.61 &    ...&  0.66 &    ...&  2.01 &  1.4  &    ...\\
N0972 &   1.97 &  ... &  1.5  &    ...&  0.9  &    ...&  2.29 &  0.5  &    ...\\
N1055 &   2.82 &  ... &  1.4  &    ...&  0.95 &    ...&  1.68 &  0.4  &    ...\\
MAF2  &  23.25 & 12.0 &  7.54 &  2.40 & 11.02 &  3.51 & 14.31 &  6.10 &  2.19 \\
N1068 &  32.75 & 14.9 & 13.44 &  8.92 & 12.18 &  3.84 & 19.39 &  5.05 &  3.85 \\
N1084 &   0.85 &  ... &  1.7  &    ...&  0.3  &    ...&  0.74 &  1.2  &    ...\\
N1097 &  11.17 &  ... &  5.50 &    ...&  3.67 &    ...&  8.00 &  4.20 &    ...\\
IC342 &  14.69 &  ... &  2.93 &  1.48 &  6.46 &  1.22 & 11.44 &  4.60 &  2.88 \\
N1365 &  18.60 &  ... &  8.68 &    ...&  9.48 &  3.39 & 15.33 &  7.40 &    ...\\
N2146 &   6.12 &  4.3 &  4.28 &  4.42 &  2.44 &  2.32 &  8.78 &  4.70 &    ...\\
N2559 &   3.55 &  ... &  2.30 &    ...&  1.24 &    ...&  3.80 &   ... &    ...\\
N2623 &   2.46 &  ... &  1.77 &  1.2  &    ...&  1.78 &  2.60 &  4.00 &    ...\\
N2903 &   4.19 &  ... &  1.70 &    ...&  1.54 &    ...&  2.26 &  1.50 &    ...\\
N3034 &  35.67 & 14.2 & 11.50 &  4.58 & 14.76 &  3.1  & 49.45 & 25.90 & 16.49 \\
N3079 &   9.46 &  ... &  5.77 &  1.92 &  2.7  &  2.6  &  8.18 &  4.97 &  1.91 \\
N3310 &   0.38 &  ... &  1.3  &    ...&  0.4  &    ...&  0.84 &  0.7  &    ...\\
N3627 &   2.79 &  ... &  1.90 &    ...&  0.8  &  0.5  &  2.93 &   ... &    ...\\
N3628 &   7.60 &  ... &  3.46 &  2.08 &  4.83 &    ...&  7.95 &  4.9  &    ...\\
N3690 &   3.29 &  ... &  0.89 &  0.3  &  1.18 &    ...&  4.74 &  3.00 &    ...\\
N4030 &   4.42 &  ... &  1.26 &  1.20 &  2.30 &    ...&  2.25 &  0.6  &    ...\\
N4038 &   3.37 &  ... &  0.96 &  1.1  &  1.48 &    ...&  4.62 &  1.42 &    ...\\
N4102 &   4.80 &  ... &  1.77 &    ...&  3.60 &    ...&  2.92 &  1.1  &    ...\\
N4321 &   6.16 &  ... &  2.08 &  1.4  &  2.05 &    ...&  3.99 &  1.5  &  1.7  \\
N4414 &   4.63 &  ... &  1.0  &    ...&  1.27 &    ...&  3.30 &   ... &    ...\\
N4527 &   4.09 &  ... &  1.23 &    ...&  1.85 &    ...&  3.68 &  1.30 &    ...\\
N4569 &   4.85 &  1.0 &  0.59 &  0.3  &  1.10 &    ...&  3.78 &  0.70 &    ...\\
N4666 &   2.98 &  ... &  0.47 &  0.62 &  1.85 &    ...&  1.93 &  0.2  &    ...\\
N4826 &   8.99 &  2.5 &  1.98 &  1.3  &  3.79 &  1.3  &  5.58 &  0.7  &    ...\\
N5033 &   1.28 &  ... &  1.2  &    ...&  1.07 &    ...&  0.95 &  1.90 &    ...\\
N5055 &   3.19 &  ... &  0.77 &  0.4  &  0.9  &    ...&  1.76 &   ... &    ...\\
N5194 &   9.51 &  1.4 &  1.23 &  1.0  &  3.64 &  0.4  &  4.47 &  0.90 &    ...\\
N5236 &  10.50 &  ... &  3.50 &    ...&  5.29 &    ...& 12.79 &   ... &  2.52 \\
N5775 &   1.64 &  ... &  ...  &    ...&  0.3  &  1.3  &  1.53 &   ... &    ...\\
N6240 &   5.62 &  5.6 &  4.21 &  3.2  &  1.15 &    ...&  9.43 &  5.25 &    ...\\
N6701 &   2.32 &  ... &  0.91 &  1.0  &  0.96 &    ...&  2.37 &  3.70 &    ...\\
N6946 &  12.97 &  5.3 &  3.18 &  0.83 &  5.24 &  1.6  & 11.13 &  2.57 &    ...\\
N6951 &   3.56 &  3.1 &  1.00 &    ...&  2.00 &  0.6  &  2.59 &  1.16 &    ...\\
N7469 &   3.28 &  ... &  1.88 &  0.6  &  1.69 &  1.0  &  3.43 &  1.41 &    ...\\
N7771 &   8.08 &  ... &  0.95 &    ...&  3.49 &  1.0  &  8.44 &  2.61 &    ...\\
A220  &   9.79 &  ... & 11.54 &  3.01 &  9.36 &   ... &  4.52 &  3.20 &    ...\\
\noalign{\smallskip}    
  \hline
\end{tabular}
}%
\end{center}
Note: $^a$ Based on IRAM measurements by Krips $\etal$ (2008)
\end{table*}

\begin{figure*}[h!]
  \begin{minipage}{16cm}
  \end{minipage}  
\vspace{-2cm}
\begin{minipage}{4.5cm}
\resizebox{4.88cm}{!}{\rotatebox{0}{\includegraphics*{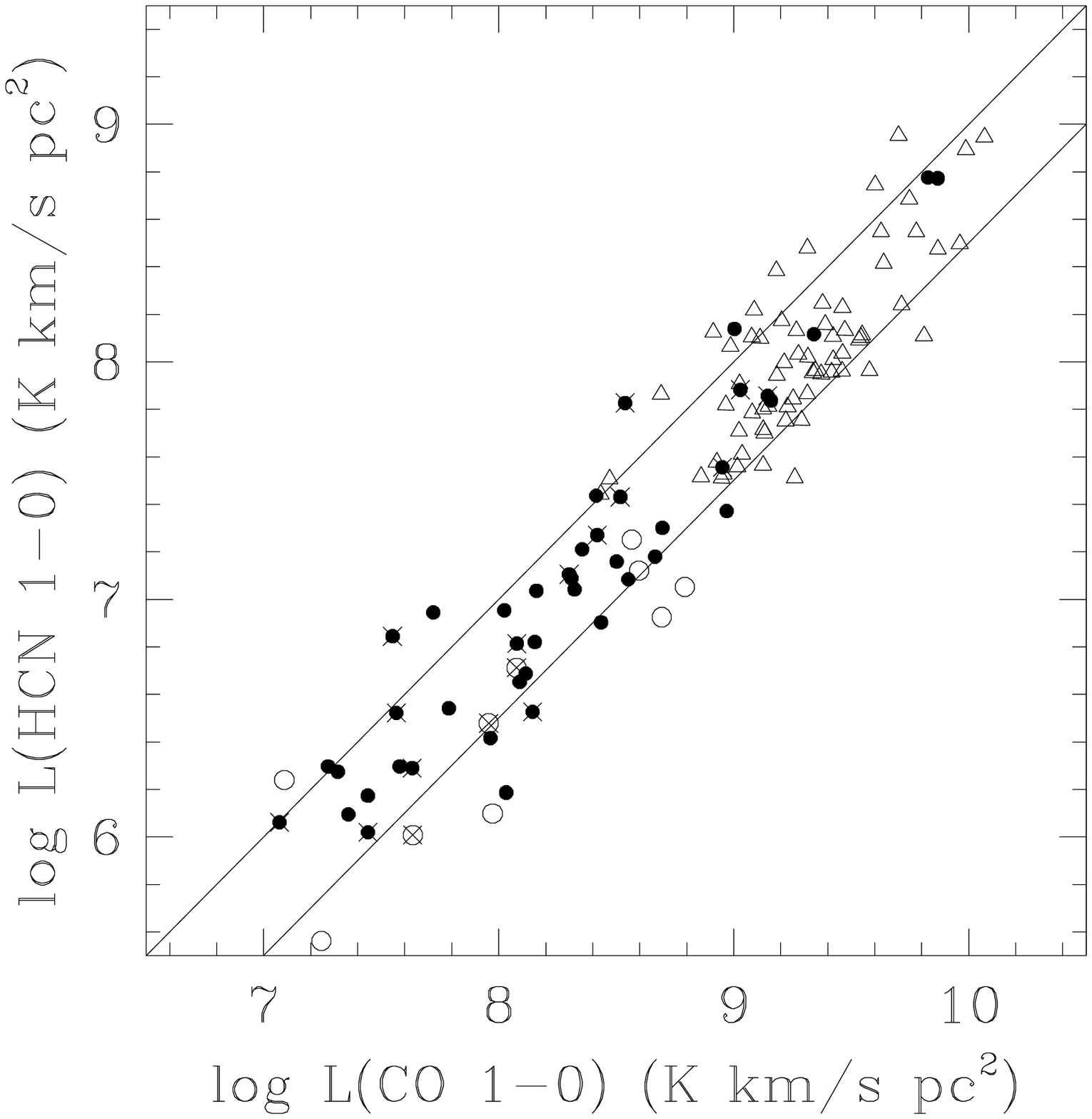}}}
\end{minipage}
\begin{minipage}{4.5cm}
\resizebox{4.88cm}{!}{\rotatebox{0}{\includegraphics*{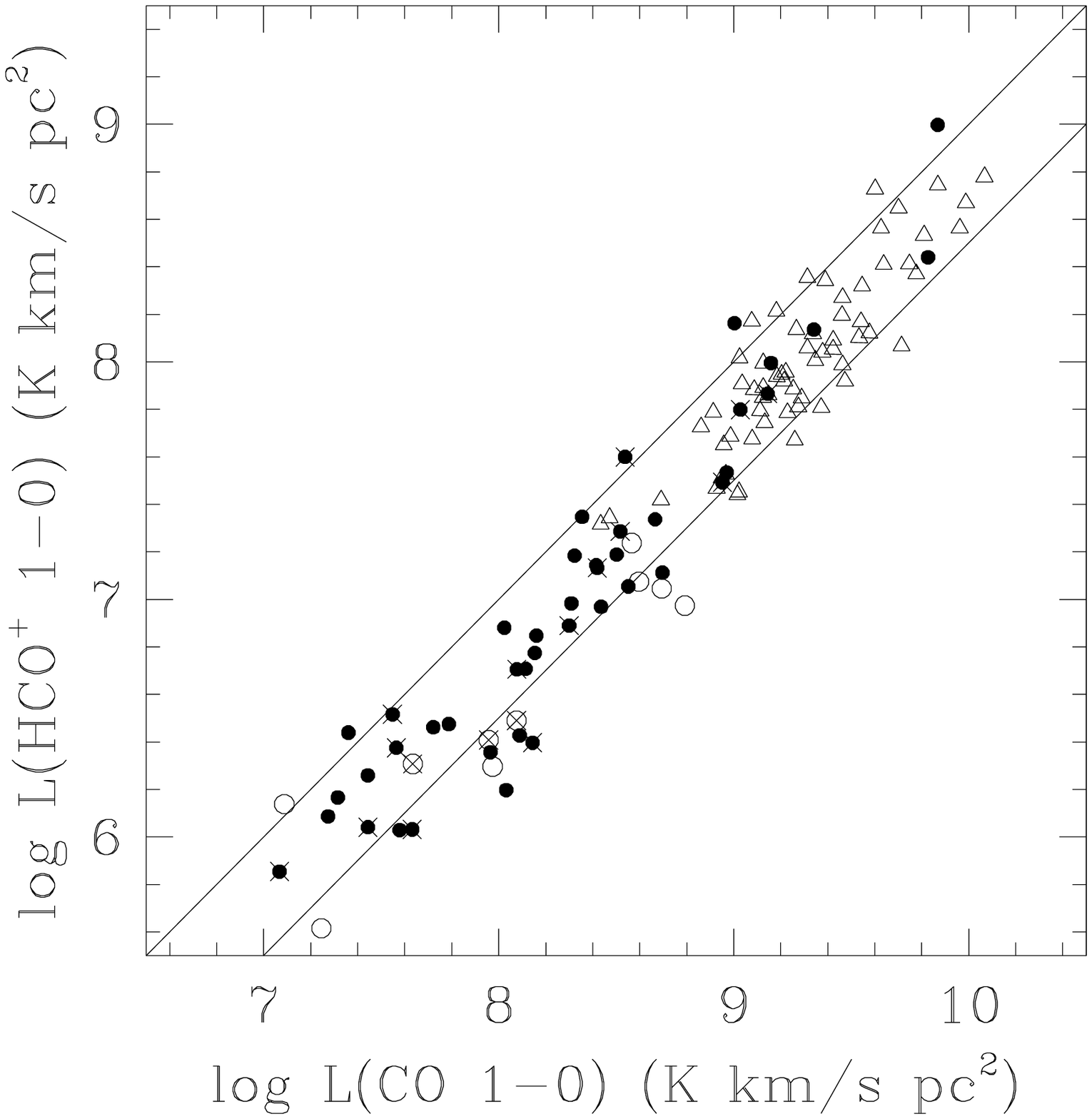}}}
\end{minipage}
\begin{minipage}{4.5cm}
\resizebox{4.88cm}{!}{\rotatebox{0}{\includegraphics*{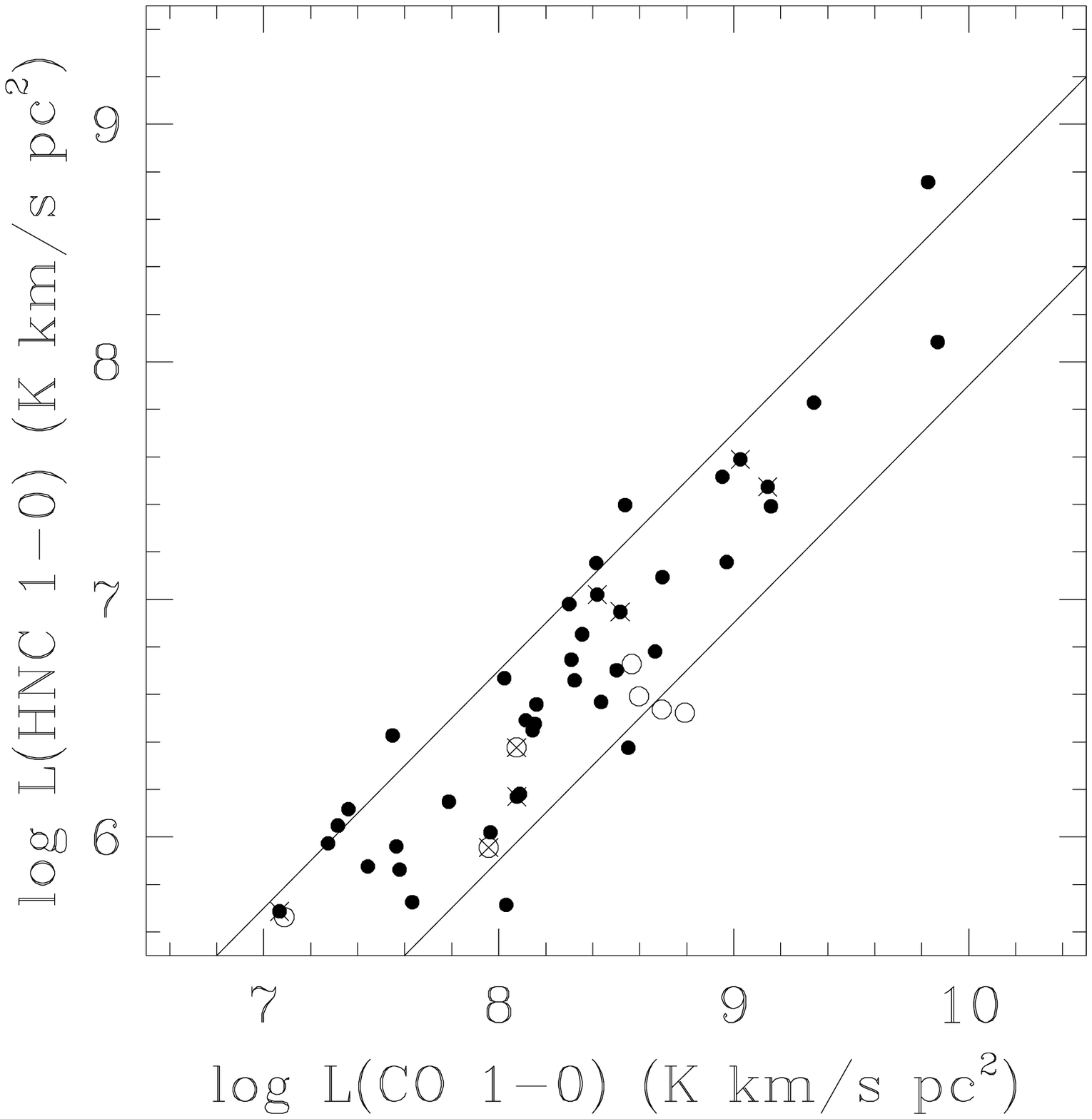}}}
\end{minipage}
\begin{minipage}{4.5cm}
\resizebox{4.88cm}{!}{\rotatebox{0}{\includegraphics*{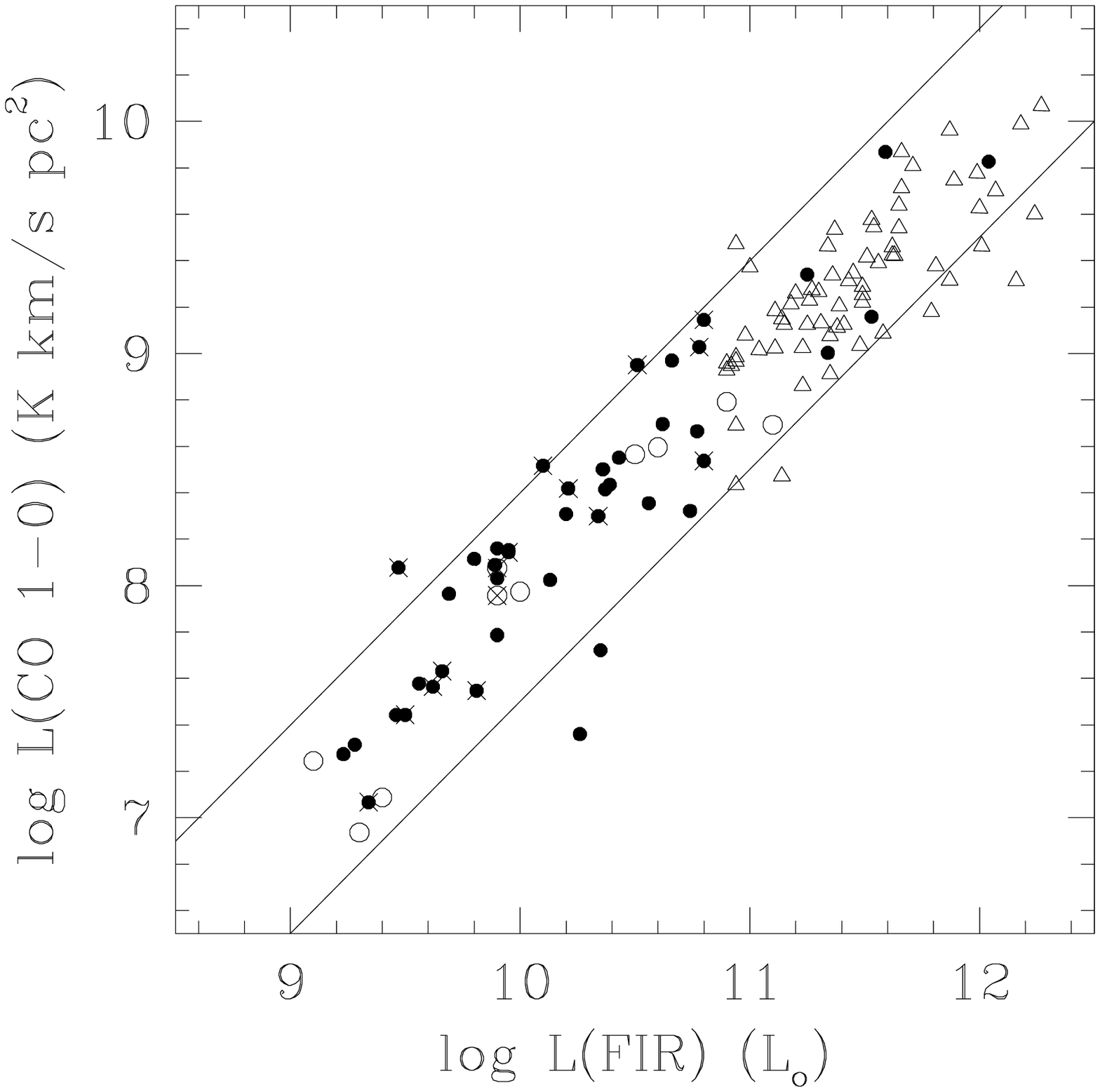}}}
\end{minipage}
\begin{minipage}{4.5cm}
\resizebox{4.88cm}{!}{\rotatebox{0}{\includegraphics*{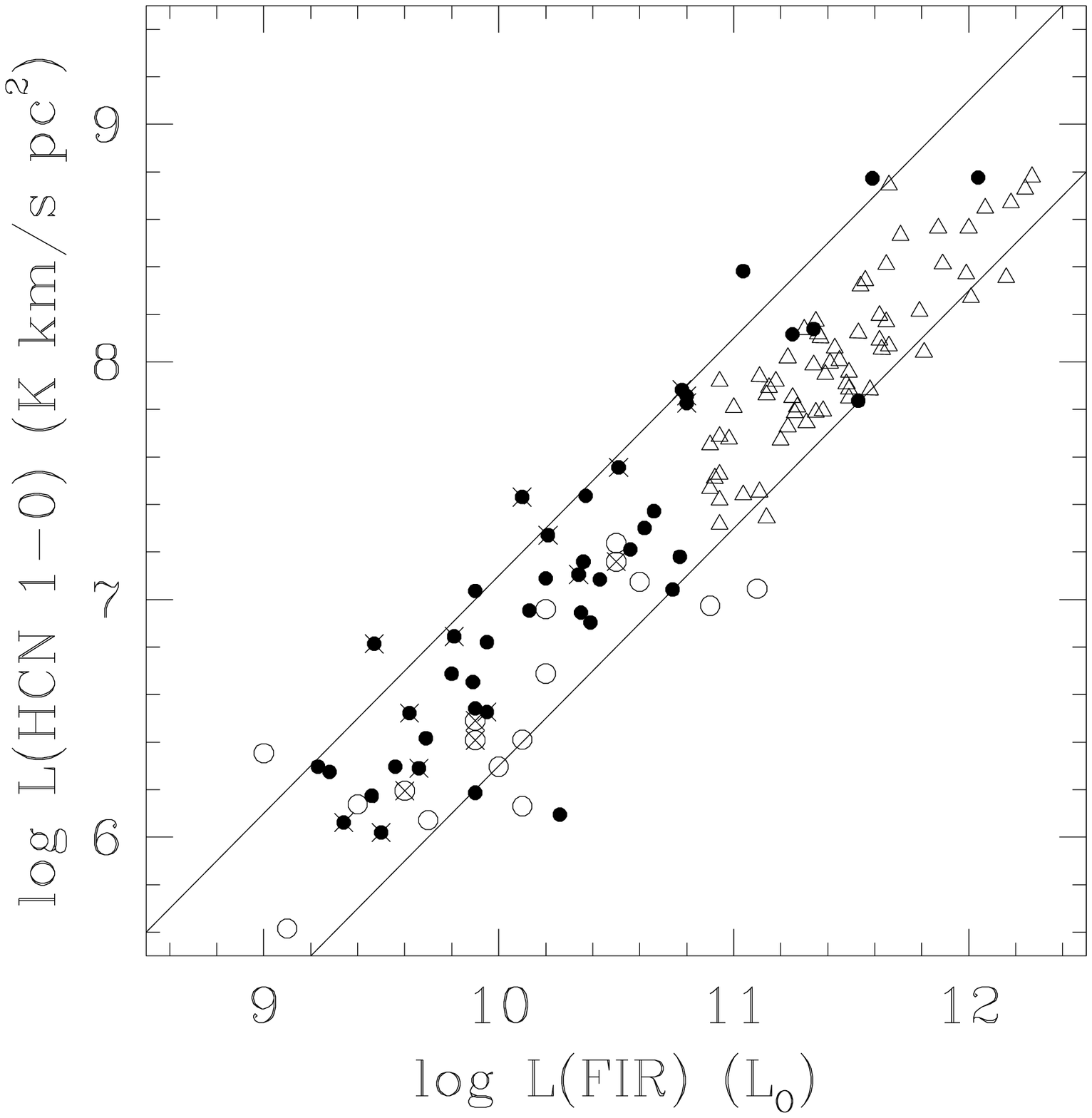}}}
\end{minipage}
\begin{minipage}{4.5cm}
\resizebox{4.88cm}{!}{\rotatebox{0}{\includegraphics*{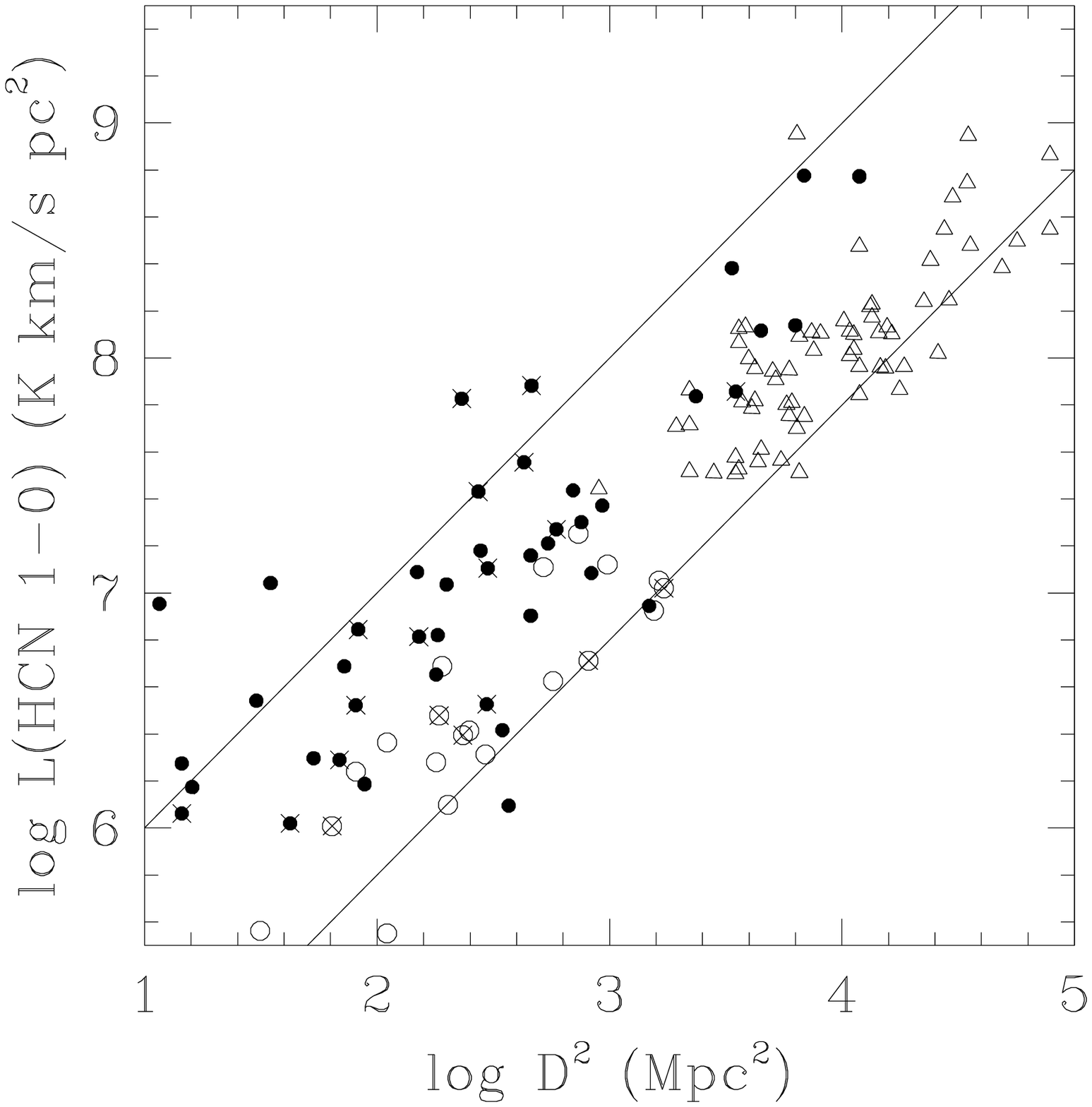}}}
\end{minipage}
\begin{minipage}{4.5cm}
\resizebox{4.88cm}{!}{\rotatebox{0}{\includegraphics*{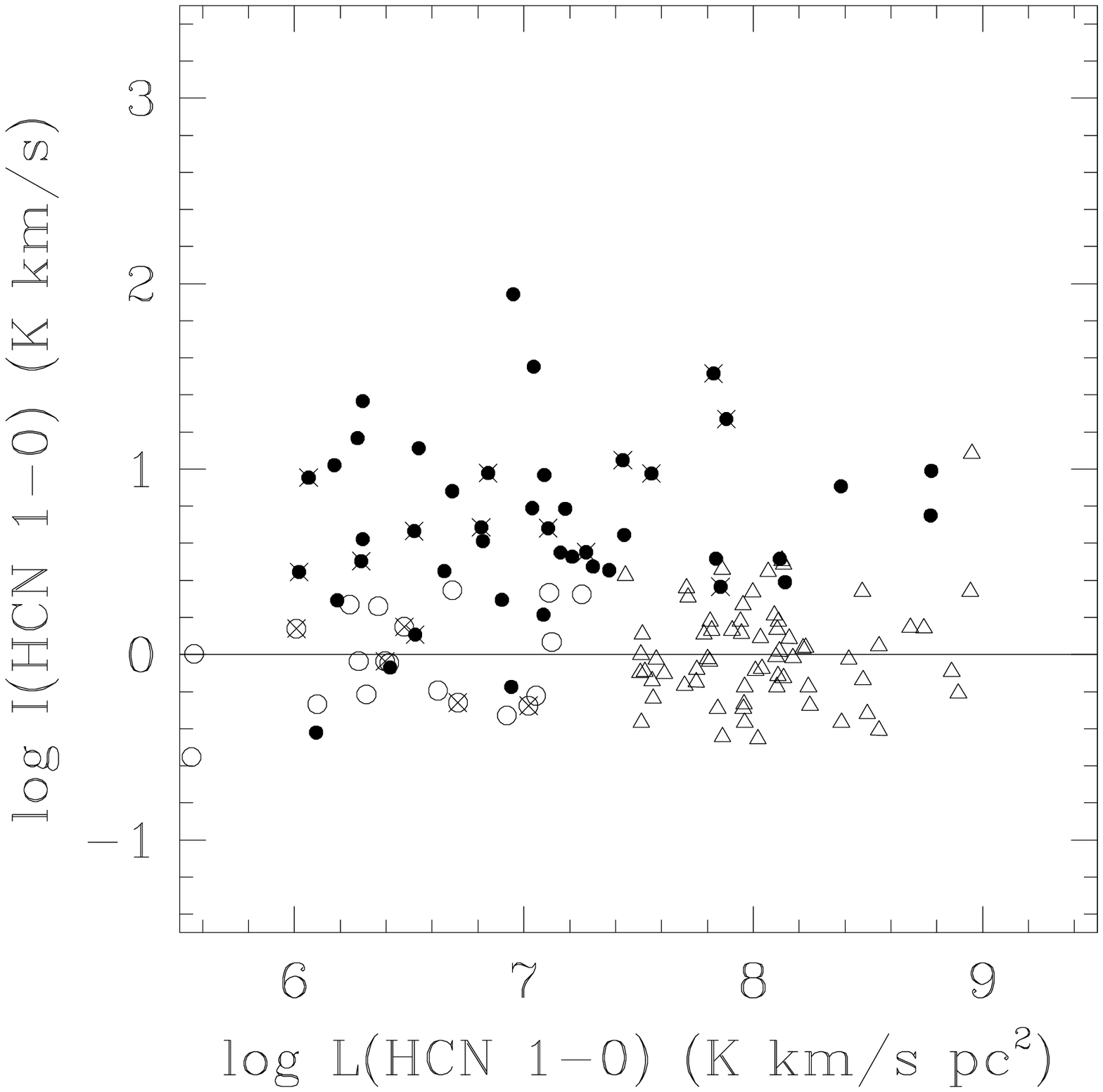}}}
\end{minipage}
\begin{minipage}{4.5cm}
\resizebox{4.88cm}{!}{\rotatebox{0}{\includegraphics*{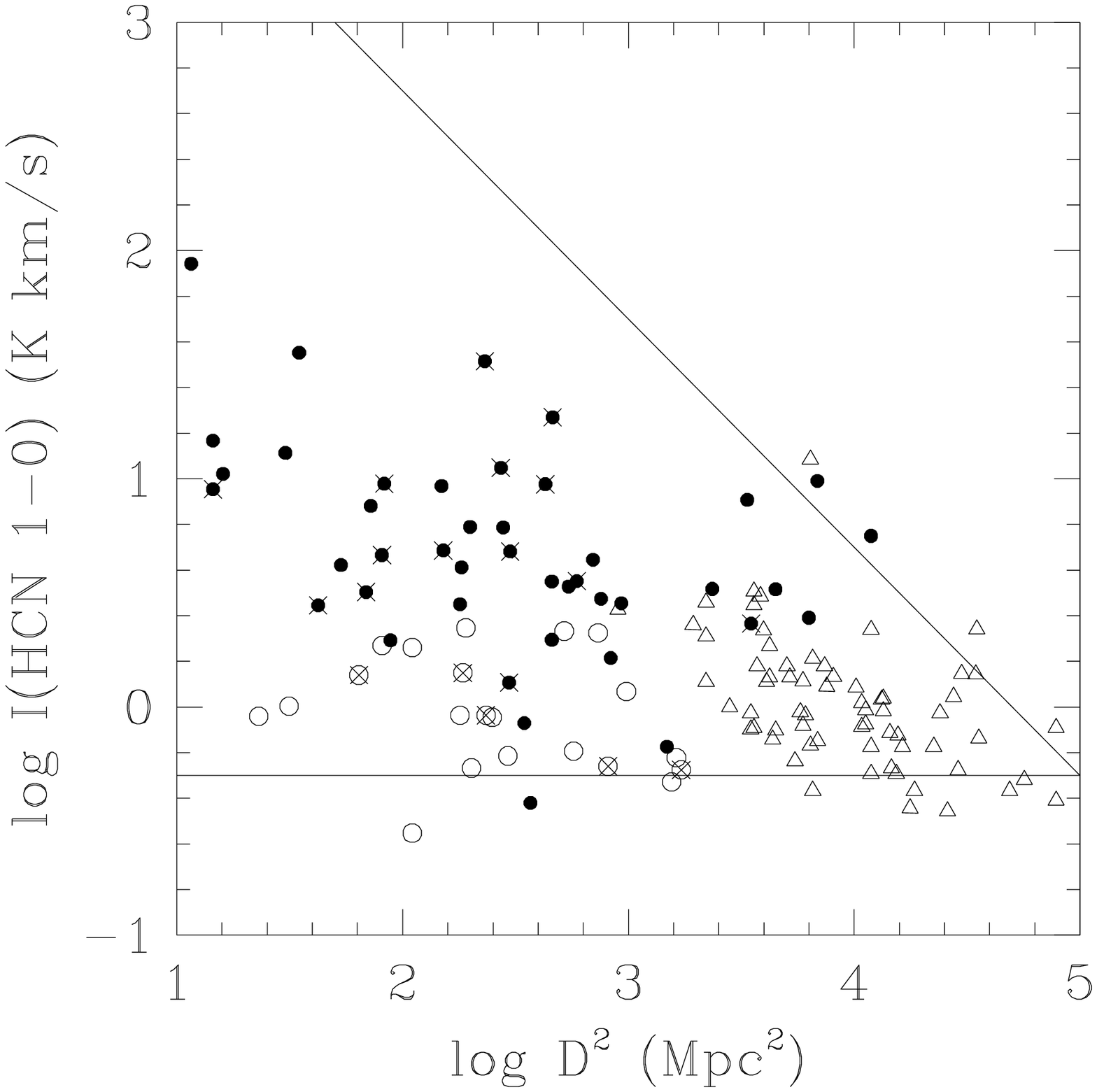}}}
\end{minipage}
\caption[] {\label{lummol} Top row: galaxy center luminosity-luminosity
  relations in identical $22"$ beams. Leftmost panels: $J$=1-0 HCN and
  $\hco$ versus $\co$ with constant ratios $0.10$ and $0.03$ marked by
  solid lines. Rightmost panels: HNC(1-0) versus $\co$ with constant
  ratios $0.050$ and $0.008$, and central $\co$ versus global far-infrared
  (FIR) continuum with constant ratios $2.5\times10^{-2}$ and
  $3.2\times10^{-3}$ K km/s pc$^2$/L$_O$  marked.
  Bottom row, leftmost panel: central HCN(1-0) versus global
  FIR continuum with constant ratios of $2\times10^{-3}$ and
  $1.25\times10^{-4}$ marked, very similar to the HCN/CO (top left)
  and HCN/FIR (top right) diagrams. Center left panel: relation
  between HCN luminosity ($L$(HCN)$\propto\,$I(HCN)$\times$D$^2$) and
  distance ($D^2$).  Center right panel: relation between luminosity
  $L$(HCN) and intensity $I$(HCN); vertical line marks constant
  surface brightness of 1 K km/s.  Rightmost panel: relation between
  $I$(HCN) and distance ($D^2$), with a horizontal line of resolved
  surface brightness of 0.5 K km/s and a diagonal line of unresolved
  point-like emission (cf. Table 1). In these and following diagrams, 
  filled circles denote the new data from this paper, open circles literature
  data on similar galaxies, crosses galaxies with an identified AGN
  and  triangles literature survey data on (U)LIRGs. \\
  The results of least-squares regression fits, of the form log(y) = a
  log(x) + b, to the data in this figure and the following figures are
  given in Table\,\ref{fittable}.
} 
\end{figure*}

\begin{table*}[h!]
\caption[]{\label{rats}Galaxy centers: molecular line ratios} 
\begin{center}
{\small %
\begin{tabular}{l|rrrrr|rr|r|rrr|r|rr} 
\noalign{\smallskip}     
\hline
\noalign{\smallskip}
NGC &\multicolumn{5}{c}{HCO$^+$/HCN}&\multicolumn{2}{c}{HNC/HCN}&HCN/CO&\multicolumn{3}{c}{HCN}&HNC&\multicolumn{2}{c}{HCO$^+$}\\
     & 1-0  &2-1$^a$& 3-2  & 3-2  & 4-3     & 1-0  & 3-2   &1-0/1-0 &2-1/1-0$^b$&3-2/1-0&4-3/1-0&3-2/1-0 &3-2/1-0&4-3/1-0\\
     &27.6''& 34.7''&9.7''&19.0''&13.7''   &27.6''& 19''  & 22''  & 22''  & 22''  & 22''  & 22''   & 22'' &22'' \\
(1)  & (2)  & (3)  & (4)  & (5)     & (6   & (7)   & (8)   & (9)   & (10)  & (11)  & (12)   & (13) & (14) & 15  \\
\noalign{\smallskip}     
\hline
\noalign{\smallskip}
253   & 0.85 & ...  & 0.93 & 0.81 & 1.21$^c$ & 0.52 & 0.45 & 0.085 &...   & 0.80 & 0.51 & 0.69 & 0.74 & 0.59 \\
520   & 1.46 & ...  & ...  & 1.27 &...       & 0.61 &...   & 0.025 &...   & 0.60 & 0.53 &...   &...   &...   \\
660   & 0.78 & ...  & 0.63 & 1.03 & 1.04$^d$ & 0.45 & 0.34 & 0.060 & 0.69 & 0.43 & 0.23 & 0.32 & 0.63 &...   \\
891   & 1.02 & ...  & 0.68 & 0.82 &...       & 0.34 &...   & 0.014 &...   & 0.82 &...   &...   & 0.69 &...   \\
972   & 1.17 & ...  & ...  & 0.4: &...       & 0.46 &...   & 0.029 &...   & 0.76 &...   &...   & 0.18 &...   \\
1055  & 0.60 & ...  & ...  & 0.3: &...       & 0.34 &...   & 0.037 &...   & 0.50 &...   &...   & 0.24 &...   \\
Maf2  & 0.62 & ...  & 0.77 &...   & 0.92     & 0.47 & 0.47 & 0.106 & 0.52 & 0.32 & 0.10 & 0.32 & 0.43 &...   \\
1068 *& 0.59 & 0.74 & 0.40 & 0.38 &0.46$^{cd}$& 0.40 & 0.29 & 0.195 & 0.46 & 0.41 & 0.27 & 0.32 & 0.26 & 0.20 \\
1084  & 0.87 & ...  & ...  & 0.68 &...       & 0.39 & 0.2: & 0.028 &...   &...   &...   &...   &...   &...   \\
1097 *& 0.72 & ...  & ...  & 0.75 &...       & 0.33 &...   & 0.082 &...   & 0.49 &...   &...   & 0.53 &...   \\
I342  & 0.78 & ...  & 0.90 & 1.43 & 1.41$^c$ & 0.59 & 0.42 & 0.091 & 0.20 & 0.10 & 0.40 & 0.14 & 0.25 &...   \\
1365 *& 0.82 & 0.74 & 0.81 &...   &1.11$^{cd}$& 0.51 & 0.39 & 0.072 &...   & 0.47 &...   & 0.36 & 0.48 &...   \\
1808 *& 0.94 & 1.04 & ...  &...   &0.96$^{cd}$& 0.41 &...   &   ... &...   &...   &...   &...   &...   &...   \\
2146  & 1.44 & ...  & 1.18 &...   &...       & 0.40 & 0.5: & 0.033 & 0.70 & 0.70 & 0.72 &...   & 0.54 &...   \\
2559  & 1.07 & ...  & ...  &...   &...       & 0.35 &...   & 0.046 &...   & 0.65 &...   &...   &...   &...   \\
2623  & 1.06 & ...  & 1.54 &...   &...       &...   & 0.99 & 0.137 &...   & 0.72 & 0.48 &...   & 1.5  &...   \\
2903  & 0.54 & ...  & ...  & 0.86 &...       & 0.37 &...   & 0.052 &...   & 0.41 &...   &...   & 0.66 &...   \\
3034  & 1.39 & ...  & 1.26 & 2.03 & 2.72$^c$ & 0.41 & 0.27 & 0.052 & 0.40 & 0.32 & 0.13 & 0.21 & 0.52 & 0.33 \\
3079 *& 0.86 & ...  & 1.03 & 0.89 & 1.10     & 0.91 & 0.2: & 0.040 &...   & 0.61 & 0.20 &...   & 0.61 & 0.23 \\
3627 *& 1.05 & ...  & ...  &...   & 0.88$^d$ & 0.28 & 0.3: & 0.038 &...   & 0.68 &...   &...   &...   &...   \\
3628  & 1.05 & 1.34 & 1.44 &...   & 2.78$^d$ & 0.64 &...   & 0.037 &...   & 0.46 & 0.27 &...   & 0.62 &...   \\
3690  & 1.44 & ...  & 2.99 &...   &...       & 0.36 &...   & 0.048 &...   & 0.27 & 0.10 &...   & 0.63 &...   \\
4030  & 0.51 & ...  & 0.45 &...   &...       & 0.53 &...   & 0.105 &...   & 0.29 & 0.27 &...   & 0.27 &...   \\
4038  & 1.37 & ...  & 1.68 & 1.52 &...       & 0.44 &...   & 0.072 &...   & 0.29 & 0.34 &...   & 0.31 &...   \\
4102 *& 0.61 & ...  & 0.55 &...   &...       & 0.75 &...   & 0.064 &...   & 0.37 &...   &...   & 0.38 &...   \\
4321  & 0.65 & ...  & 0.69 &...   &...       & 0.33 &...   & 0.075 &...   & 0.34 & 0.23 &...   & 0.38 &...   \\
4414 *& 0.71 & ...  & ...  &...   &...       & 0.27 &...   & 0.091 &...   & 0.22 &...   &...   &...   &...   \\
4527  & 0.90 & ...  & 0.98 &...   &...       & 0.45 &...   & 0.046 &...   & 0.30 &...   &...   & 0.35 &...   \\
4569 *& 0.78 & ...  & 0.89 &...   &...       & 0.23 &...   & 0.054 & 0.21 & 0.12 & 0.06 &...   & 0.19 &...   \\
4666  & 0.65 & ...  & 0.3: &...   &...       & 0.62 &...   & 0.040 &...   & 0.16 & 0.21 &...   & 0.10 &...   \\
4826 *& 0.62 & ...  & 0.75 &...   &...       & 0.42 & 0.68 & 0.099 & 0.28 & 0.22 & 0.15 & 0.36 & 0.13 &...   \\
5033 *& 0.74 & ...  & ...  & 0.5: &...       & 0.84 &...   & 0.024 &...   & 0.94 &...   &...   & 0.52 &...   \\
5055 *& 0.55 & ...  & ...  &...   &...       & 0.27 &...   & 0.046 &...   & 0.24 & 0.12 &...   &...   &...   \\
5194 *& 0.47 & ...  & 0.6: & ...  &...       & 0.38 & 0.4: & 0.198 & 0.15 & 0.13 & 0.10 &...   & 0.20 &...   \\
5236  & 1.22 & ...  & ...  &...   & 1.14$^c$ & 0.50 & 0.34 & 0.054 &...   & 0.33 &...   &...   &...   & 0.16 \\
5775  & 0.94 & ...  & 0.5: &...   &...       & 0.19 &...   & 0.032 &...   &...   &...   &...   &...   &...   \\
6240  & 1.68 & 1.36 & 0.92 & 0.4: & 1.65$^d$ & 0.21 &...   & 0.080 & 1.00 & 0.75 & 0.58 &...   & 0.14 &...   \\
6701 *& 1.02 & ...  & 0.95 & 0.76 &...       & 0.41 &...   & 0.052 &...   & 0.39 & 0.44 & 0.44 & 0.67 &...   \\
6946  & 0.85 & ...  & 0.93 & 0.83 & 3.21$^d$ & 0.40 & 0.51 & 0.057 & 0.41 & 0.25 & 0.06 & 0.31 & 0.23 &...   \\
6951 *& 0.74 & ...  & 1.13 & 1.15 &...       & 0.56 & 0.59 & 0.071 & 0.87 & 0.28 &...   & 0.30 & 0.45 &...   \\
7469 *& 1.05 & ...  & 0.75 & 0.75 &...       & 0.52 & 0.5: & 0.060 &...   & 0.57 & 0.18 &...   & 0.41 &...   \\
7771  & 1.05 & ...  & 0.60 & 2.08 &...       & 0.43 & 1.02 & 0.081 &...   & 0.12 &...   & 0.28 & 0.31 &...   \\
A220  & 0.46 & ...  & 0.25 &...   & 0.76     & 0.96 &...   & 0.089 &...   & 1.18 & 0.31 &...   & 0.71 & 0.21 \\
\noalign{\smallskip}     
\hline
\noalign{\smallskip}
MW$^e$& 0.60 & ...  & ...  & ...  & ...      & 0.30 & ...  & 0.10  & ...  &  ...  & ... &  ... & ...  & ...   \\   
\noalign{\smallskip}     
\hline
\noalign{\smallskip}
Ave  & 0.90 & 1.1   & 0.95 & 1.06 & 1.42     & 0.46 & 0.52 & 0.046 & 0.49 & 0.41 & 0.27 & 0.34 & 0.42 & 0.29 \\
&$\pm$0.05&$\pm$0.2 &$\pm$0.10&$\pm$0.11&$\pm$0.21&$\pm$0.03&$\pm$0.07&$\pm$0.006&$\pm$0.08&$\pm$0.04&$\pm$0.04&$\pm$0.05&$\pm$0.03&$\pm$0.07\\
MD$^f$& 0.30 & 0.40 & 0.52 & 0.47 & 0.82     & 0.17 & 0.24 & 0.039 & 0.27 & 0.25 & 0.18 & 0.16 & 0.19 & 0.16 \\
Err$^g$&0.05  & 0.15 & 0.15 & 0.2  & 0.3      & 0.05 & 0.15 &  0.15 & 0.3  & 0.2  & 0.3  & 0.2  & 0.15 & 0.3 \\ 
\noalign{\smallskip}    
\hline                          
\end{tabular}
} %
\end{center}
Notes:
a. Based on $APEX$ 12m observations by Zhou $\etal$ (2022).
b. Based on $IRAM$ observatioins(HPBW $14"$) by Krips $\etal$ (2008).
c. Includes $JCMT$ observations by Tan $\etal$ (2018). 
d. From $APEX$ 12m observations (HPBW $18"$) by F.P. Israel
(unpublished) and by Zhang $\etal$ (2014).
e. Milky Way Circumnuclear Molecular Zone values from $MOPRA$
22m observations (effective resolution $39"$) by Jones $\etal$ (2012).
f. Mean deviation of sample.
g. Logarithm of typical observational error of ratios plotted in
Figures\,\ref{lummol}-\ref{32Bratspec}; individual errors vary
depending on observed intensity (cf. Table 3).\\
\end{table*}

\begin{table}
\begin{center}
{\small %
\caption[]{\label{ratincrease} Line transition ratios as a function of CO luminosity}
\begin{tabular}{cllll}
\noalign{\smallskip}     
 \hline
\noalign{\smallskip} 
log $L$(CO)$^a$& \multicolumn{3}{c}{HCN$^b$}    & $\hco$ $^b$\\
$\kkms$   &(2-1)/(1-0)&(3-2)/(1-0)&(4-3)/(1-0)&(3-2)/(1-0)\\
(1)       &    (2)    &   (3)     &  (4)      & (5)       \\
\noalign{\smallskip}      
\hline
\noalign{\smallskip} 
   7.4    &  0.31 (5) & 0.25  (9) & 0.16  (6) &  0.32  (6) \\
   8.3    &  0.53 (5) & 0.41 (16) & 0.26  (9) &  0.37 (14) \\
   8.8    &   ...     & 0.68 (14) & 0.42  (4) &  0.45  (3) \\
   9.3    &   ...     & 0.35  (9) & 0.30  (4) &  0.45 (10) \\
   9.9    &   ...     & 0.67  (6) &  ...      &  0.61  (7) \\
7.0-9.0$^c$& 0.49 (12)& 0.40 (29) & 0.26 (19) &  0.37 (23) \\
7.0-10.0$^c$&   ...   & 0.43 (44) & 0.27 (24) &  0.43 (40) \\
\noalign{\smallskip}     
\hline
\end{tabular}
}%
\end{center} 
Notes: a. Center of luminosity interval considered; b. Number of
points in interval is given between parentheses; entries for intervals
with only one point are deleted; c. Range considered for overall
average.
\end{table}   

\section{Results}

\subsection{Galaxy sample}

The new survey results are listed as velocity-integrated line
intensities in Tables\,\ref{fluxes} and \ref{morefluxes}.  The headers
identify molecular species, line transition, telescope used, and the
resolution at the observing frequency.  Additional $IRAM$ and $JCMT$
line intensities from the published literature complementing the
survey results are included in the tabulation, with the appropriate
reference at the bottom.

The database of 46 observed galaxies is further expanded by
adding the ground-state HCN, $\hco$ and HNC intensities of galaxies
.not included in our observing program, extracted from published $IRAM$
30m surveys. First, we selected from the surveys by Costagliola
$\etal$ (2011), Jiang $\etal$ (2011) and Li $\etal$ (2021) an
additional 21 galaxies in which the HCN(1-0), $\hco$(1-0), and
HNC(1-0) lines were all well-detected, with associated $\co$ and
$\thirco$ data from Li $\etal$ (2015) and Israel (2020). There are no
comparable $IRAM$ 30m surveys covering the higher transitions.

Secondly, we considered the luminous infrared galaxies specifically
targeted in large $IRAM$ 30m surveys by Graci\'a-Carpio $\etal$ 2006,
2008, Garc\'ia-Burillo $\etal$ 2012, Privon $\etal$ 2015, and
Herrero-Illana $\etal$ 2019).  We found 63 galaxies with well-measured
intensities of both HCN(1-0) and $\hco$(1-0), 27 of which are also
detected in HNC(1-0) and 10 in HCN(3-2) and $\hco$(3-2). About half of
these have corresponding $\co$ and $\thirco$ data, published by
Graci\'a-Carpio (2009), Li $\etal$ (2015), and Herrero-Illana $\etal$
(2019).

In total, there are well-established intensities of the three $J$=1-0
HCN, $\hco$ and HNC lines for 94 galaxies covering a large luminosity
range ($9.2 \leq log L_{FIR}\leq 12.3$). An additional 36 galaxies have
good HCN(1-0) and $\hco$(1-0) line intensities only. About 50 galaxies
have good HCN(3-2) and $\hco$(3-2) data, mostly provided by our new
survey that also contributes a smaller number of additional HNC(3-2),
HCN(4-3) and $\hco$(4-3) data.

\subsection{Normalized intensities}

Because the $J$=1-0 line frequencies of HCN, HNC and $\hco$ are very
close together, their intensities measured with the $IRAM$ 30m
telescope are directly comparable. This not true, however, for
measurements in other transitions, for measurements with a different
telescope aperture, and for measurements of species such as CO, which
are all observed at a different spatial resolution.  Molecular line
intensities observed at different resolutions must be reduced
(normalized) to the same resolution before they can be compared in a
meaningful way. The resolution of $22"$ is conveniently intermediate
between the resolutions of the $IRAM$ $J$=1-0 ($27"$) and $JCMT$
$J$=3-2 ($19"$) lines, and it is also the reference for the CO
transition ladders published by Israel (2020). Only minor
extrapolations are required to bring the observed $J$=1-0 and $J$=3-2
intensities to values corresponding to this resolution.

For $J$=1-0 HCN we used the intensities of 36 sample galaxies
measured in larger beams (NRAO $\theta\,=\,72"$, 25 galaxies; SEST
  $\theta\,=\,57"$, 14 galaxies; FCRAO $\theta\,=\,50"$, 6 galaxies;
  OSO $\theta\,=\,44"$, 4 galaxies) in combination with the IRAM
  measurements ($\theta\,=\,27.6"$) of the present survey to
  extrapolate the latter to the equivalent flux at $22"$. On average,
  the extrapolation is almost linear with the resolution, intermediate
  between the extremes expected for infinitely large (extended) and
  infinitely small (pointlike) sources. The average extrapolation
  factor $f_{10}\,=\, 1.2$ (corresponding to a power law extrapolation
  with index 0.8) was used for the remaining galaxies lacking
  observations at other resolutions. As the modest change of
resolution limits possible multiplication factors to
$1.00\leq\,f_{10}\,\leq1.57$, extrapolation uncertainties add
relatively little to the overall errors.

We applied the HCN(1-0) normalization factors to the $\hco$(1-0) and
HNC(1-0) intensities because these species have, unlike HCN, few
observations at other resolutions. The implicit assumption of
identical spatial emission distributions likewise does not introduce
more than small errors.

The paucity of published HCN, $\hco$, or HNC measurements in higher
transitions at other resolutions than presented here was the rationale
for observing $J$=3-2 HCN and $\hco$ transitions with both the JCMT
15m and $IRAM$ 30m facilities. Their resolutions differ by a factor of
two ($9.6"$ and $19"$), from which the normalized intensity at $22"$
is once again derived by a modest extrapolation, with correction
factors limited to $1.00\,\geq\,f_{32}\,\geq0.75$.  The actual average
reduction factors derived from the data in Table\,\ref{fluxes} are
$f_{HCN32}\,=\,0.91$ (33 galaxies) and $f_{HCO32}\,=\,0.94$ (14
galaxies). The difference is not significant and justifies the
assumption of very similar HCN and HCO$^{+}$ emission distributions.
No duplicate $J$3-2 data were obtained for the HNC lines but the
spatial distributions of HNC and HCN are even more likely to be
identical so that the same factors apply. Finally, we assumed the
resolution dependence of the $J$=4-3 HCN and $\hco$ intensities to be
the same as that of the $J$=3-2 lines and interpolated the latter to
obtain the normalization factors for the $J$=4-3 intensities, with
resulting averages $f_{HCN43}\,=\,0.65$ (24 galaxies) and
$f_{HCO4-3}\,=\,0.74$ (8 galaxies). In Table\,\ref{norm} all
intensities are normalized to resolutions of $22"$ by the appropriate
factors. This table also includes $IRAM$ $J$=2-1 HCN intensities
($14"$) from Krips $\etal$ 2008.

\subsection{Line ratios and transition ladders}

In columns 2 through 7 of Table\,\ref{rats}, intensities of $\hco$ and
HNC relative to HCN are listed. The resolutions change with each
transition, but they are near-identical for lines observed in the same
transition.  The $IRAM$ HCO$^{+}$/HCN and HNC/HCN intensity ratios are
each determined from lines in the same spectrum and are thus quite
accurate. The JCMT-derived line ratios are less accurate as they are
determined from separate single-line $JCMT$ observations made on
different occasions. The HCN/CO intensity ratios in Column 8 are based
on the normalized $J$=1-0 HCN intensities from Table\,\ref{norm} and
the $J$=1-0 CO intensities from Israel(2020) and refer to identical
beams of $22"$. The transition ladder ratios of HCN, HNC and HCO$^{+}$
in columns 9 through 14 are likewise derived from the normalized values
in Table\,\ref{norm}.

\section{Analysis}

This section identifies what observational patterns exist for the
various galaxies and in particular which luminosities and luminosity
ratios, if any, show a degree of correlation. Possible explanations
for the resulting findings and their meaning are discussed in the
following Section 5.

The $IRAM 30m$ line measurements of both HCN(1-0) and $\hco$(1-0) in
130 galaxies constitute a sample with a size and homogeneity that
renders it highly suitable to the large-scale investigation of the
molecular gas in galaxies of a diverse nature and wide range of
luminosities. Such investigations have been carried out before, first
by Gao $\&$ Solomon (2004b), by Baan $\etal$(2008), and most recently
by Li $\etal$ (2021) but there are major differences between their
surveys and ours. The Gao $\&$ Solomon data set is half the size, has
a poorer resolution ($NRAO$ 12m versus $IRAM$ 30m), contains only
ground-state HCN and CO transitions, but attempts to cover entire
galaxies instead of just central regions. The data set used by Baan
$\etal$ covers the HCN, $\hco$ and HNC ground state only, but includes
CN and CS lines, it is smaller, less complete, and combines data from
telescopes with different and generally larger beams making it very
inhomogeneous and less accurate. The Li $\etal$ $IRAM$ 30m data set
contains a similar number of detected galaxies (124 versus 130 with
both HCN and $\hco$, and 85 versus 94 with also HNC) but it also
includes only ground state transitions.

In the following, molecular line luminosities\footnote{Throughout this
  paper, line luminosities are defined as the product of line flux in
  K km/s and beam surface area in pc$^2$} and intensities all refer to
the normalized ($22"$) central aperture discussed above. The
far-infrared intensities and luminosities, however, refer to the
$IRAS$ aperture which is a factor of ~100 greater and should be
considered more representative for the galaxy as a whole.

\subsection{HCN(1-0) and $\hco$(1-0)  versus  FIR and CO(1-0) luminosities}

Comparison shows that the normalized central HCN(1-0), $\hco$(1-0), and 
HNC(1-0) luminosities are all linearly correlated with each other, as
well as the CO luminosity, with slopes corresponding to average ratios
HCN/CO = 0.046, $\hco$/CO = 0.041, and HNC/CO = 0.021
(Fig\,\ref{lummol}).  In each relation, the dispersion does not change
with luminosity. It exceeds the observational uncertainty and reflects
variation between individual galaxies. Solomon $\etal$ (1992) were the
first to draw attention to the tight relation between global HCN and
FIR luminosities, further elaborated by Gao $\&$ Solomon (2004a), and
rediscovered or confirmed by almost all later authors for HCN(1-0),
$\hco$(1-0), HNC(1-0), CS(3-2), and far-infrared luminosity in
matched beams.

The CO line and FIR continuum luminosities are also linearly related
(rightmost panel of Fig.\,\ref{lummol}) and it makes little difference
whether the other luminosities are plotted as a function of CO or of
FIR luminosity.  This is illustrated by the similarity of the CO-FIR,
HCN-CO and HCN-FIR diagrams in Fig.\,\ref{lummol} (top right, top left
and bottom left panels).

In the bottom row of Fig.\,\ref{lummol} we further investigate the
relations between the derived luminosity $L$, the observed surface
brightness $I$, and the assumed distance $D$. Measured in the same
normalized beam, luminosities vary with surface brightness and
distance ($L\,\propto\,I\times D^2$) only. The luminosity $L$(HCN) is
strongly correlated with distance $D^2$ (center left panel) and very
weakly or not at all with observed intensity $I$(HCN) (center right
panel). Thus, the luminosities are essentially determined by the
distance and not by the intensity of the observed galaxies.  The
intensity $I$(HCN) and distance $D^2$ are weakly anti-correlated
(rightmost panel). The surface brightness goes down with increasing
distance as the constant aperture covers an increasingly larger part
of the galaxy. As Gao $\&$ Solomon (2004b) already surmised, the HCN
surface brightness peaks at the nucleus and drops when it is averaged
over increasingly larger surface area. This is well-illustrated by HCN
line maps (Green $\etal$ 2016, Tan $\etal$ 2018, Jim\'enez-Donaire
$\etal$ 2019) that include several of the galaxies in our
sample. Because of the anticorrelation between $I$(HCN) and $D^2$, the
relation between luminosity $L$(HCN) and distance $D^2$ is modestly
sub-linear. The smaller departure from linearity, and the larger
dispersion in the $L$(CO)-$L$(FIR) and $L$(HCN)-$L$(FIR) diagrams is due
to the aperture mismatch between the FIR continuum and the molecular
line measurements. In matched apertures, the discrepancy disappears
(Li $\etal$ 2021). Nevertheless, comparison of the results by Gao $\&$
Solomon (2004b) and Li $\etal$ (2021) shows that the use of global
instead of central luminosities leads to similar results.

As long as the derived luminosities are dominated by the distance
factor $D^2$ and no large changes occur in the underlying line
intensities, the luminosity plots will show (almost) linear
relationships. As we will show, the line intensities in this paper
tend to be characterized by constant ratios. The observed linearity of
luminosity-versus-luminosity plots thus reflects geometric factors
unrelated to intrinsic galaxy properties.

\subsection{$J$=1-0 HCN/CO and FIR/CO as a function of luminosity}

Whereas absolute luminosities are dominated by galaxy distances and
shed little light on the physical conditions in the central regions of
galaxies, this may be different for luminosity\footnote{In identical
  apertures, there is no meaningful distinction between intensity
  ratios and luminosity ratios, and we will use both terms
  interchangeably} ratios.  First, we turn to relations between the
HCN(1-0)/CO(1-0) ratio and the CO, HCN, $\hco$, and FIR
luminosities. With some reservations, Gao $\&$ Solomon (2004a)
interpreted the ratio $L$(HCN)/$L$(CO) as an indicator for the
fraction of dense molecular gas in galaxies. They found it to be
rather low ($\leq0.06$) in low-luminosity normal spiral galaxies but
dramatically increasing to $L$(HCN)/$L$(CO)$\sim0.25$ in high
luminosity galaxies $L$(FIR)$\geq11$), all galaxies with
$L$(HCN)/$L$(CO)$\geq0.06$ being (ultra)luminous (Gao $\&$ Solomon
2004b). We do not confirm these findings. The more accurate HCN/CO
ratios at galaxy central positions shown in Fig.\,\ref{gaocomp} (top
left panel) do not exhibit such a jump at high values of $L$(FIR) and
show the same dispersion at low and high luminosities. The behaviour
of HCN/CO as a function of $L$(CO) is no different and this is also
true if we substitute $\hco$(1-0) for HCN(1-0). In
Fig.\,\ref{gaocomp}, there is no systematic correlation between the
HCN/CO ratio and either $L$(FIR) (top left) or $L$(CO) (top
right). Similar plots of infrared and molecular line measurements in
matching beams presented by Li $\etal$ (2021 likewise fail to reveal
correlations between the infrared luminosity and the HCN/IR or
$\hco$/IR ratios.  The center left panel of Fig\,\ref{gaocomp} shows
at best a weak correlation between $L$(HCN)/$L$(CO) and $L$(HCN), with
$L$(HCN)/$L$(CO)$\,\propto\,L$(HCN)$^{0.1}$. Substitution of $\hco$
for HCN yields the same result with a slightly lower dispersion.  A
direct comparison of the present results and those by e.g. Li $\etal$
(2021) with the results published by Gao $\&$ Solomon (2004a, b) is
not easy because the latter constructed spatially integrated
luminosities from heterogeneous data obtained at different
resolutions, at lower sensitivities in much narrower bands.  It is not
entirely clear how they constructed luminosity ratios such as those of
HCN to CO. In the combined database presented here, luminous galaxies
have on average significantly lower HCN/CO ratios than those listed by
Gao $\&$ Solomon (2004b), whereas many lower-luminosity galaxies have
significantly higher HCN/CO ratios.  The two data sets have 39
galaxies in common. For seven (18$\%$) galaxies Gao $\&$ Solomon list
lower HCN/CO ratios, but for twenty-four (62$\%$) they find
(significantly) higher ratios. It is noteworthy that almost all of
these refer to (luminous) galaxies with wide lines of low-amplitude,
filling most of the observed spectral window and leaving little room
for accurate baseline-fitting. We suspect that, as a result, Gao $\&$
Solomon have overestimated the corresponding HCN luminosities, and
that the more recently obtained data supersede these older values.

In a similar way, the ratio $L$(FIR)/$L$(CO) is often presented as a
proxy for the star formation efficiency (more correctly: the inverse
of the molecular gas depletion rate), under the assumption that
$L$(FIR) is proportional to the star formation rate. We find, however,
that the ratio $L$(FIR)/$L$(CO) is uncorrelated with the luminosities
$L$(CO), $L$(HCN) or $L(\hco)$, and only shows a weak correlation with
$L$(FIR) with a modest logarithmic slope $\sim$0.13
(Fig.\,\ref{gaocomp}, center right). It is not clear whether this is a
slow increase over the full range of log $L$(FIR) or a rise only at
values log $L$(FIR)$\geq11$ of the luminous galaxies.  Grac\'ia-Carpio
$\etal$ (2009) found a similar weak correlation between
$L$(FIR)/$L$(HCN) and $L$(FIR) with a logarithmic slope 0.24 for
(ultra)luminous galaxies. The larger sample studied here does not show
such a correlation between either $L$(FIR)/$L$(HCN) or
$L$(FIR)/$L$($\hco$) and $L$(FIR).

Gao $\&$ Solomon (2004b) found the `star formation efficiency'
$L$(FIR)/$L$(CO) to be strongly correlated with the `dense gas
fraction' $L$(HCN)/$L$(CO). We find a similar but weaker relation in
the bottom left panel of Fig.\,\ref{gaocomp}. Gao $\&$ Solomon did not
find a relation between $L$(FIR)/$L$(CO) and $L$(FIR)/$L$(HCN), but
our data reveal these quantities to be well-correlated
(Fig.\,\ref{gaocomp}, bottom right), even though $L$(FIR)/$L$(HCN) and
$L$(CO) are not. In these bottom panels there is no systematic
difference between normal galaxies, starburst galaxies and AGNs.

Neither $L$(HCN)/$L$(CO) nor $L$(FIR)/$L$(CO) is correlated with
distance $D$. Hence, they also do not depend on the extent of the
central area sampled.

\subsection{Dense and very dense molecular gas: $\hco$(1-0) and HCN(1-0)}

\begin{figure}
  \begin{minipage}{8.94cm}
  \end{minipage}  
\vspace{-2cm}
\begin{minipage}{8.94cm}
\begin{minipage}{4.4cm}
\resizebox{4.7cm}{!}{\rotatebox{0}{\includegraphics*{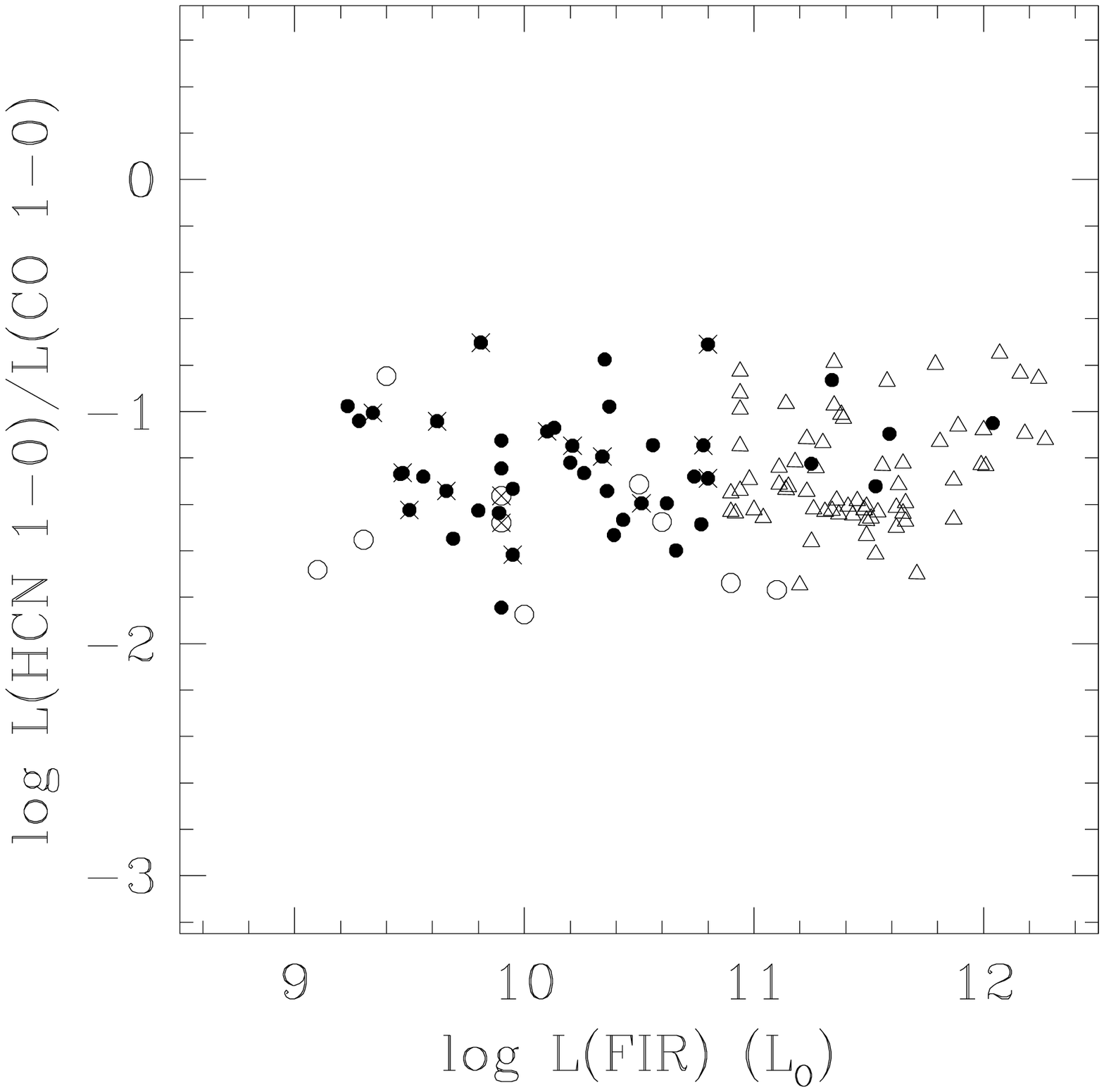}}}
\end{minipage}
\begin{minipage}{4.4cm}
\resizebox{4.7cm}{!}{\rotatebox{0}{\includegraphics*{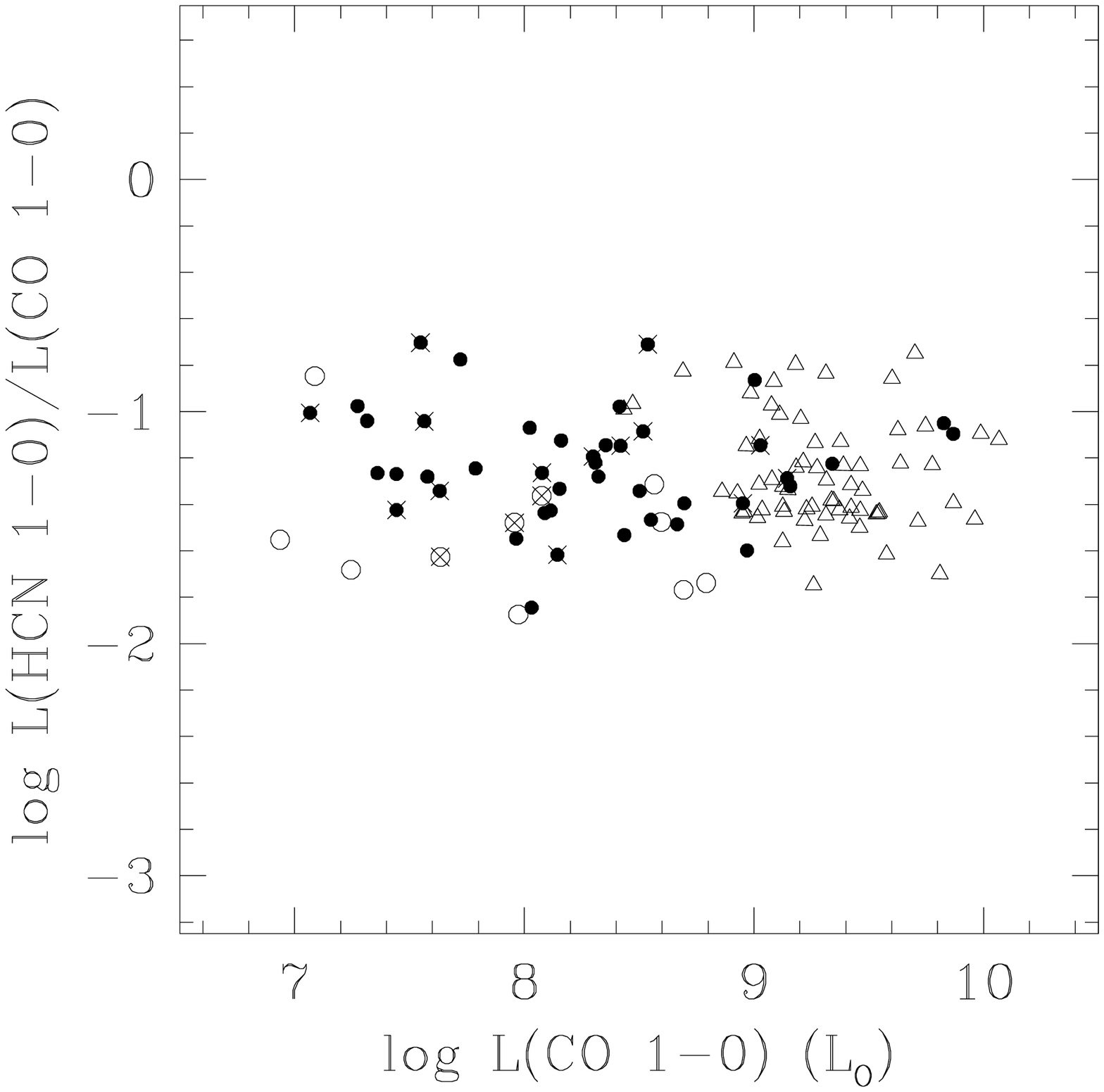}}}
\end{minipage}
\end{minipage}
\vspace{-2cm}
\begin{minipage}{8.94cm}
\begin{minipage}{4.4cm}
\resizebox{4.7cm}{!}{\rotatebox{0}{\includegraphics*{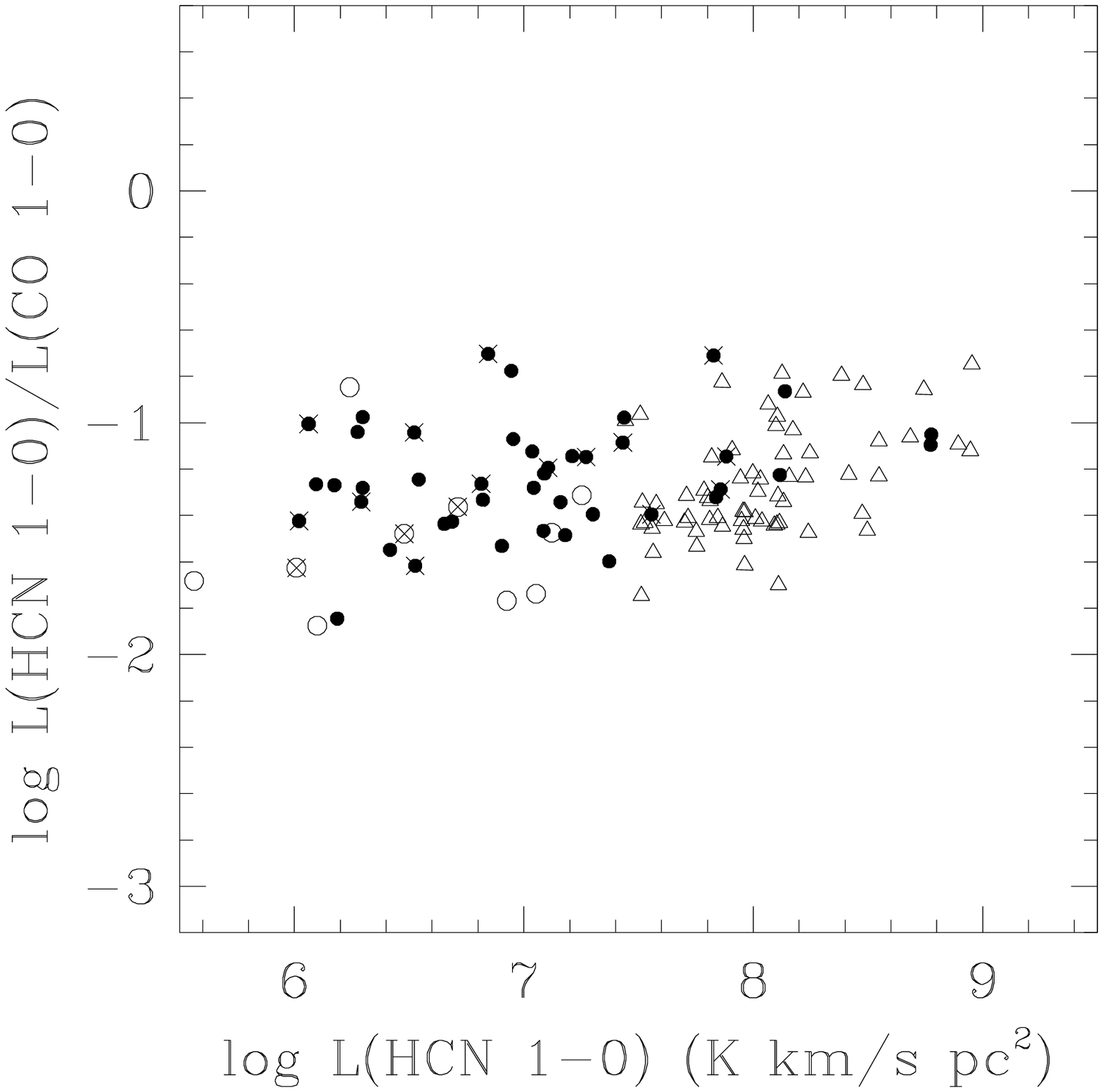}}}
\end{minipage}
\begin{minipage}{4.4cm}
\resizebox{4.7cm}{!}{\rotatebox{0}{\includegraphics*{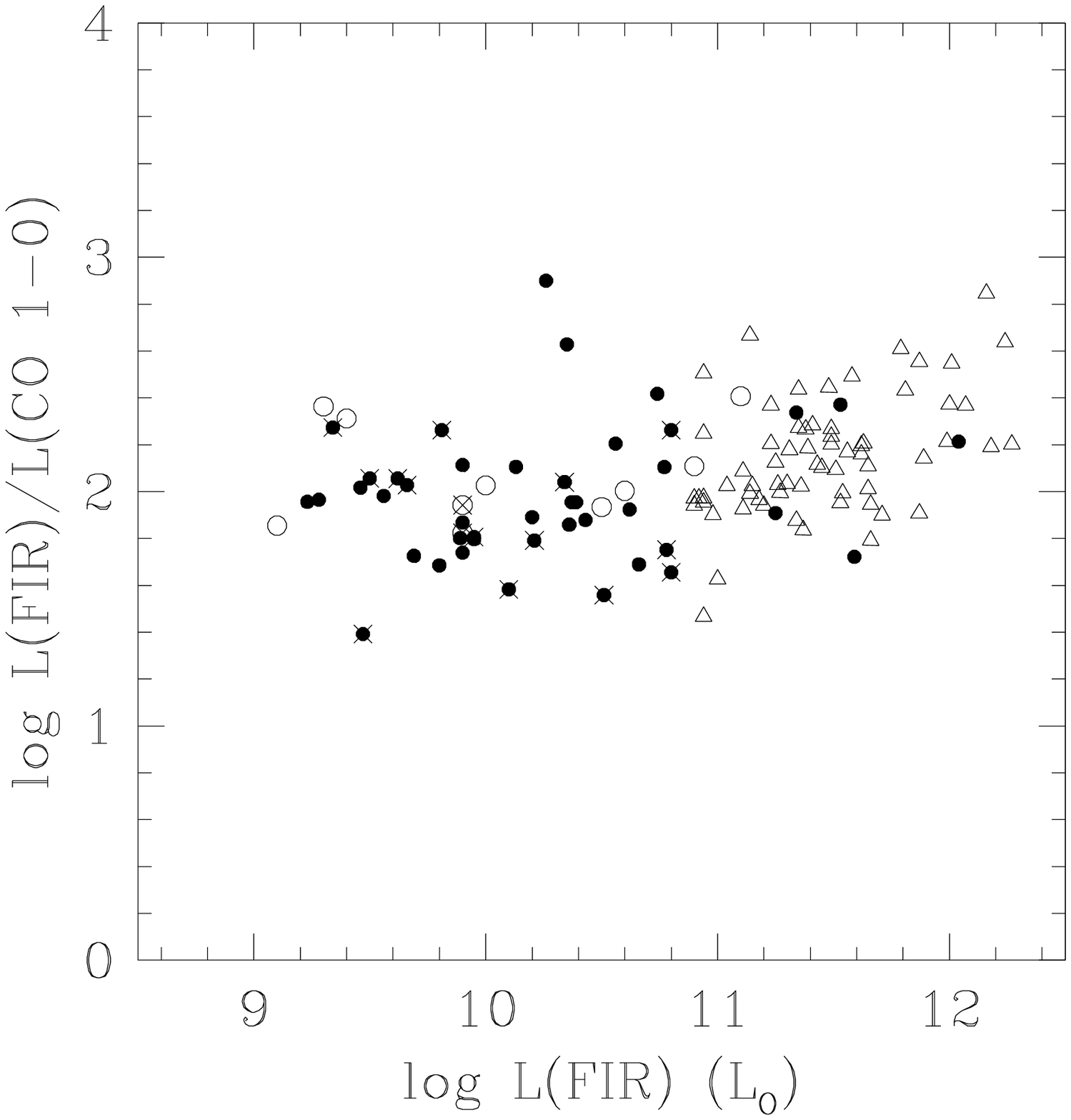}}}
\end{minipage}
\end{minipage}
\begin{minipage}{8.94cm}
\begin{minipage}{4.4cm}
\resizebox{4.7cm}{!}{\rotatebox{0}{\includegraphics*{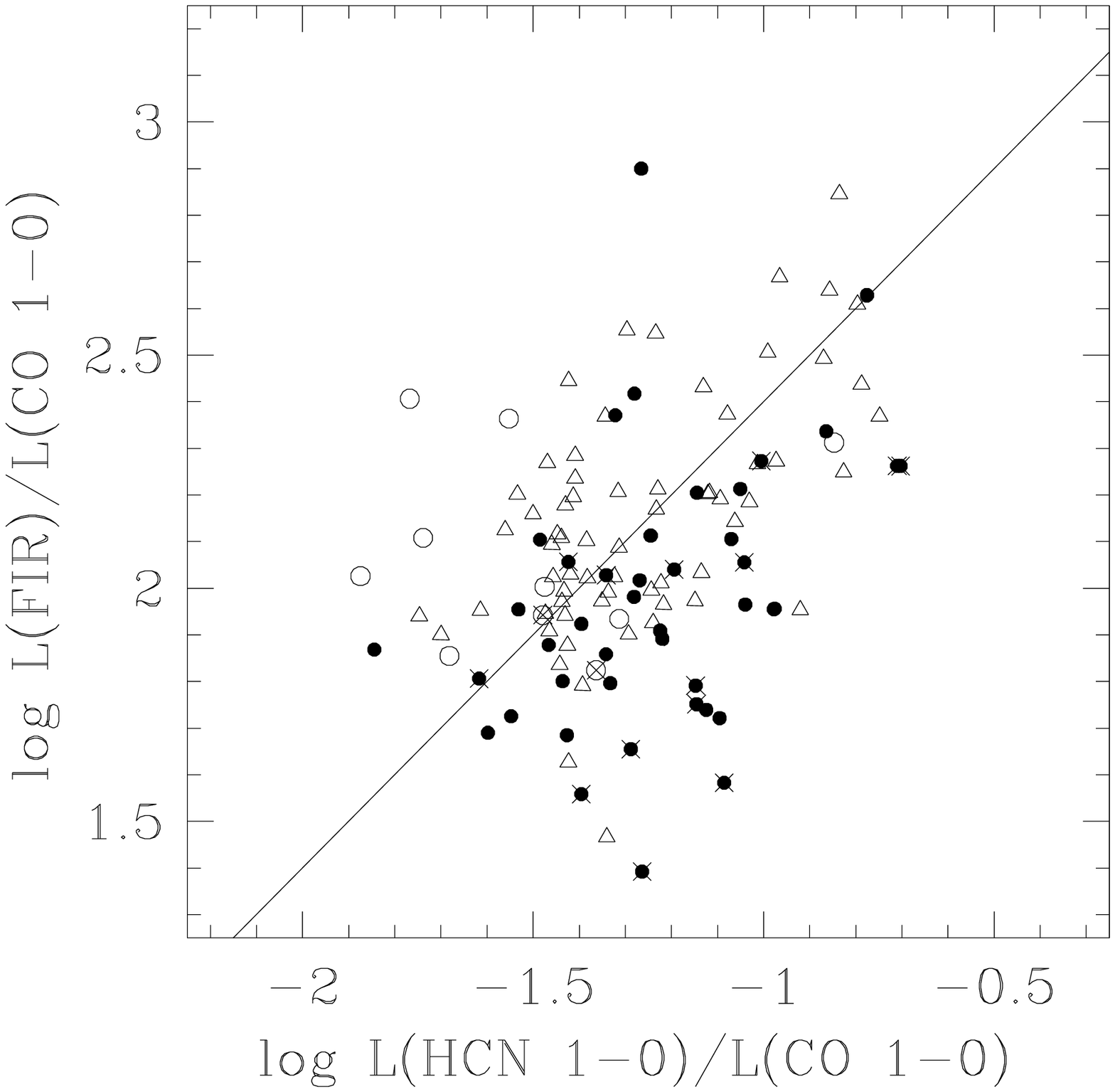}}}
\end{minipage}
\begin{minipage}{4.4cm}
\resizebox{4.7cm}{!}{\rotatebox{0}{\includegraphics*{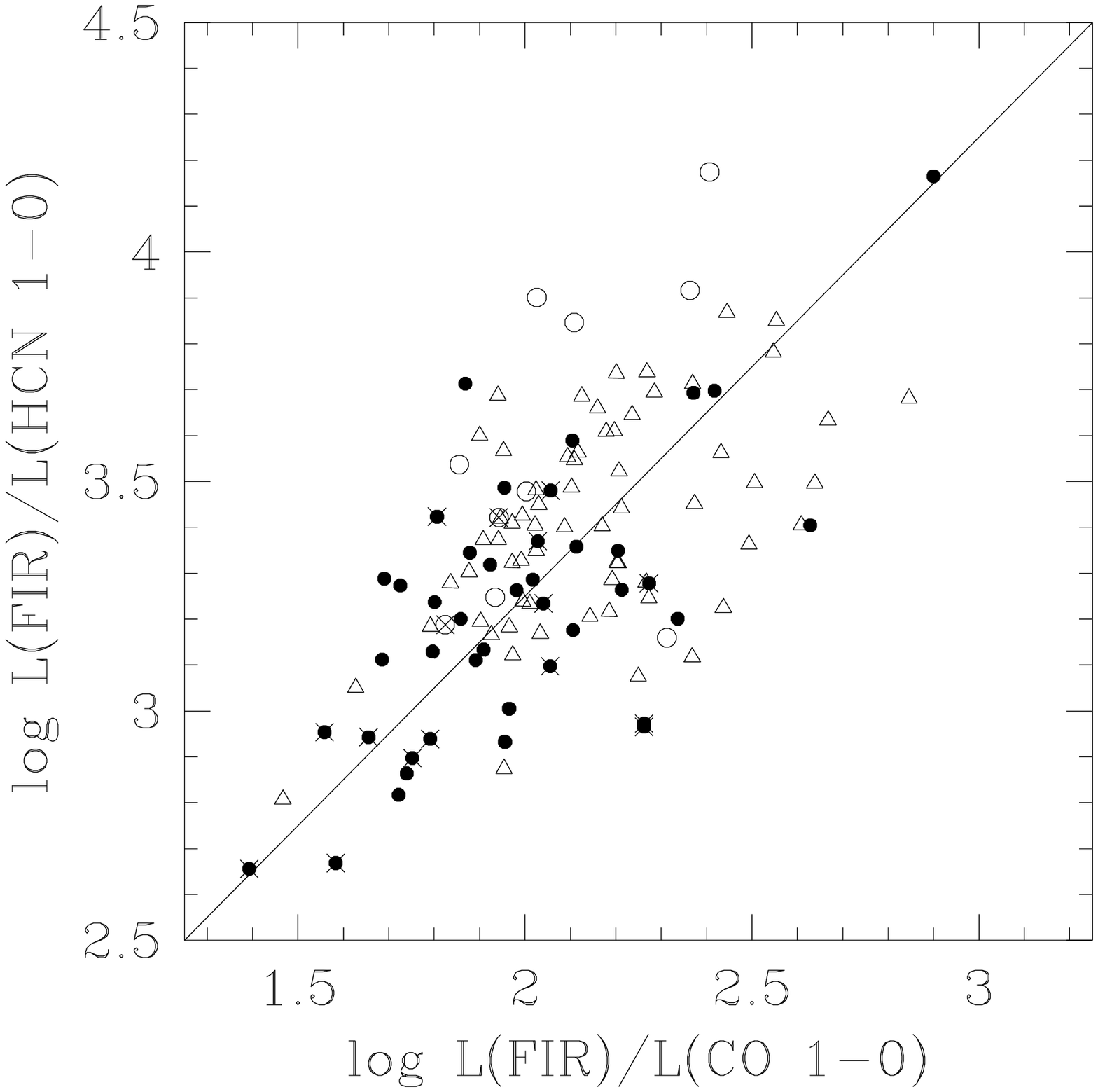}}}
\end{minipage}
\end{minipage}
\caption[]{\label{gaocomp} Top row: the J=1-0 HCN/CO ratio as a function of
  $IRAS$ FIR luminosity (left) and J=1-0 CO luminosity
  (right). The horizontal line in the panel at left marks the ratio
  separating normal and luminous galaxies in the paper by Gao $\&$
  Solomon (2004b). Neither panel shows any increase in HCN/CO at
  (U)LIRG luminosities, nor any change in the dispersion of the ratio.
  Center row: The HCN/CO ratio as a function of L(HCN) (left) and the
  FIR/CO ratio as a function of L(FIR) (right). Bottom row: The
  FIR/CO ratio versus HCN/CO ratio (left); the FIR/HCN ratio versus
  the FIR/CO ratio (right). In both panels, straight lines mark unity
  slopes. Fit parameters are listed in Table\,\ref{fittable}.\\
}
\end{figure}

\begin{figure*}
\begin{minipage}{4.45cm}
  \resizebox{4.8cm}{!}{\rotatebox{0}{\includegraphics*{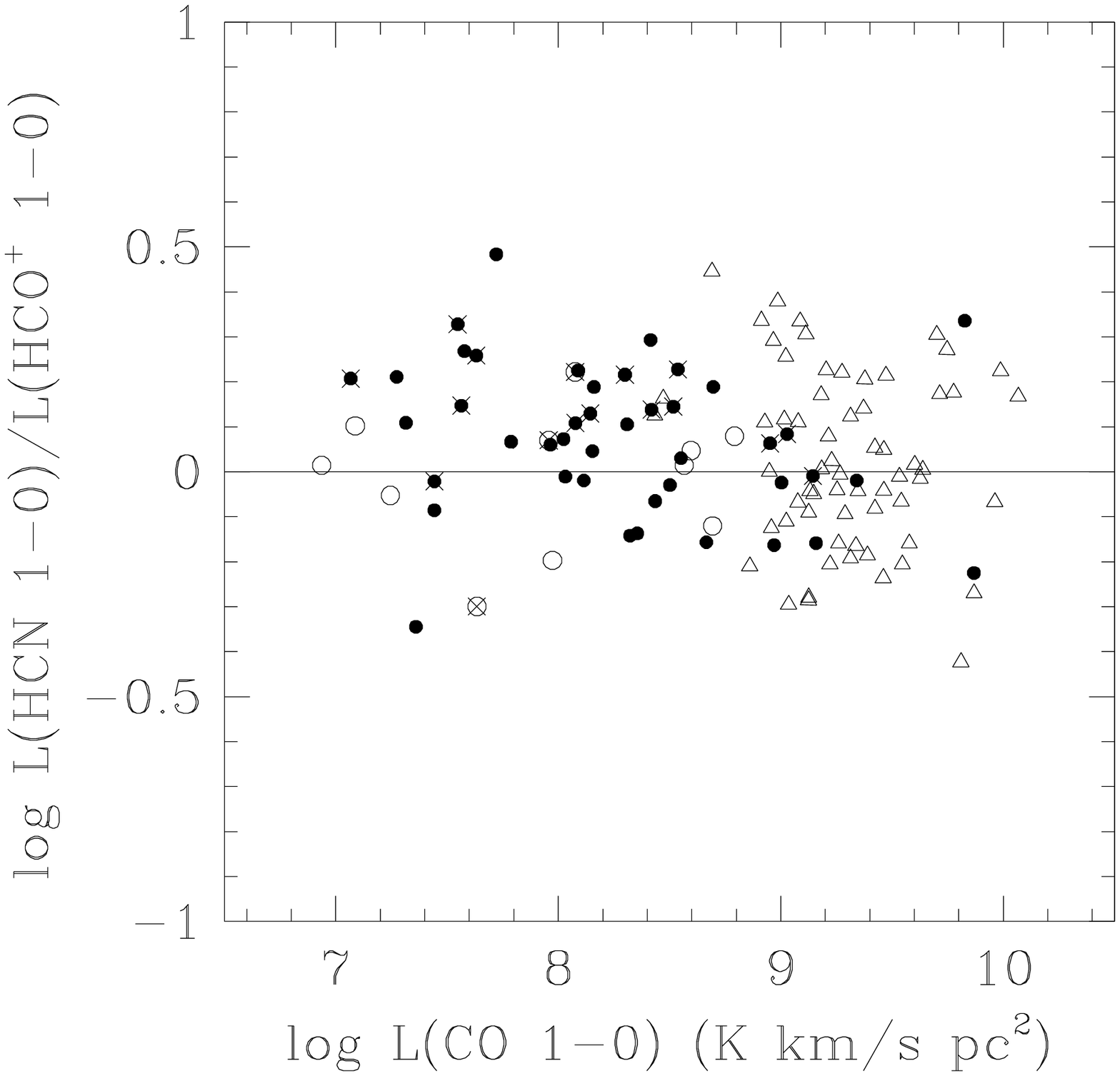}}}
\end{minipage}
\begin{minipage}{4.45cm}
  \resizebox{4.8cm}{!}{\rotatebox{0}{\includegraphics*{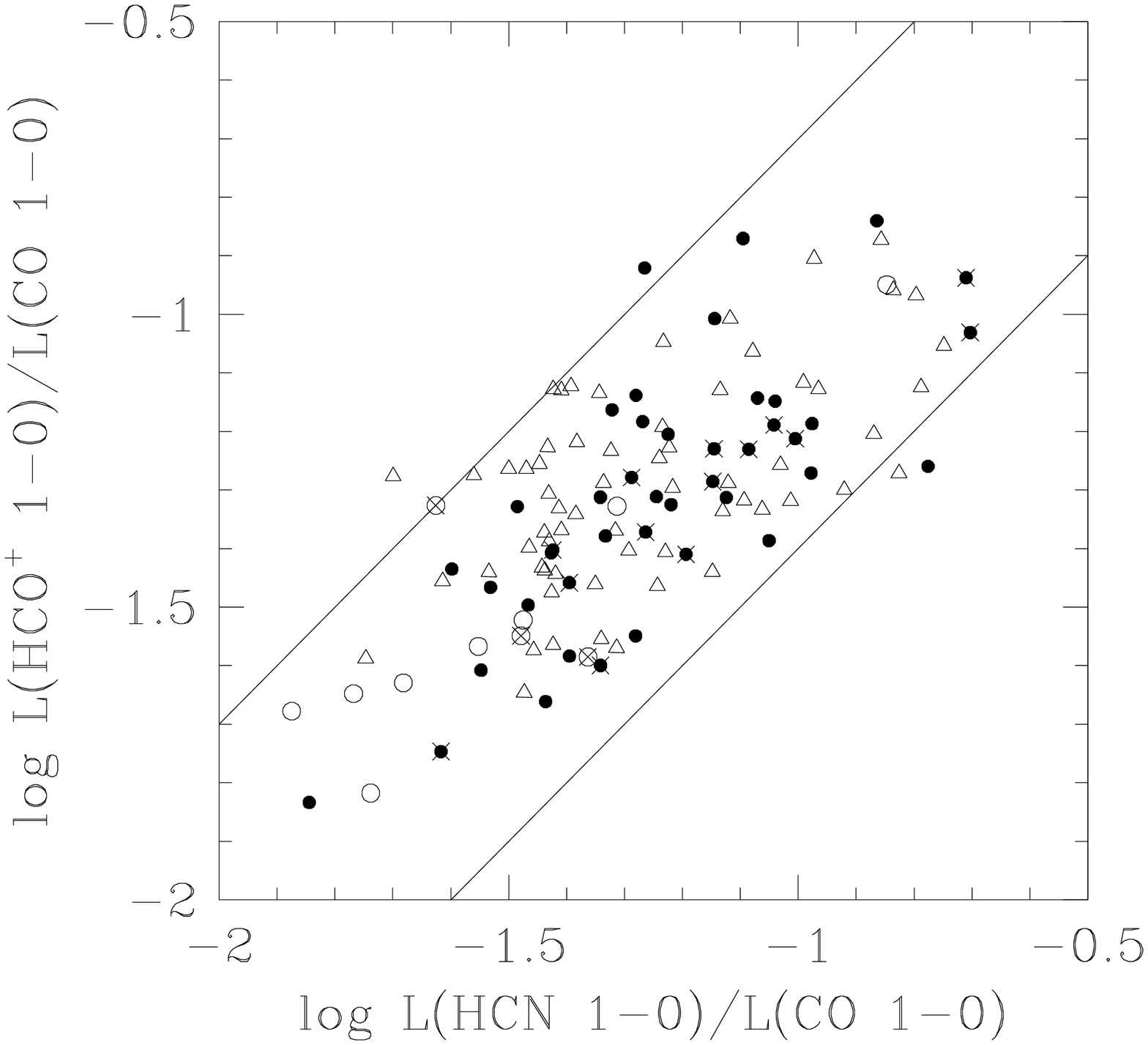}}}
\end{minipage}
\begin{minipage}{4.45cm}
  \resizebox{4.8cm}{!}{\rotatebox{0}{\includegraphics*{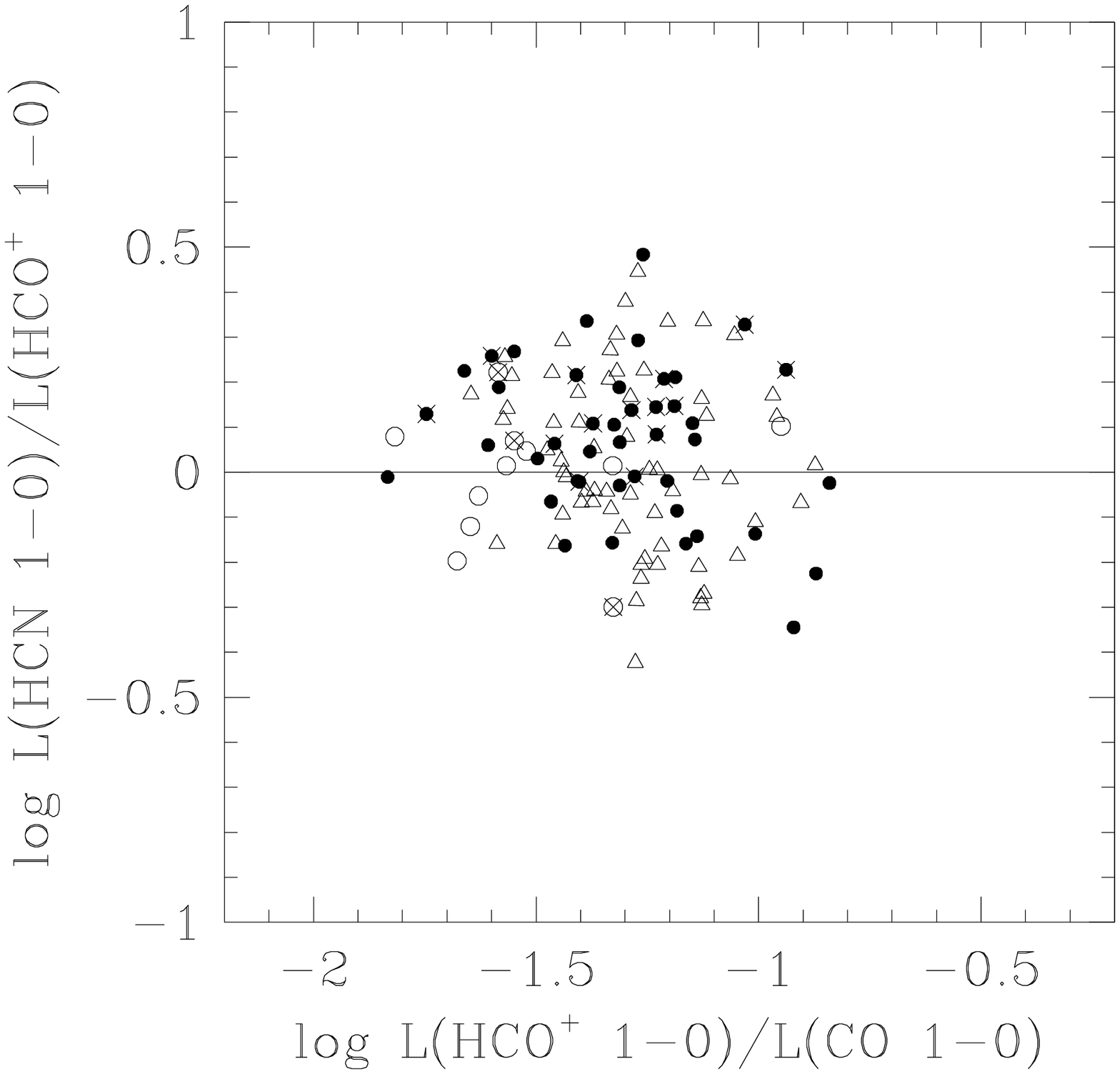}}}
\end{minipage}
\begin{minipage}{4.45cm}
  \resizebox{4.8cm}{!}{\rotatebox{0}{\includegraphics*{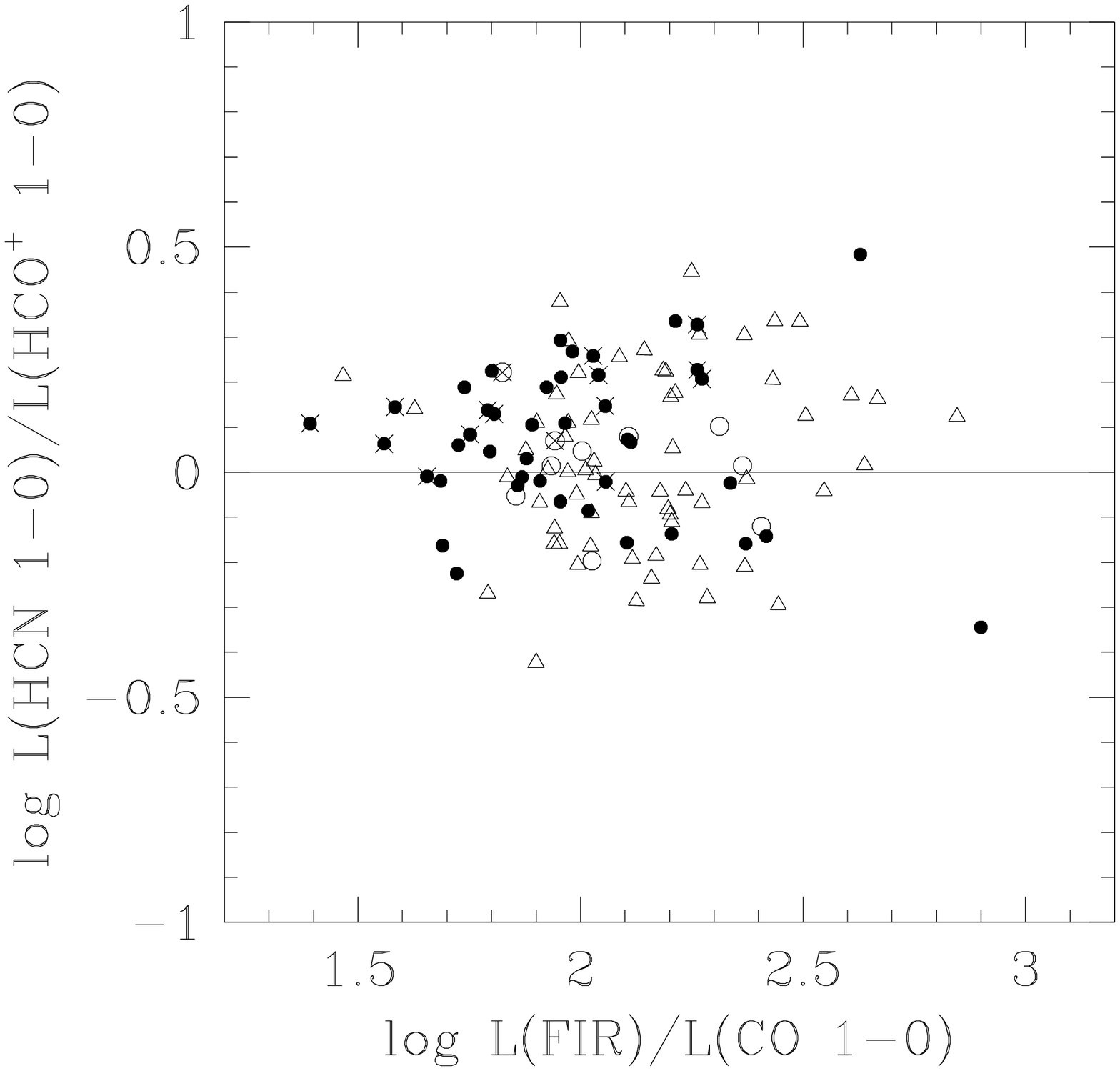}}}
\end{minipage}
\caption[] {Left: The HCN-to-$\hco$ versus the FIR-to-CO luminosity
  ratio. The horizontal line marks equal HCN and $\hco$
  luminosities. Center: The $\hco$/CO versus the HCN/CO luminosity
  ratio. Right: The $\hco$/HCN versus the HCN/CO luminosity
  ratio. The solid line has slope -0.55.  Fit parameters are listed
  in Table\,\ref{fittable}.
}
\label{gasprop}
\end{figure*}

The relation between the dense gas supposedly traced by $\hco$ and the
very dense gas traced by HCN is of further interest, especially since
their ground state intensities are derived from the same observed
profiles so that their ratio is free of systematic uncertainties and
thus quite accurate.  Despite critical densities an order of magnitude
apart, the $J$=1-0 $\hco$ and HCN luminosities are closely related and
practically interchangeable.  Grac\'ia-Carpio $\etal$ (2008) claimed a
correlation between the ground-state HCN-to-$\hco$ ratios and the
infrared luminositiies of (U)LIRGs but Privon $\etal$ (2015) could not
confirm this, nor can we. The present data extending over a three
times larger luminosity range clearly establish that HCN/$\hco$ ratios
are not related to either FIR or CO luminosities (Fig.\,\ref{gasprop}
left panel), nor to CO intensities.  The average HCN-to-$\hco$ ratio
of 1.1 is constant and slightly above unity. The same behavior is also
found on smaller scales, such as the central kilo-parsec of NGC~253
(Knudsen $\etal$ 2007). In the $J$=2-1 (Zhou $\etal$ 2022), $J$=3-2
and $J$=4-3 (This paper, Zhang $\etal$ 2014, Imanishi $\etal$ 2018)
transitions the average ratios are 0.9, 1.0, and 0.7 ,
respectively. The ratios of the first three transitions are within
each others errors but the $J$=4-3 ratio is significantly lower.  When
the luminosity distance bias is eliminated, for instance by weighting
both HCN and $\hco$ by their corresponding CO luminosities, $\hco$/CO
increases somewhat more slowly than HCN/CO with a logarithmic slope
$\sim 0.6$ as shown in Fig\,\ref{gasprop} (center left panel). There,
low-luminosity normal galaxies and high-luminosity LIRGs do not define
distinct or separate groups, in line with the absence of correlations
between CO luminosity and the ratios of HCN to CO and $\hco$ to CO in
Fig.\,\ref{gaocomp}.  Finally, the center right panel shows that the
`very-dense-to-dense gas' ratio $J$=1-0 HCN/$\hco$ is unrelated to the
`dense-to-modestly-dense gas' ratio $\hco$/CO, and the rightmost panel
shows that there is no correlation between the $\hco$/HCN and FIR/CO
either.

\subsection{HCN, $\hco$, and the $\co$/$\thirco$ ground-state ratio} 

\begin{figure*}
\begin{minipage}{4.45cm}
\resizebox{4.8cm}{!}{\rotatebox{0}{\includegraphics*{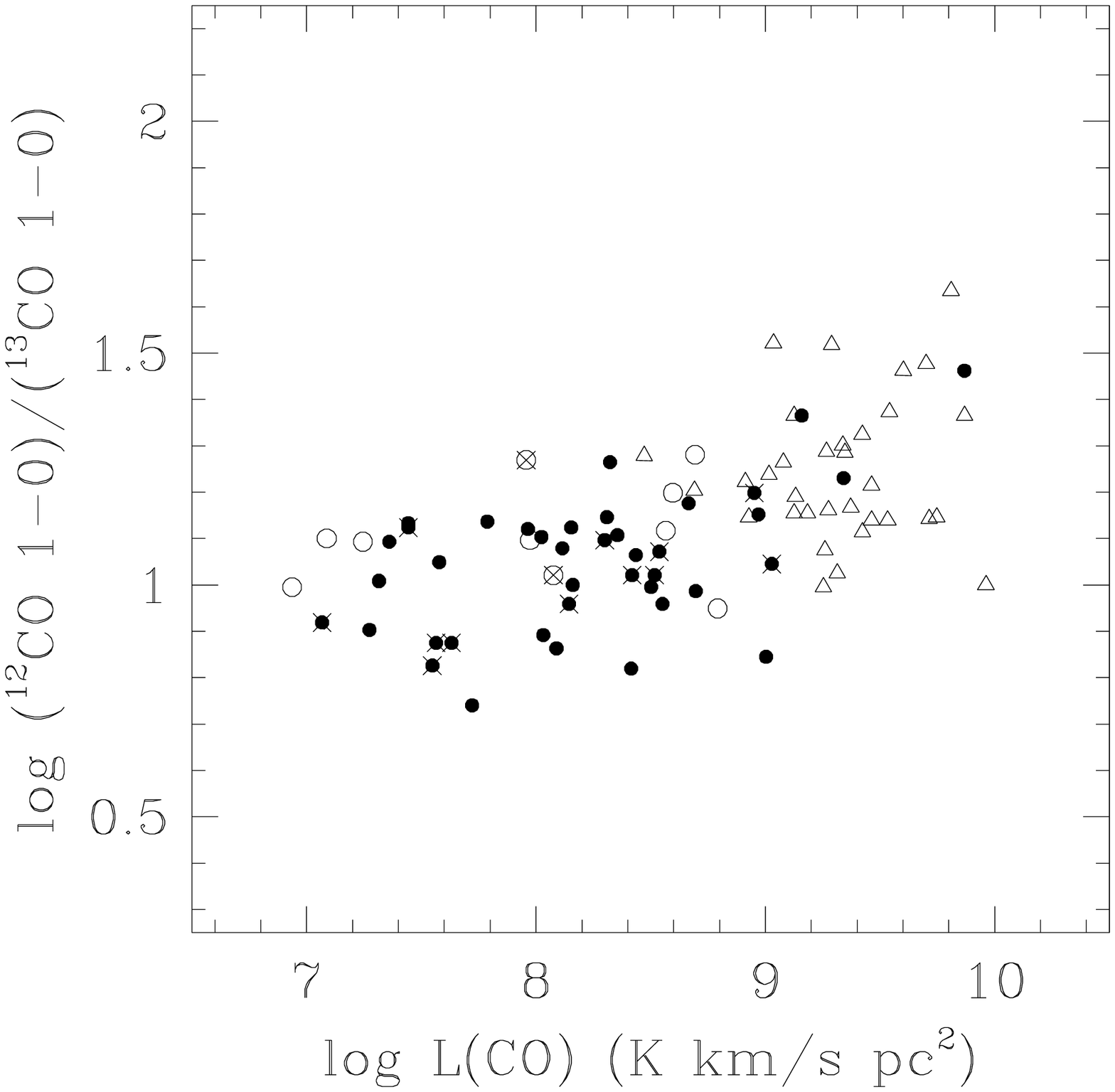}}}
\end{minipage}
\begin{minipage}{4.45cm}
\resizebox{4.8cm}{!}{\rotatebox{0}{\includegraphics*{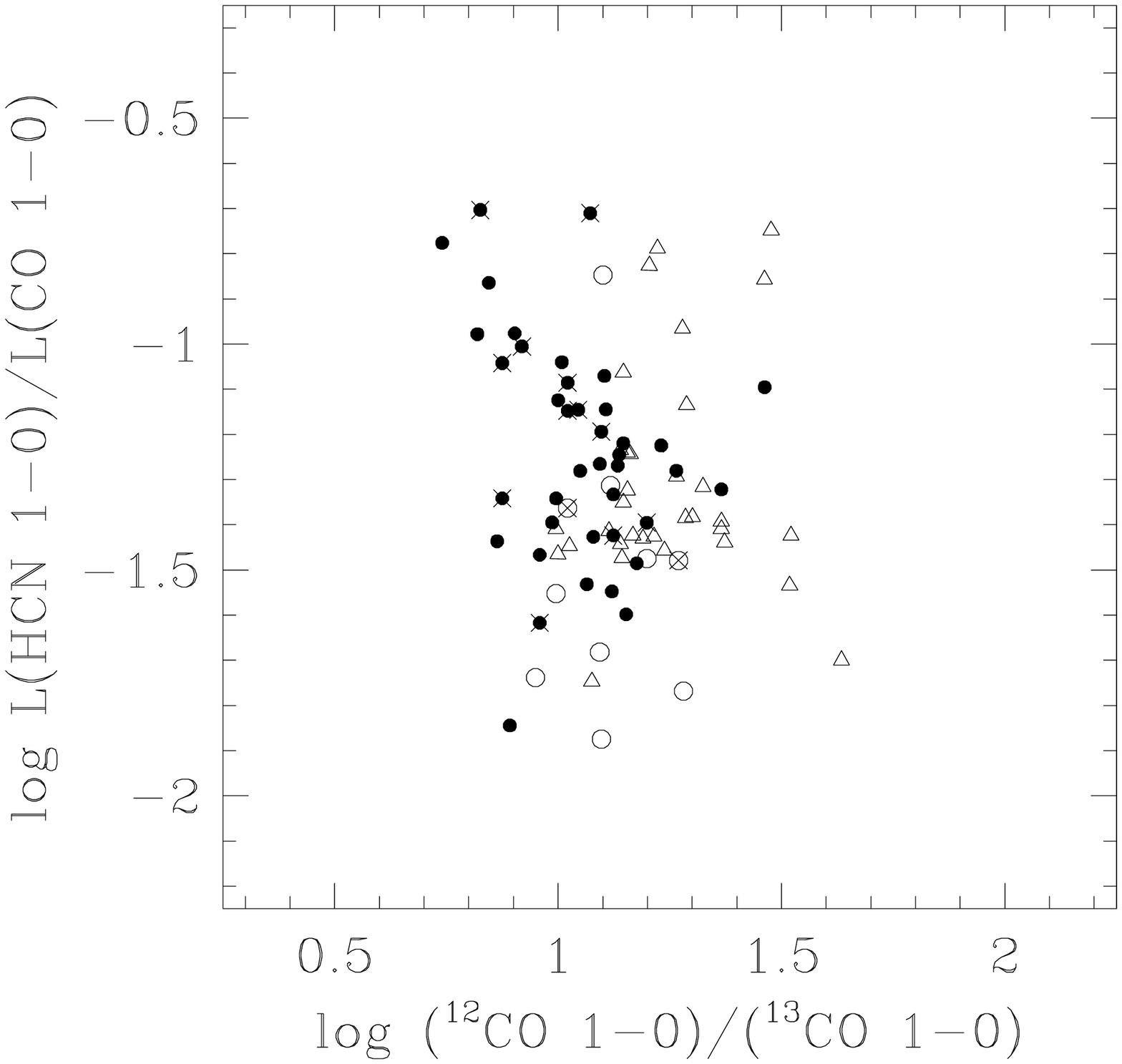}}}
\end{minipage}
\begin{minipage}{4.45cm}
\resizebox{4.8cm}{!}{\rotatebox{0}{\includegraphics*{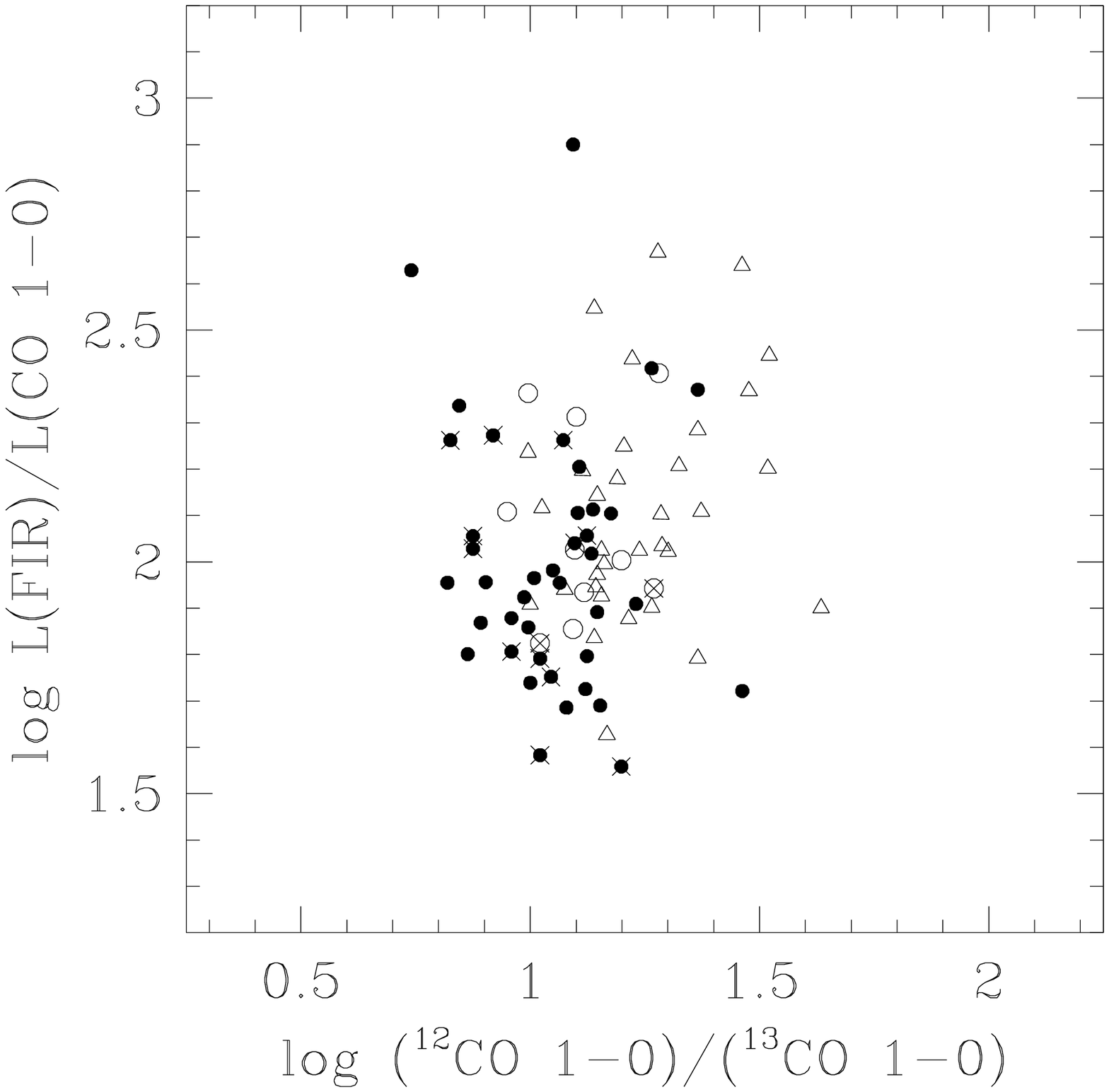}}}
\end{minipage}
\begin{minipage}{4.45cm}
\resizebox{4.8cm}{!}{\rotatebox{0}{\includegraphics*{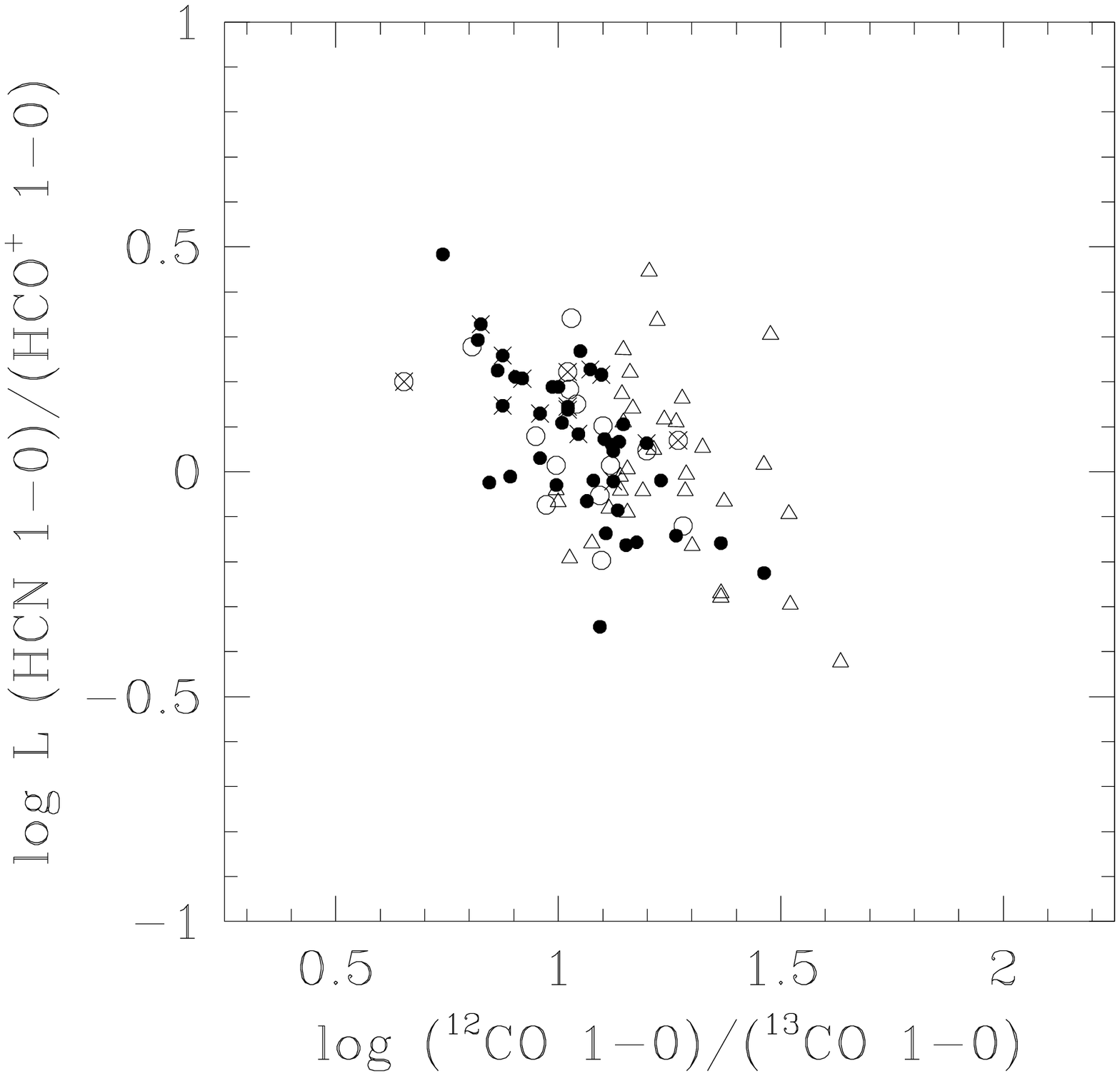}}}
\end{minipage}
\caption[] {From left to right, the ground-state isotopologue
  $\co$/$\thirco$ ratio as a function CO luminosity and the
  HCN/CO, FIR/CO, and HCN/$\hco$ ratios as a function of the
  $\co$/$\thirco$ ratio. Fit parameters are listed
  in Table\,\ref{fittable}.
}
\label{isorat}
\end{figure*}

The observed isotopological ratio $I(\co$)/I($\thirco$) depends on
both the intrinsic isotopic abundance ratio and the CO optical
depth. The magnitude of the isotopic abundance ratio in galaxy centers
is poorly known and estimates range from very low values of 20-40 to
high values of 70-150 (see, for instance, the compilation by Viti
$\etal$ 2020). Moreover, they may vary from galaxy to galaxy. The
present database contains measurements of the isotopological ratio for
the majority of galaxies (see Section 3.1).  $\co$(1-0)/$\thirco$(1-0)
ratios increase both with the CO luminosity $L$(CO) (Fig\,\ref{isorat}
leftmost panel) and with the far-infrared luminosity $L$(FIR). Ratios
range from 6-14 at the lowest luminosities and from 10-50 at the
highest luminosities. Fig.\,\ref{isorat} illustrates that the ratio of
HCN-to-$\hco$ decreases as a function of $\co$/$\thirco$ (rightmost
panel) whereas the `dense-gas ratio' HCN/CO does not (center left
panel) nor does the ratio $\hco$/CO. Likewise, no relation is found
between the `star-formation-efficiency' FIR/CO and the isotopological
ratio (center right panel).

\begin{figure}
  \begin{minipage}{8.94cm}
  \end{minipage}  
\vspace{-2cm}
\begin{minipage}{8.94cm}
\begin{minipage}{4.4cm}
\resizebox{4.7cm}{!}{\rotatebox{0}{\includegraphics*{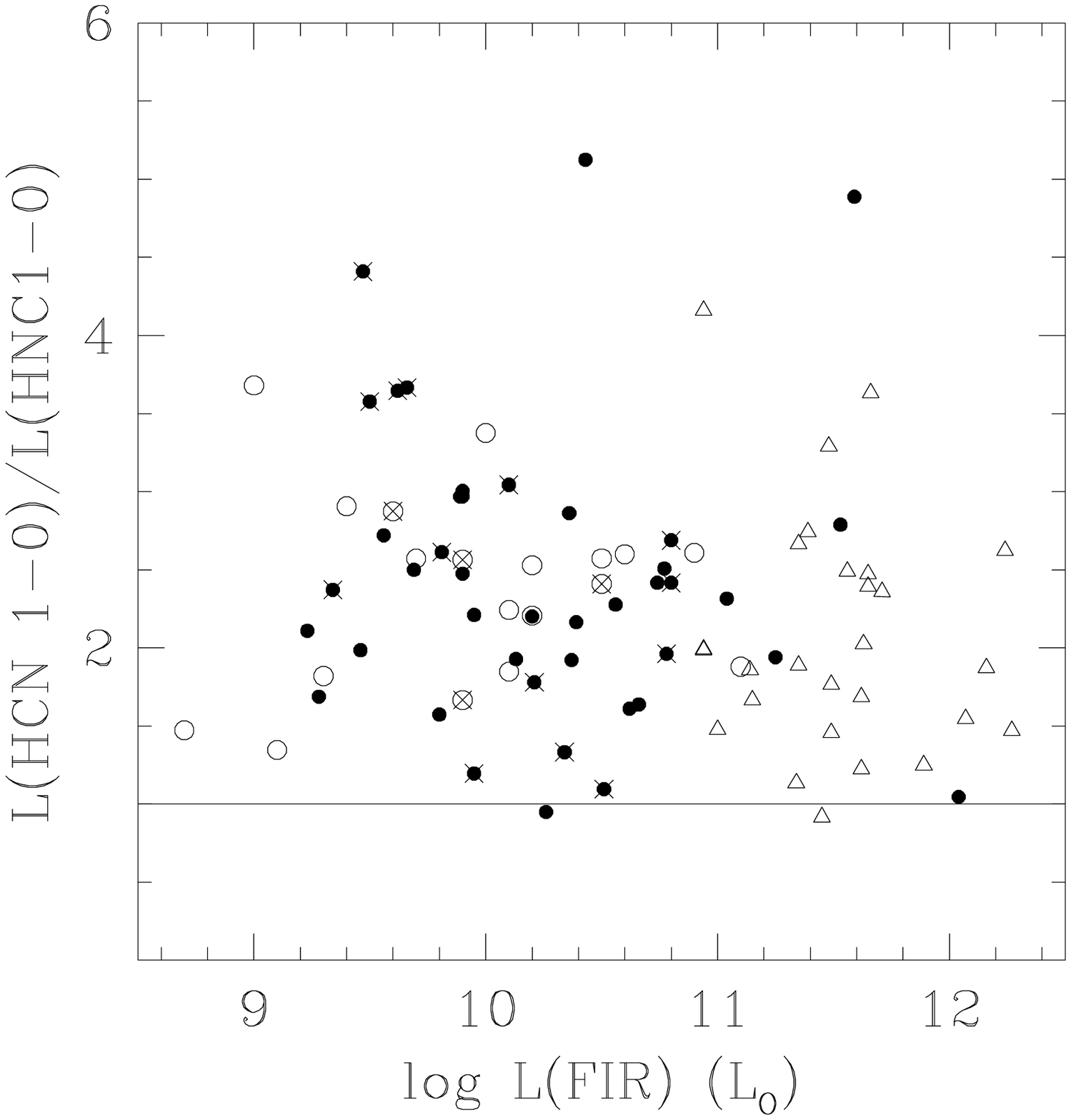}}}
\end{minipage}
\begin{minipage}{4.4cm}
\resizebox{4.7cm}{!}{\rotatebox{0}{\includegraphics*{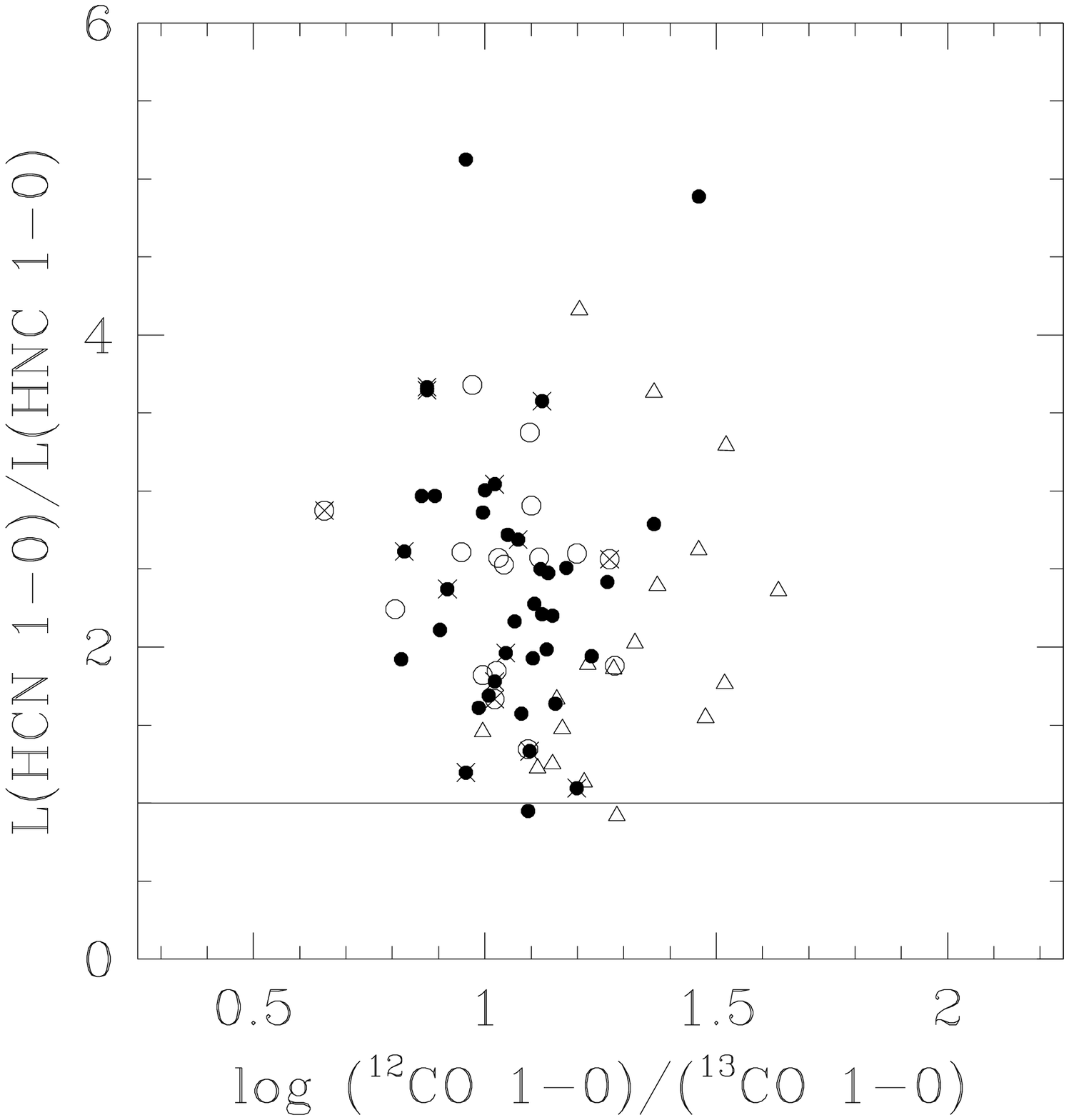}}}
\end{minipage}
\end{minipage}
\vspace{-2cm}
\begin{minipage}{8.94cm}
\begin{minipage}{4.4cm}
\resizebox{4.7cm}{!}{\rotatebox{0}{\includegraphics*{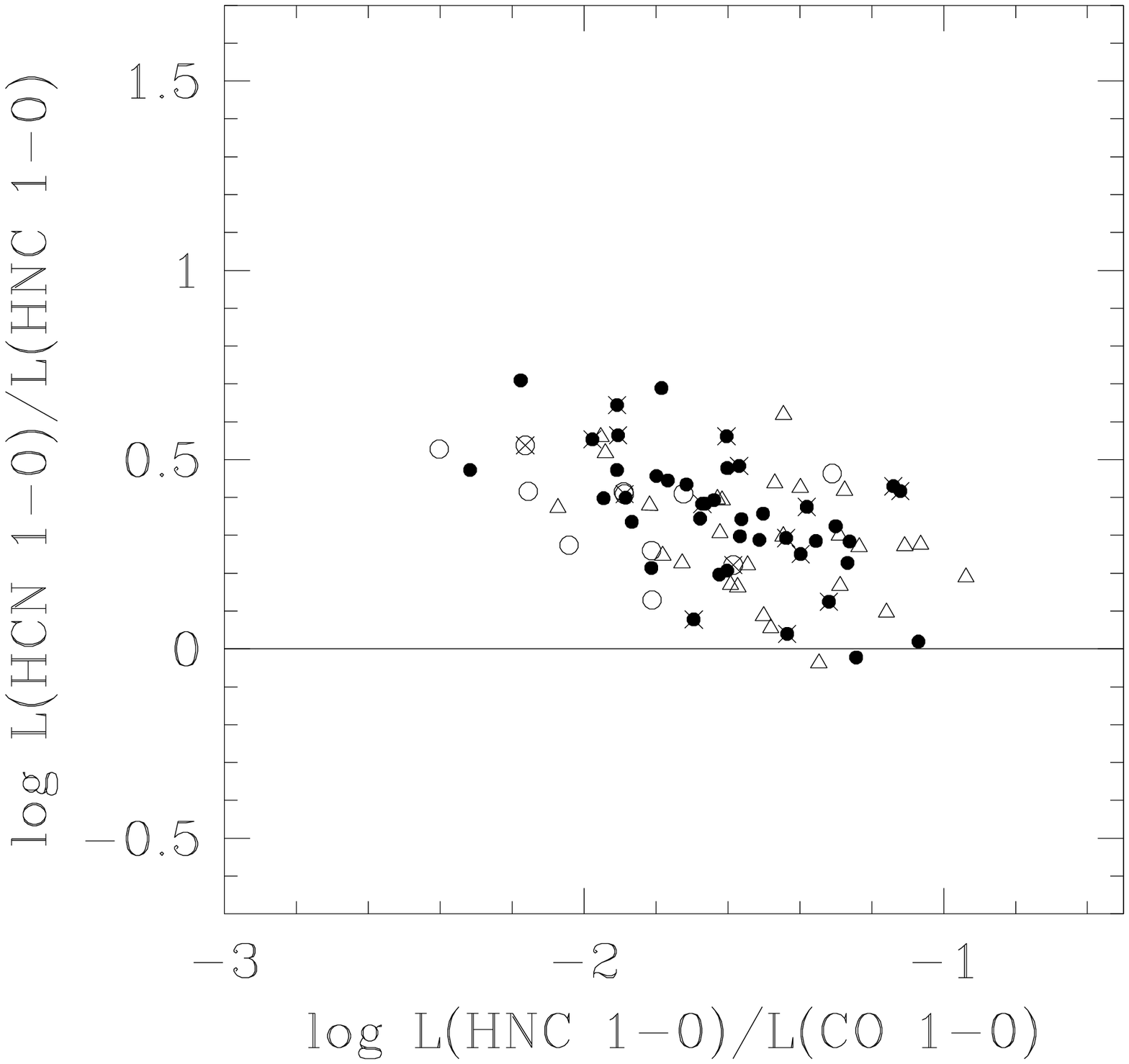}}}
\end{minipage}
\begin{minipage}{4.4cm}
\resizebox{4.7cm}{!}{\rotatebox{0}{\includegraphics*{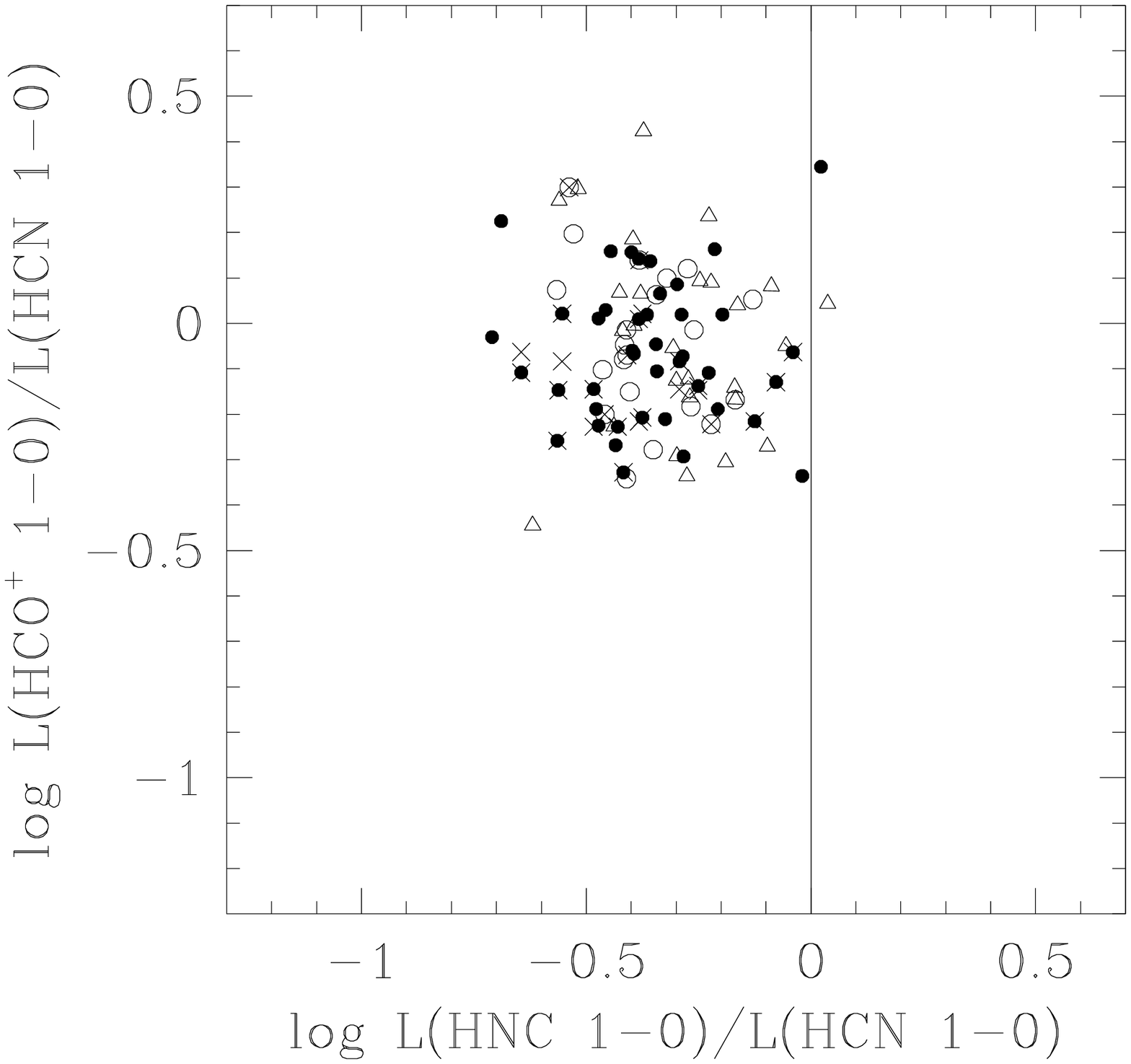}}}
\end{minipage}
\end{minipage}
\begin{minipage}{8.94cm}
\begin{minipage}{4.4cm}
\resizebox{4.7cm}{!}{\rotatebox{0}{\includegraphics*{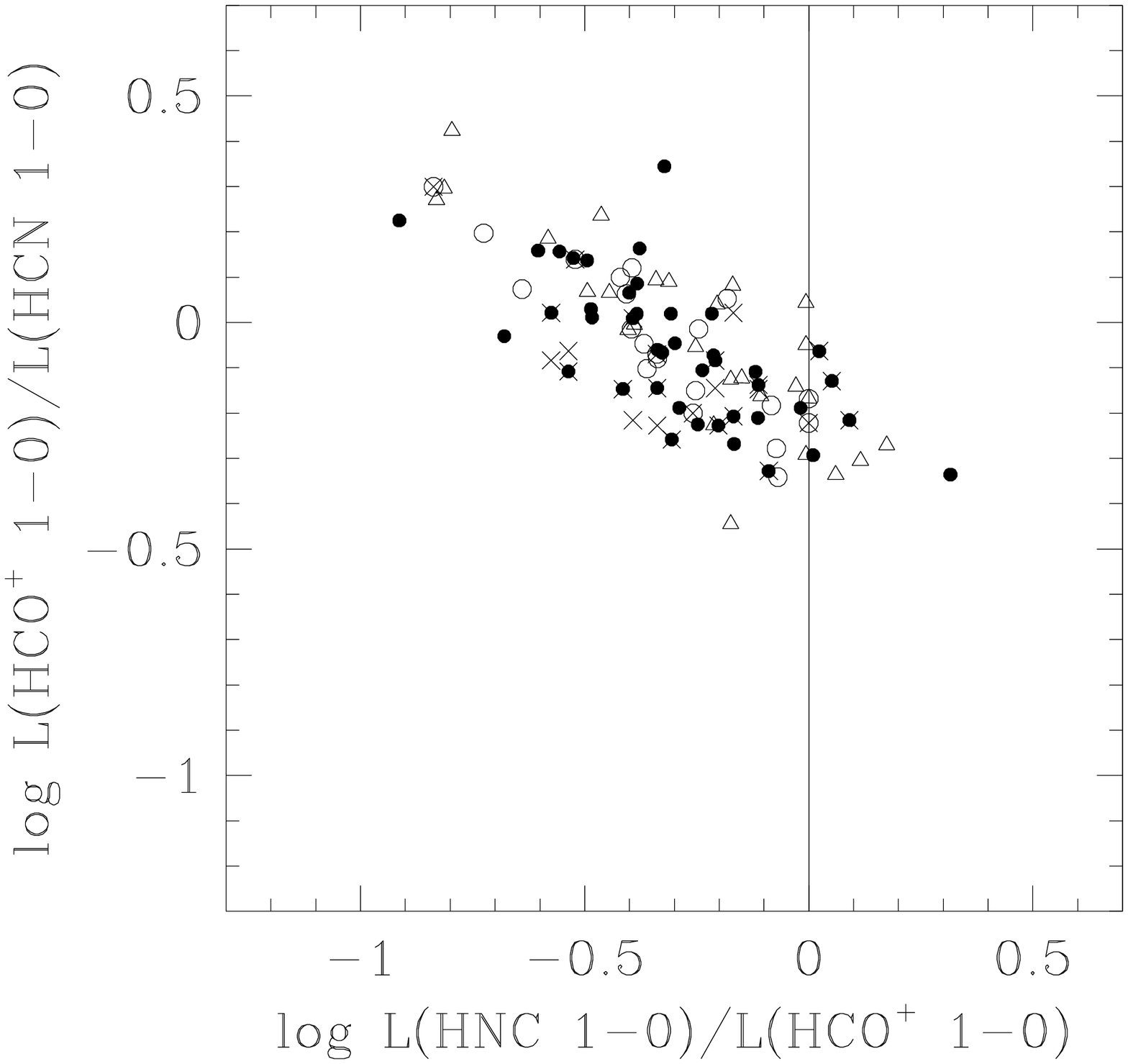}}}
\end{minipage}
\begin{minipage}{4.4cm}
\resizebox{4.7cm}{!}{\rotatebox{0}{\includegraphics*{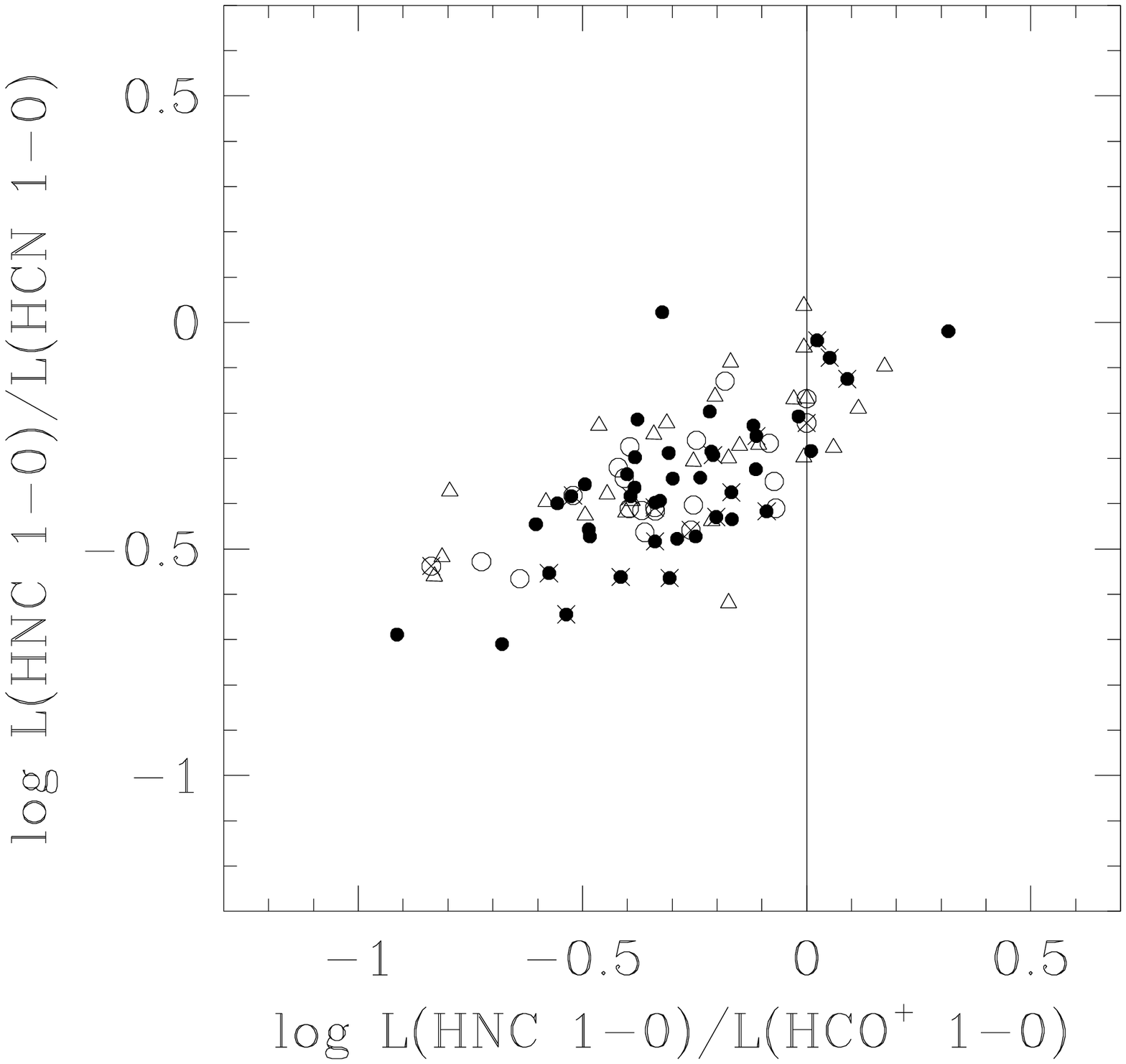}}}
\end{minipage}
\end{minipage}
\caption[] {Top: the ground-state HCN/HNC isomer ratio as a function of
  FIR luminosity (left) and the $\co$/$\thirco$ isotopologue ratio
  (right).  Center: the ground-state HCN/CO (left) and $\hco$/HCN
  (right) ratios as a function of the (inverse) HNC/HCN isomer rasio. Bottom:
  the ground-state $\hco$/HCN (left) and HNC/HCN (right) ratios as a function
  of the HNC/$\hco$ ratio. Solid lines mark unity isomer ratios. Fit
  parameters are listed in Table\,\ref{fittable}.
}
\label{hncrat}
\end{figure}

\begin{figure*}
\begin{minipage}{4.45cm}  
\resizebox{4.8cm}{!}{\rotatebox{0}{\includegraphics*{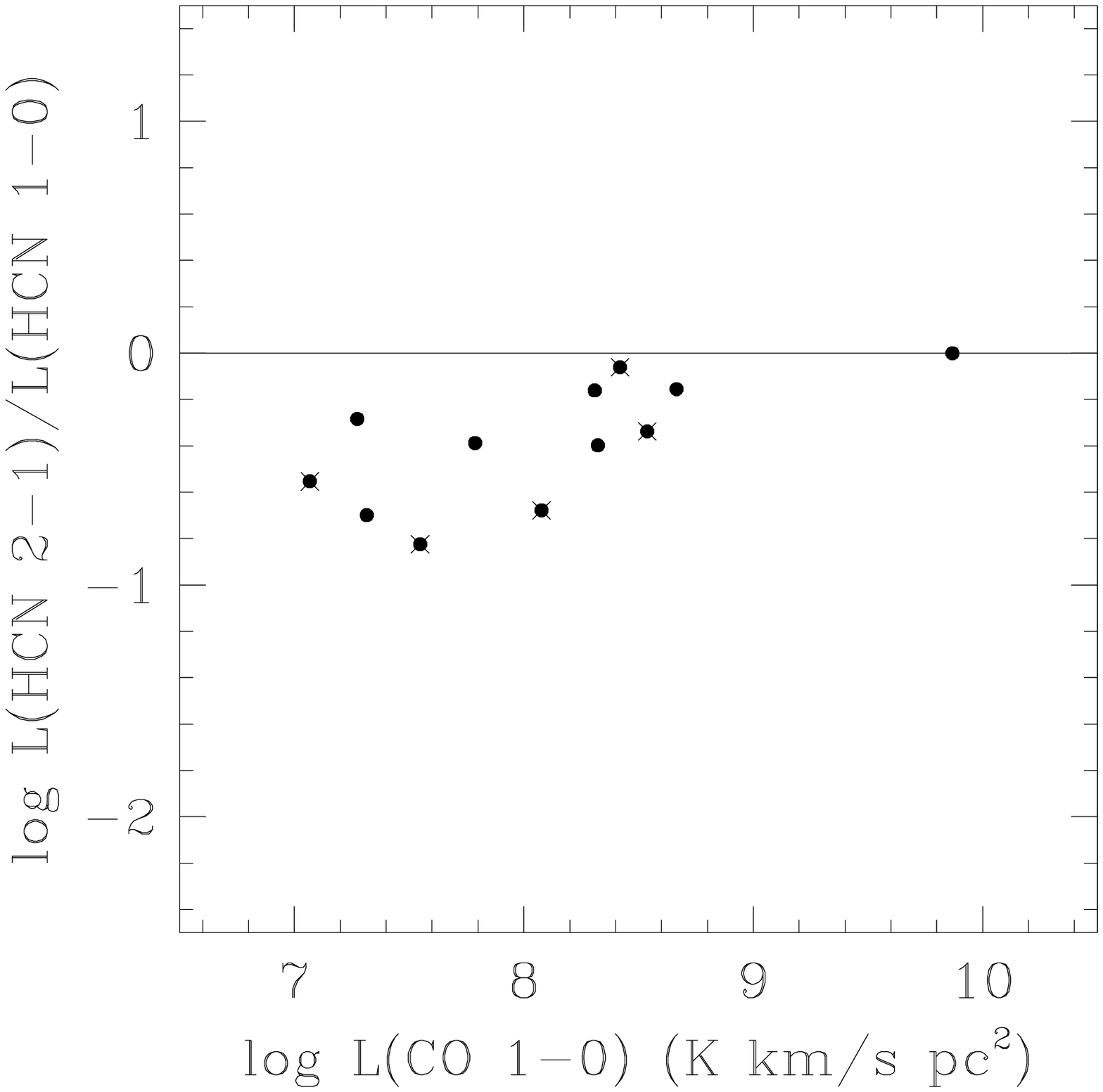}}}
\end{minipage}
\begin{minipage}{4.45cm}  
\resizebox{4.8cm}{!}{\rotatebox{0}{\includegraphics*{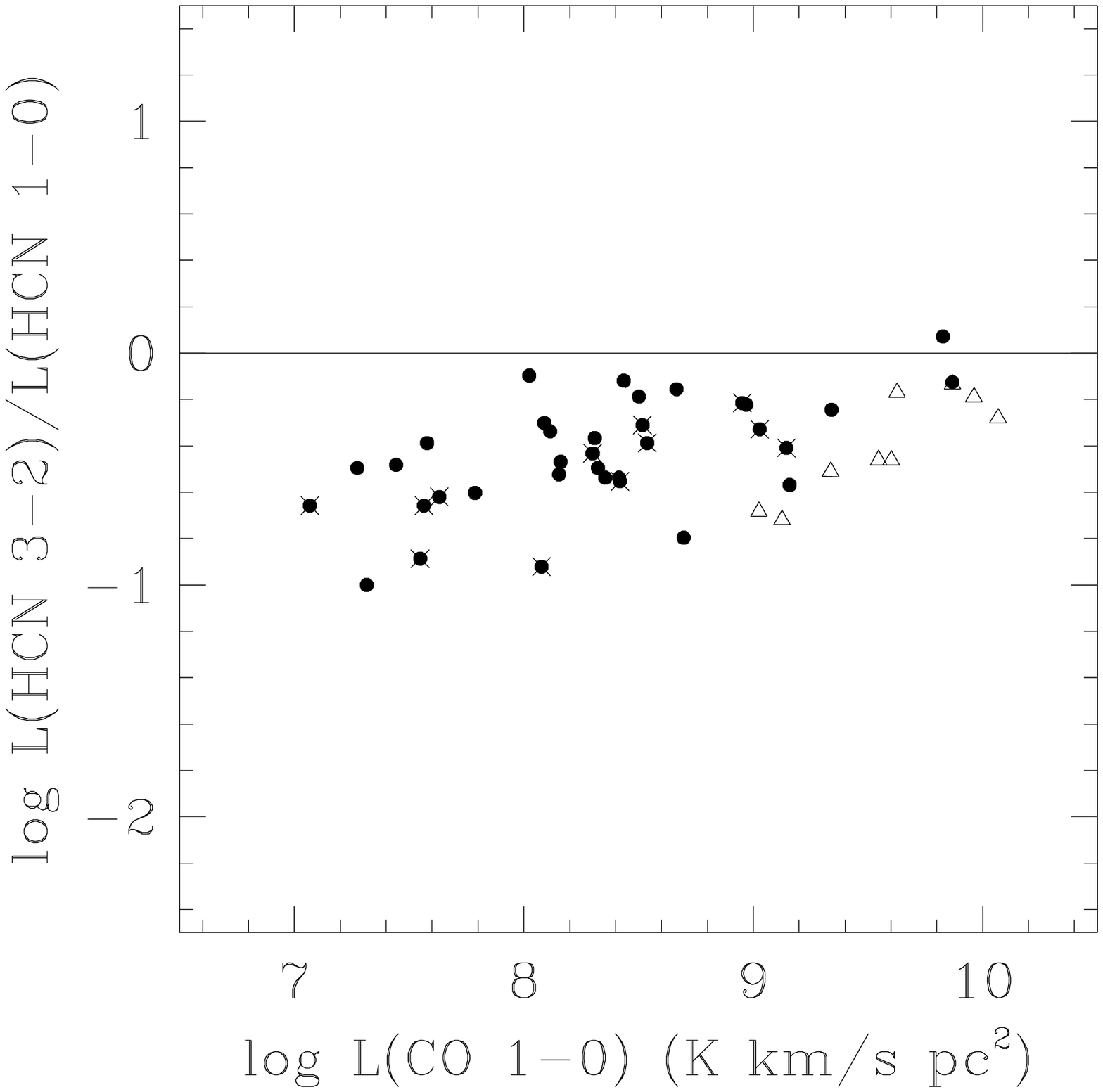}}}
\end{minipage}
\begin{minipage}{4.45cm}  
\resizebox{4.8cm}{!}{\rotatebox{0}{\includegraphics*{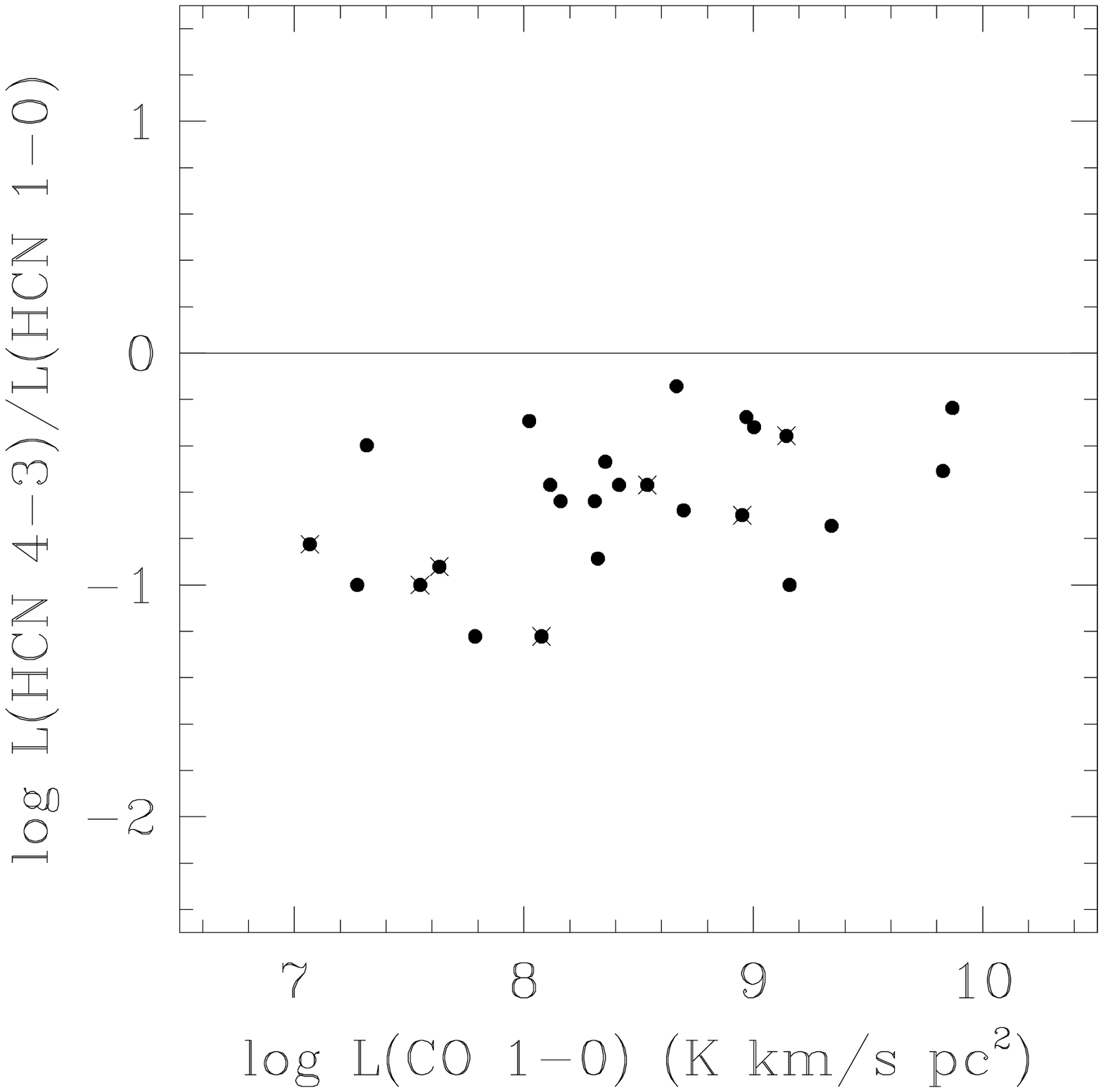}}}
\end{minipage}
\begin{minipage}{4.45cm}
\resizebox{4.8cm}{!}{\rotatebox{0}{\includegraphics*{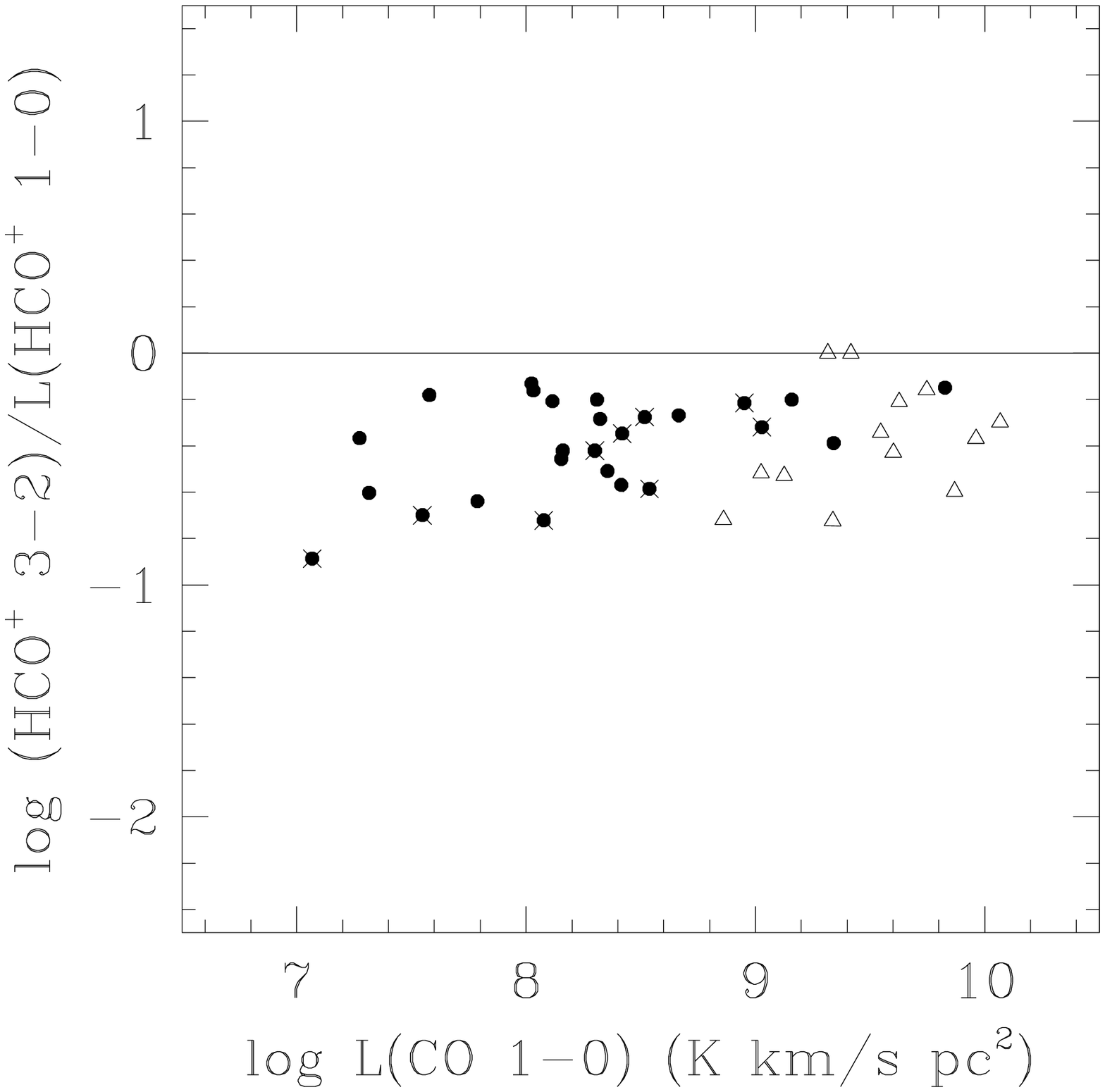}}}
\end{minipage}
\caption[] {Molecular line ladders: the HCN and $\hco$ line intensity
  ratios as a function of transition in normalized (equal) beams. Left
  to right: the HCN $J$=2-1/$J$=1-0, $J$=3-2/$J$=1-0, $J$=4-3/$J$-1-0,
  and the HCO$^+$ $J$=3-2/$J$=1-0 ratios as a function of $\co$
  $J$=1-0 luminosity. Solid lines mark ratios of unity. Symbols are as
  in Fig.\ref{lummol} but almost all data are from this paper.
  Fit parameters are listed in Table\,\ref{fittable}.
}
\label{rattrans}
\end{figure*}

\begin{figure*}
\begin{minipage}{4.45cm}
\resizebox{4.8cm}{!}{\rotatebox{0}{\includegraphics*{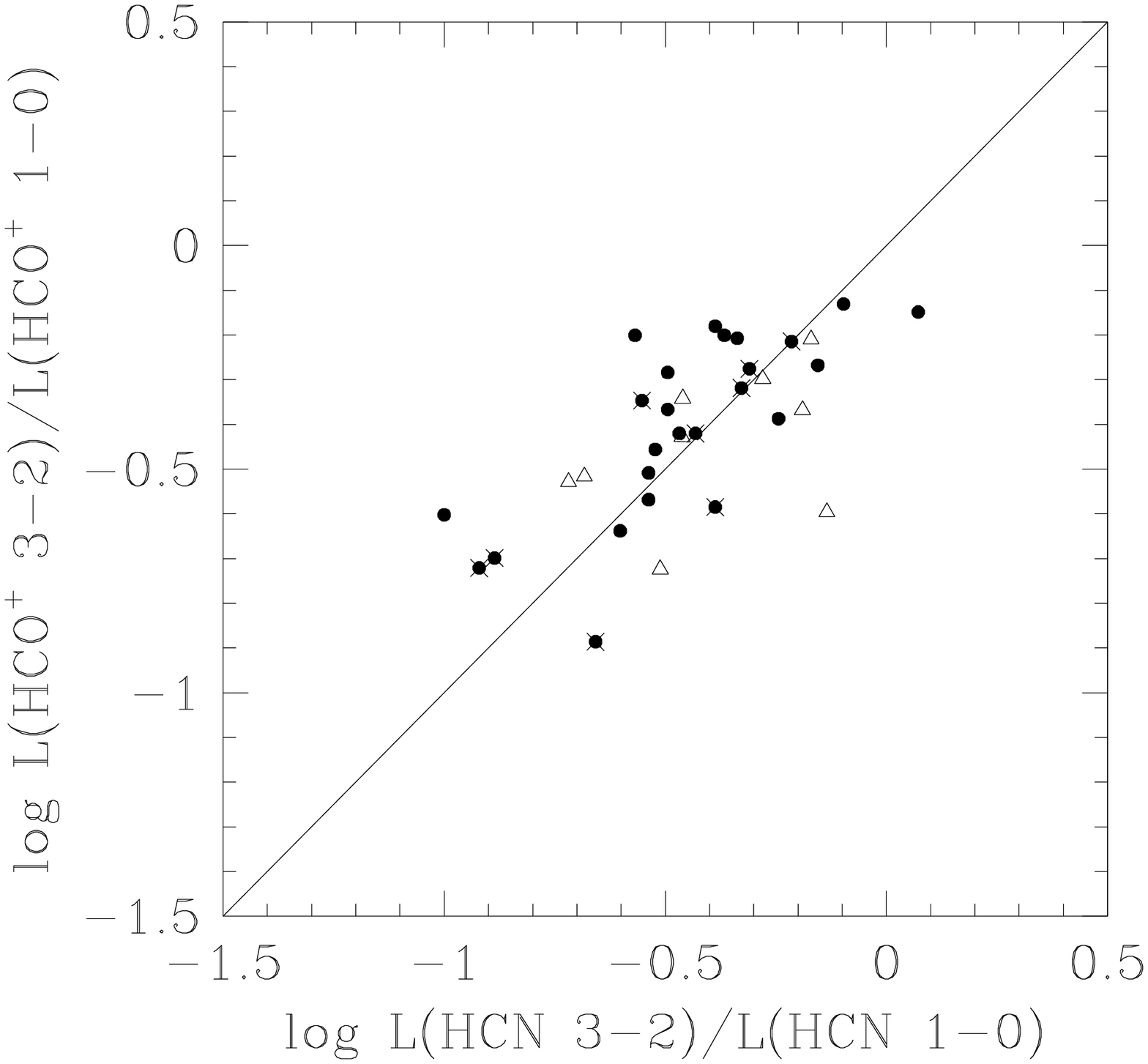}}}
\end{minipage}
\begin{minipage}{4.45cm}
\resizebox{4.8cm}{!}{\rotatebox{0}{\includegraphics*{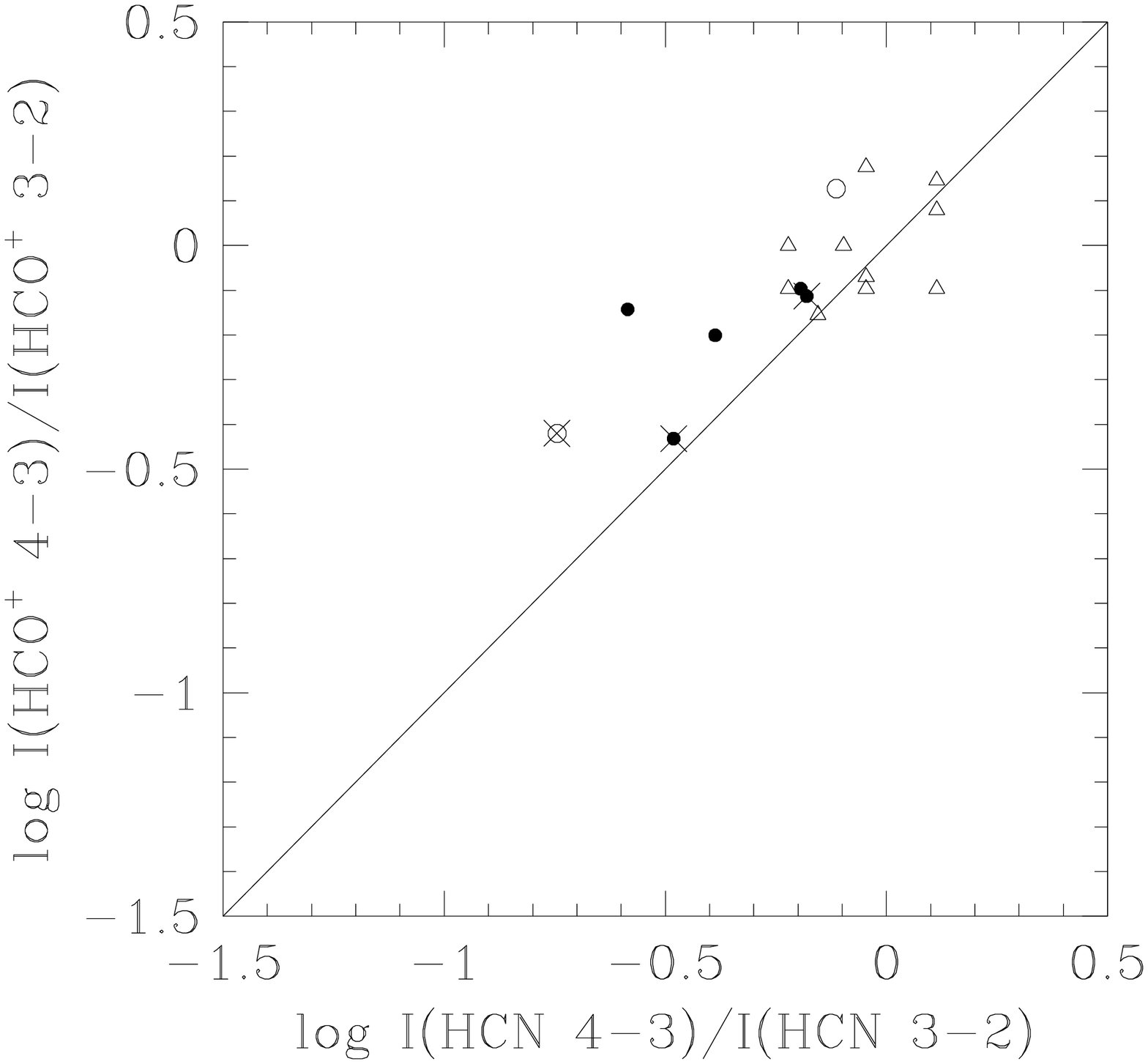}}}
\end{minipage}
\begin{minipage}{4.45cm}  
\resizebox{4.8cm}{!}{\rotatebox{0}{\includegraphics*{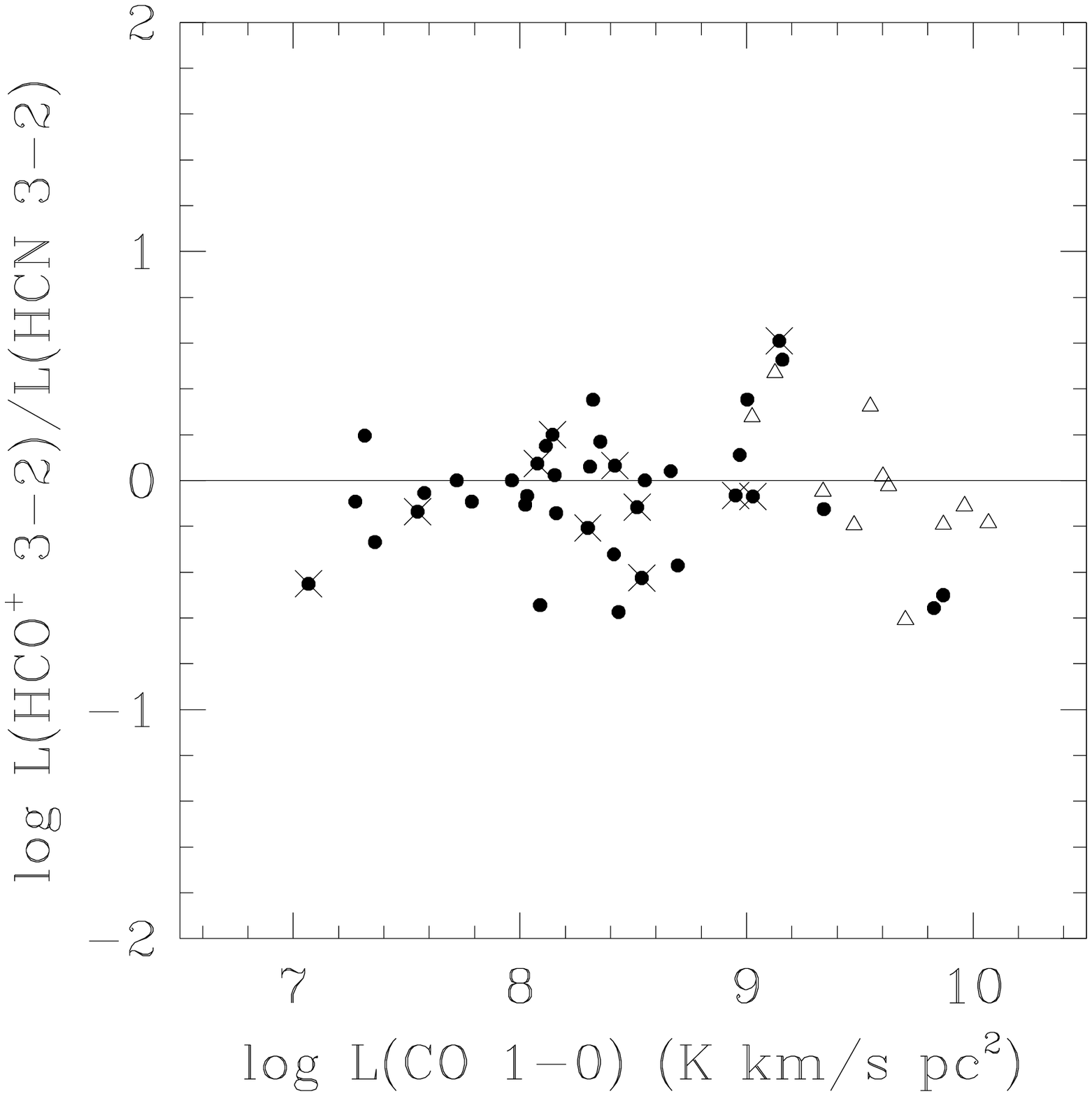}}}
\end{minipage}
\begin{minipage}{4.45cm}  
\resizebox{4.8cm}{!}{\rotatebox{0}{\includegraphics*{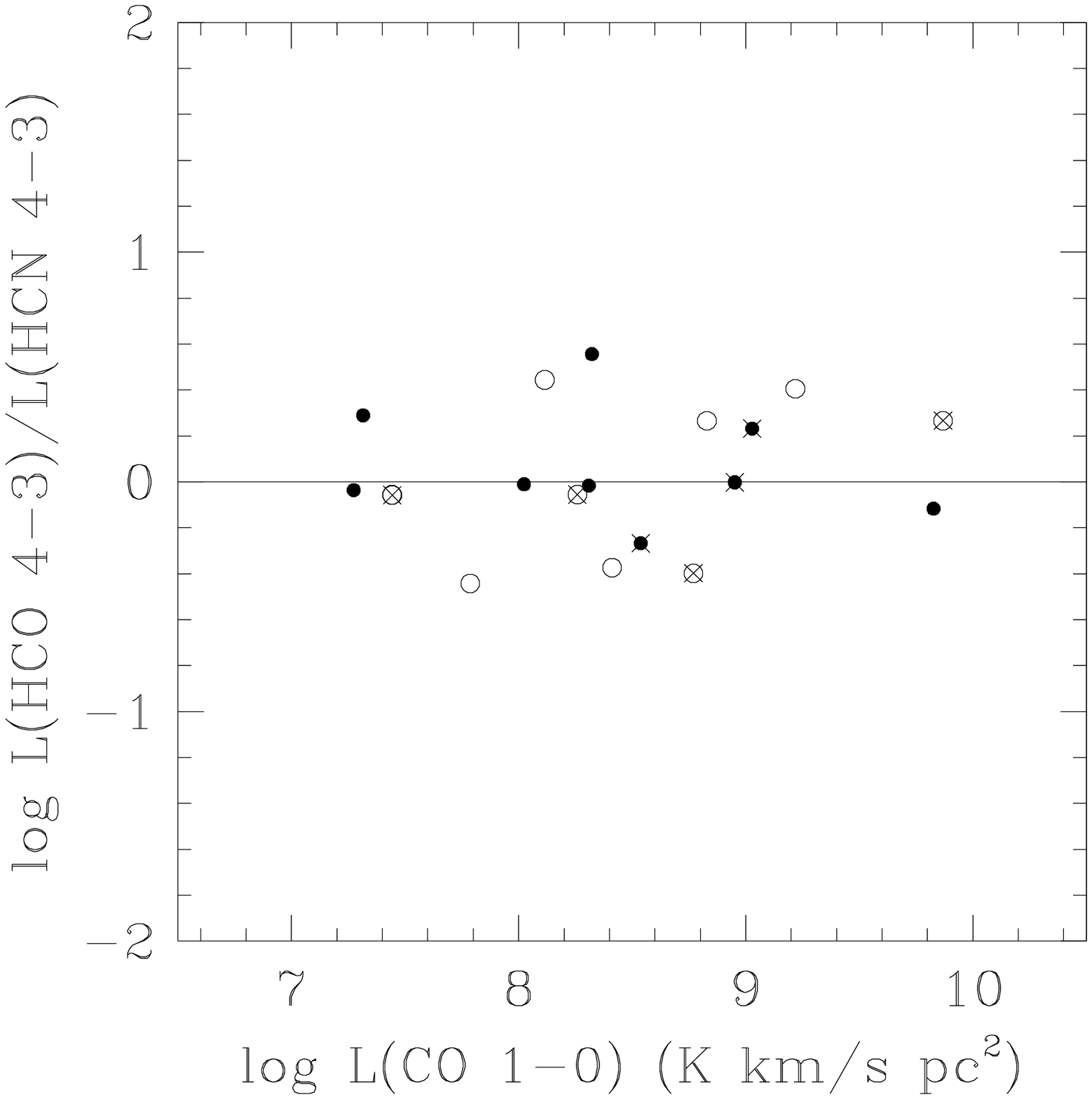}}}
\end{minipage}
\caption[] {The leftmost two panels ratios illustrate the similarity
  of the $\hco$ and HCN ladders. The rightmost panels show that the
  $\hco$/HCN intensity ratios remain constant over the full range of
  $J$=1-0 CO luminosities. The $J$=4-3 data shown include data from APEX
  and ALMA in addition to the IRAM data from this paper. Straight lines
  mark ratios of unity.  Fit parameters are listed
  in Table\,\ref{fittable}.
}
\label{ratspec}
\end{figure*}

\begin{figure}
\begin{minipage}{8.9 cm}
\begin{minipage}{4.4cm}
\resizebox{4.8cm}{!}{\rotatebox{0}{\includegraphics*{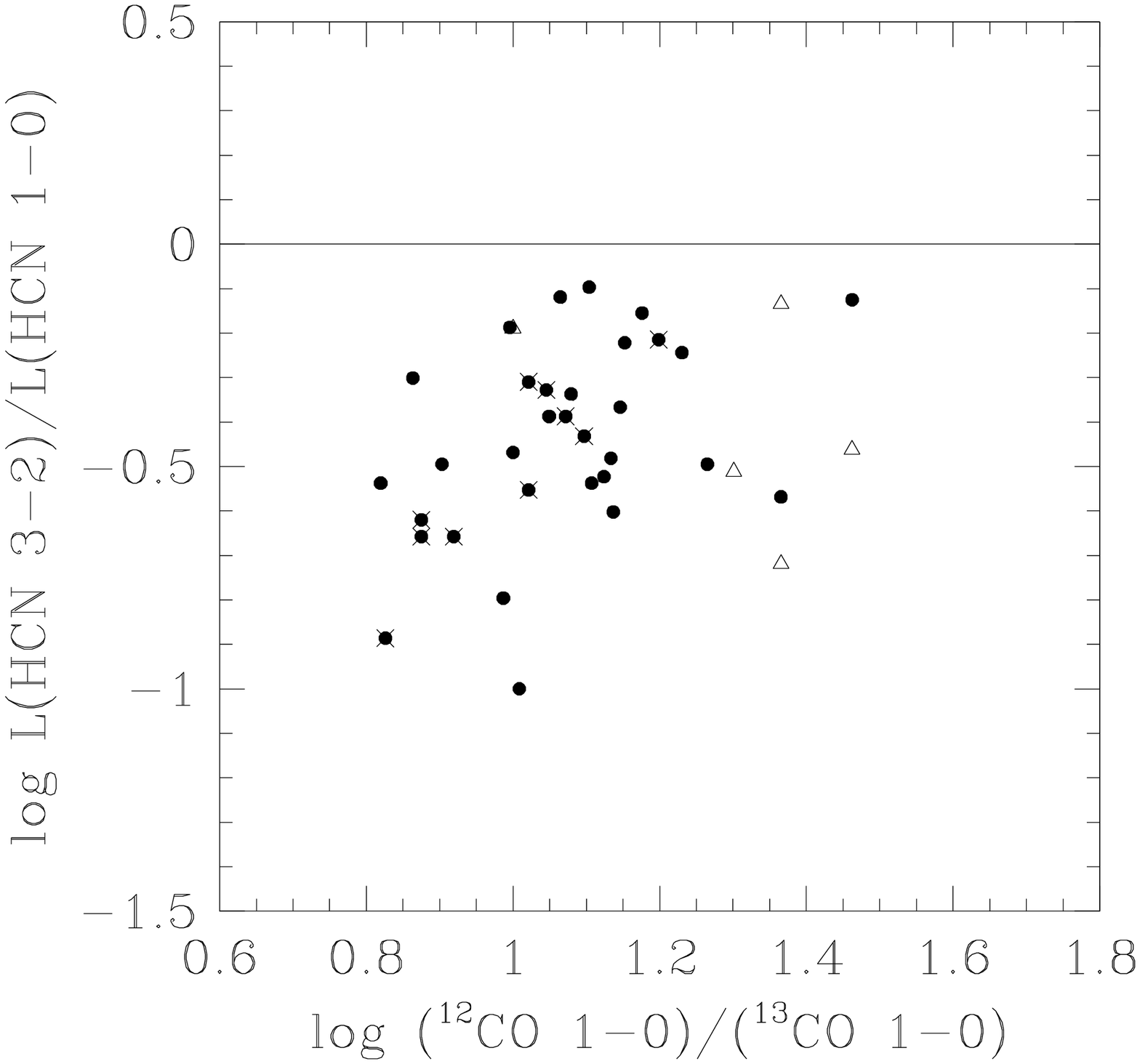}}}
\end{minipage}
\begin{minipage}{4.4cm}
\resizebox{4.8cm}{!}{\rotatebox{0}{\includegraphics*{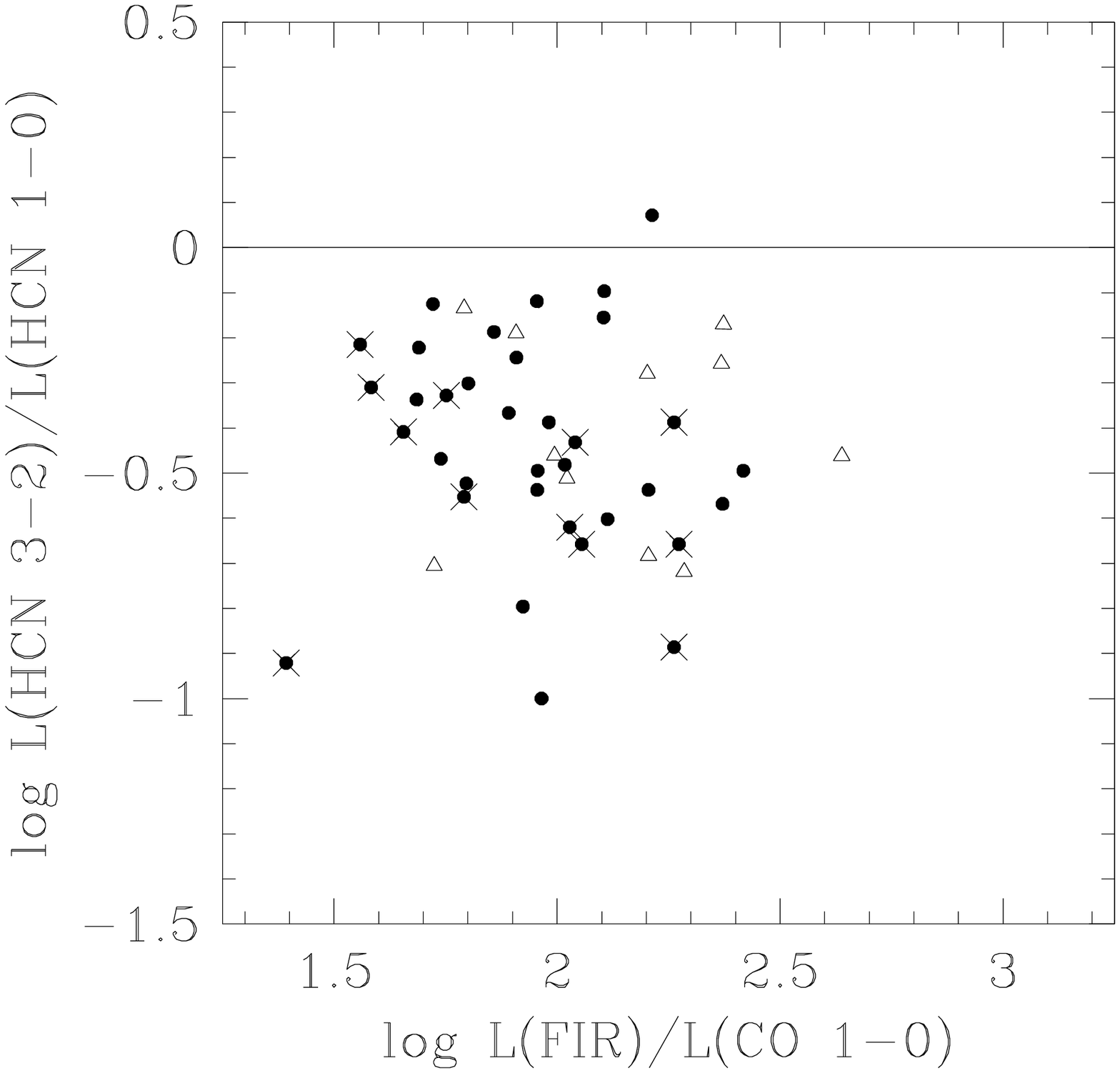}}}
\end{minipage}
\end{minipage}
\caption[] {Comparison of the $J$=3-2/$J$=1-0 HCN ratio as a
  function of the isotopological ratio $\co$/$\thirco$ (leftmost panel)
  and FIR/CO (rightmost panel). Solid lines mark ratios of unity.
  Symbols are as in Fig.\ref{lummol}.  Fit parameters are listed
  in Table\,\ref{fittable}.
}
\label{32Aratspec}
\end{figure}

\begin{figure}
\begin{minipage}{8.9cm}  
\begin{minipage}{4.4cm}
\resizebox{4.8cm}{!}{\rotatebox{0}{\includegraphics*{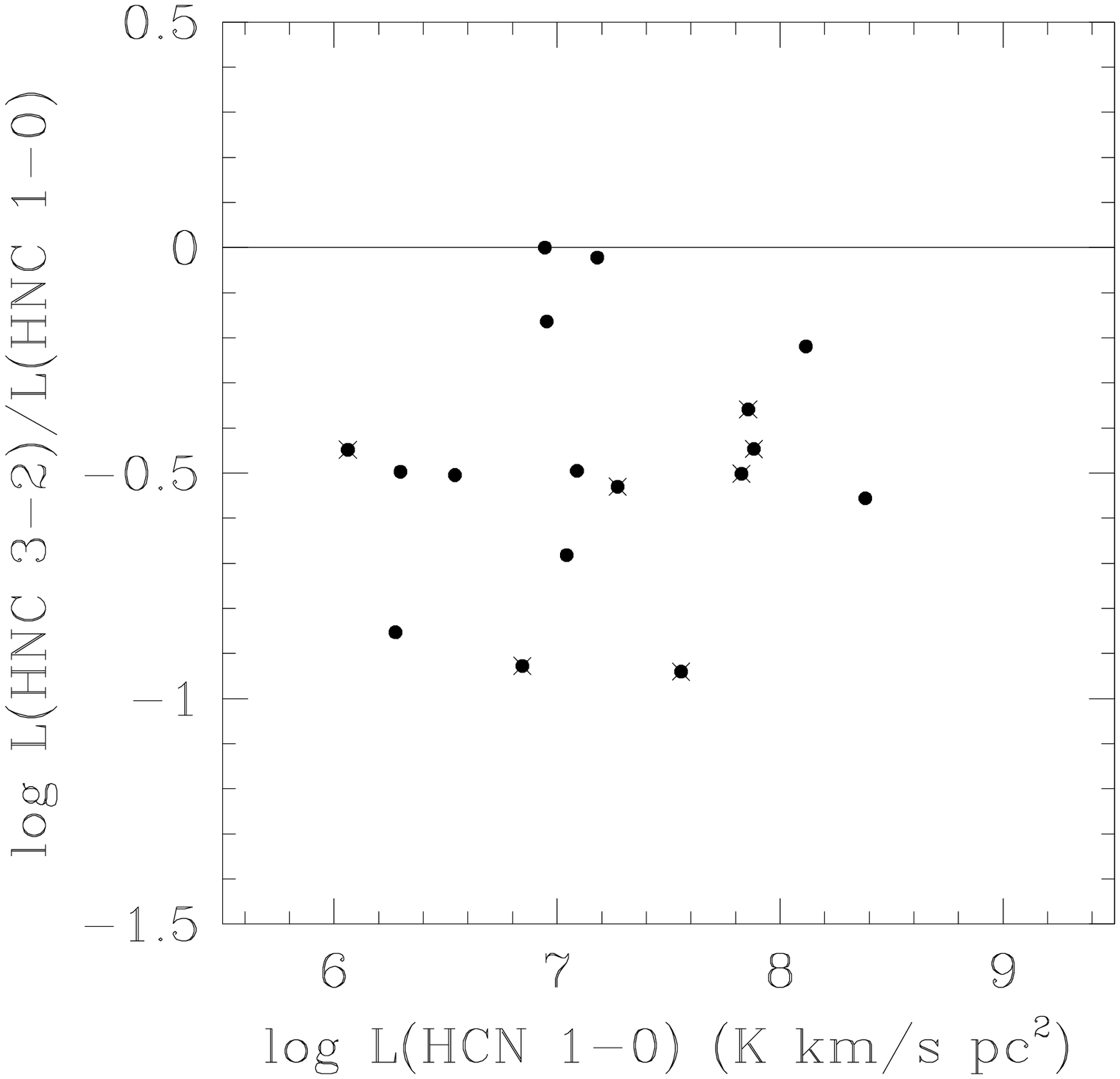}}}
\end{minipage}
\begin{minipage}{4.4cm}
\resizebox{4.8cm}{!}{\rotatebox{0}{\includegraphics*{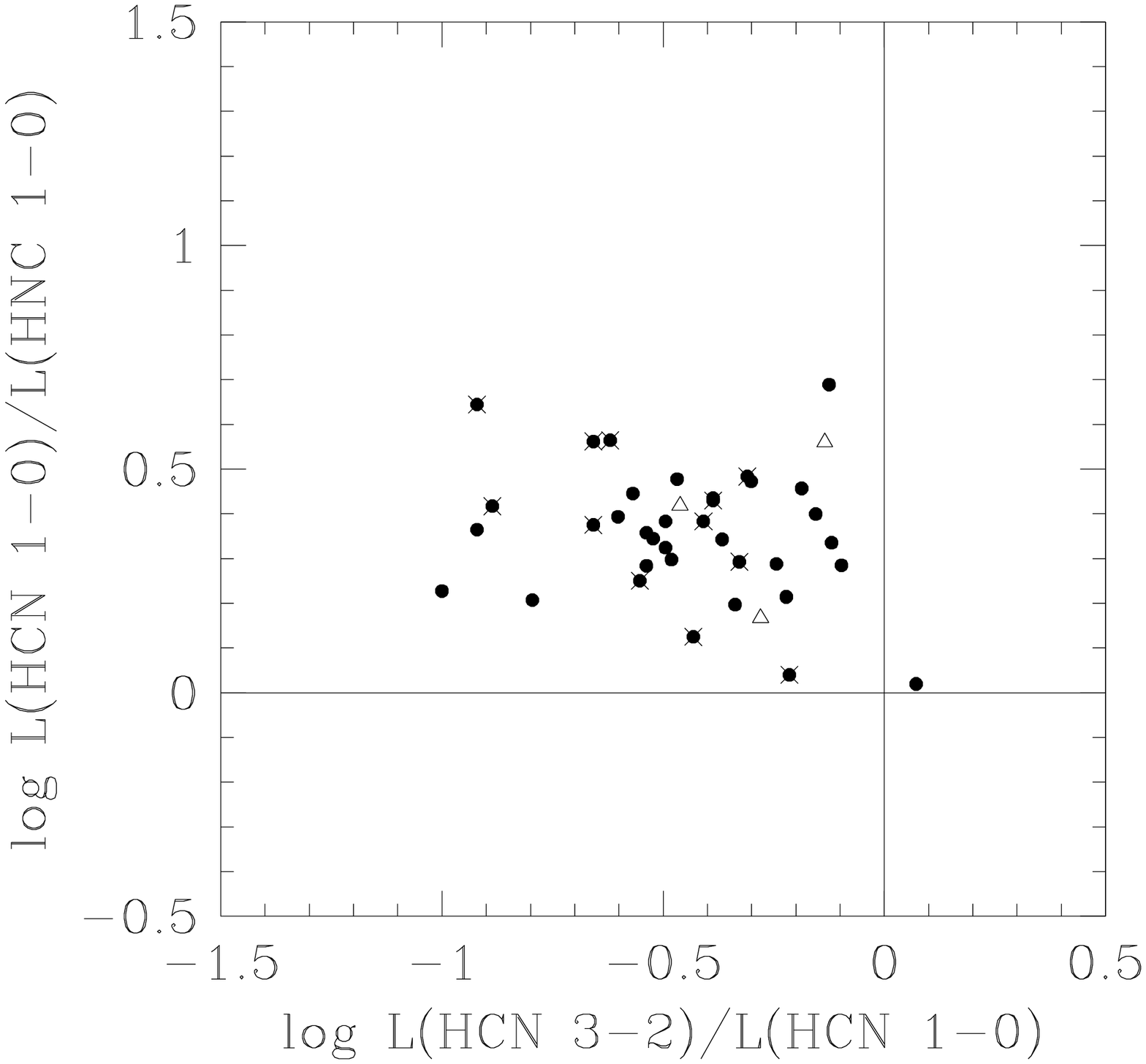}}}
\end{minipage}
\end{minipage}
\caption[] {Left: the HNC(3-2)/HNC(1-0) excitation ratio as a function
  of the HCN(1-0) luminosity. The ground-state HCN/HNC ratio as a
  of the HCN(3-2)/HCN(1-0) excitation ratio. Solid lines mark ratios of
  unit. Symbols are as in Fig.\ref{lummol}; most data are from this
  paper.  Fit parameters are listed in Table\,\ref{fittable}.
}
\label{32Bratspec}
\end{figure}

\begin{table*}
\begin{center}
{\small %
  \caption[]{\label{fittable}Linear regression fits to luminosities
    and ratios plotted in Figs. 3 through 11}
\begin{tabular}{llrrrrl}
\noalign{\smallskip}     
\hline
\noalign{\smallskip} 
  y          & x              & a     & b     & rms  & sfcc & Reference \\
  
 (1)         & (2)            & (3)   & (4)   & (5)  & (6)  & \\
\noalign{\smallskip}      
\hline
\noalign{\smallskip} 
L(CO)        & L(HCO$^+$)   &  1.03 & -1.60 & 0.21 &   0.95 & Fig. 3 \\
L(CO)        & L(HCN)       &  0.99 & -1.16 & 0.25 &   0.92 & \\
L(CO)        & L(HNC)       &  1.02 & -1.83 & 0.29 &   0.94 & \\
L(FIR)       & L(HCN)       &  0.90 & -2.26 & 0.28 &   0.92 & \\
L(FIR)       & L(CO)        &  0.88 & -0.72 & 0.25 &   0.93 & \\
L(HCN)       & L(HCO$^+$)   &  1.01 & -0.11 & 0.18 &   0.97 & \\
L(HNC)       & L(HCN)       &  0.95 &  0.71 & 0.15 &   0.98 & \\
I(HNC)       & I(HCN)       &  1.97 &  0.43 & 1.87 &   0.91 & \\
I(HCN)       & I(HCO$^+$)   &  0.86 &  0.15 & 2.24 &   0.88 & \\
D$^2$        & L(HCN)       &  0.71 &  5.23 & 0.34 &   0.91 & \\
D$^2$        & I(HCN)       & -0.29 &  1.28 & 0.34 & - 0.61 & \\
L(HCN)       & I(HCN)       & -0.14 &  1.39 & 0.43 & - 0.27 & \\  
L(CO)        & L(HCN)/L(CO) & -0.02 & -1.08 & 0.25 & -0.06 & Fig. 4 \\
L(HCN)       & L(HCN)/L(CO) &  0.09 & -1.92 & 0.24 &  0.27 & \\
L(FIR)       & L(HCN)/L(CO) &  0.02 & -1.54 & 0.25 &  0.06 & \\
L(FIR)       & L(FIR)/L(CO) &  0.13 &  0.72 & 0.25 &  0.39 & \\
L(HCN)/L(CO) & L(FIR)/L(CO) &  0.43 &  2.63 & 0.25 &  0.38 & \\
L(FIR)/L(CO) & L(FIR)/L(HCN)&  0.64 &  2.03 & 0.23 &  0.53 & \\
L(CO)          & L(HCN)/L(HCO$^+$) & -0.05 &  0.48 & 0.17 & -0.22 & Fig. 5 \\
L(HCN)/L(CO)   & L(HCO$^+$)/L(CO)  &  0.59 & -0.51 & 0.15 &  0.69 & \\
L(HCO$^+$)/L(CO)& L(HCN)/L(CO)     & -0.14 & -0.15 & 0.18 & -0.17 & \\
L(FIR)/L(CO)   & L(HCN)/L(HCO$^+$) & -0.00 &  0.04 & 0.18 & -0.01 & \\
L(CO)        & $^{12}$CO/$^{13}$CO  &  0.12 &  0.08 & 0.14 &  0.58 & Fig. 6 \\
$^{12}$CO/$^{13}$CO & L(HCN)/L(CO)  & -0.29 & -0.98 & 0.25 & -0.12 & \\
$^{12}$CO/$^{13}$CO & L(FIR)/L(CO)  &  0.18 &  1.85 & 0.25 &  0.15 & \\
$^{12}$CO/$^{13}$CO&L(HCN)/L(HCO$^+$)& -0.55 &  0.66 & 0.15 & -0.43 & \\
$^{12}$CO/$^{13}$CO&L(HCN)/L(HNC)   &  0.06 &  2.26 & 0.87 & -0.07 & Fig. 7 \\
L(FIR)         & L(HCN)/L(HNC)     & -0.08 &  3.20 & 0.87 & -0.14 & \\
L(CO)          & L(HNC)/L(HCO$^+$) & -0.06 &  2.82 & 0.87 & -0.13 & \\
L(HNC)/L(CO)   & L(HCN)/L(HNC)     & -0.26 &  0.08 & 0.14 & -0.47 & \\
L(HNC)/L(HCN)  & L(HCO$^+$)/L(HCN) & -0.05 & -0.06 & 0.17 & -0.02 & \\
L(HNC)/L(HCO$^+$&L(HCO$^+$)/L(HCN) & -0.54 & -0.20 & 0.12 & -0.74 & \\
L(HNC)/L(HCO$^+$& L(HNC)/L(HCN)    &  0.46 & -0.20 & 0.12 &  0.66 & \\
L(CO)   & HCO$^+$(3-2)/(1-0)&  0.12 & -1.37 & 0.25 &  0.32 & Fig. 8 \\
L(CO)   & HCN(2-1)/(1-0)        &  0.21 & -2.08 & 0.18 &  0.64 & \\
L(CO)   & HCN(3-2)/(1-0)        &  0.17 & -1.81 & 0.21 &  0.46 & \\
L(CO)   & HCN(4-3)/(1-0))       &  0.23 & -2.61 & 0.25 &  0.47 & \\
HCN(3-2)/(1-0) & HCO$^+$(3-2)/(1-0)&  0.57 & -0.14 & 0.17 & 0.68 & Fig. 9\\
HCN(4-3)/(3-2) & HCO$^+$(4-3)/(3-2)&  0.70 &  0.03 & 0.11 & 0.70 & \\  
L(CO)          &$J$=3-2 HCO$^+$/HCN &  0.14 & -0.15 & 0.24 & 0.17 & \\
L(CO)          &$J$=4-3 HCO$^+$/HCN &  0.08 & -0.68 & 0.32 & 0.13 & \\
$^{12}$CO/$^{13}$CO&HCN(3-2)/(1-0)&  0.32 & -0.71 & 0.23 &  0.16 & Fig. 10 \\
L(FIR)/L(CO)     & HCN(3-2)/1-0)& -0.34 &  0.33 & 0.22 & -0.38 & \\
L(HCN)           & HNC(3-2)/(1-0)&  0.07 & -1.02 & 0.24 &  0.18 & Fig. 11 \\
HCN(3-2)/(1-0)   & HCN/HNC(1-0)& -0.16 &  0.30 & 0.15 & -0.18 & \\  
\noalign{\smallskip}     
\hline
\end{tabular}
}%
\end{center} 
Notes: a. The regression fits are of the form log(y) = a log(x) + b.
Columns 1 and 2 identify the data used in the fit. Column 3 and 4 list
the slope a and y-axis intercept of the fitted relation. Column 5 lists
the r.m.s. residuals of the fit; the observed dispersion exceeds the
observational error, and the physical range of the y-parameter is about
four times the quoted r.m.s. Column 6 gives the Spearman correlation
coefficient of the fit, ranging from -1.00 (perfect anti-correlation) to
1.00 (perfect correlation). Coefficients in the range -0.35 to +0.35
should be interpreted as indicating a lack of correlation, especially
when slope a is also close to zero. 
\end{table*}

\subsection{The ground-state HNC/HCN isomer ratio}

The HCN isomer HNC(1-0) was detected by H\"uttemeister $\etal$ (1995)
in fifteen normal galaxies and mapped in five with the $IRAM$ 30m
telescope. They determined the isomer ratios by using older HCN
measurements.  Aalto $\etal$ (2002) presented a complementary HNC(1-0)
survey of nine luminous infrared galaxies, obtained with the
relatively large beams of the $OSO$ ($42"$) and $SEST$ ($55"$)
telescopes. The survey of 23 galaxies by Costagliola $\etal$ (2011),
also incorporated into our database, covered a variety of molecular
lines in the 3mm window, including HCN(1-0) and HNC(1-0). In all
surveys, the ground-state isomer ratio HNC/HCN varies significantly
from galaxy to galaxy. A strong motivation for HCN and HNC
observations is the potential use of their ratio as a diagnostic for
molecular gas kinetic temperature and density (Schilke $\etal$ 1992,
Graninger $\etal$ 2014), specifically sensitive to UV but not X-ray
irradiation (Bublitz $\etal$ (2019, 2022). As in the previous smaller
samples, the HNC/HCN ratios of the more than ninety galaxies presented
here are all in the range 0.2--1.1; the average is 0.46
(Table\,\ref{rats}). There is no correlation with global far-infrared
luminosity (also noted by Aalto $\etal$ 2002), nor with CO
luminosity. H\"uttemeister $\etal$ (1995) did not find a correlation
with putative tracers of star formation rate or dust temperature. In
our much larger sample, the HCN/HNC ratio is wholly unrelated to
either the HCN/CO ratio or the HCN/FIR ratio. We confirm the
anti-correlation found by Costagliola $\etal$ (2011) between the
$\hco$/HCN and the HNC/HCN ratios when they are plotted against the
HNC/$\hco$ ratio (Fig.\,\ref{hncrat}, bottom panels) but unlike these
authors we see no relation when they are plotted against each other or
sorted by galaxy type (Fig.\,\ref{hncrat} center right panel). We did
not find any other correlation involving the isomer ratio.

\subsection{The excitation ladders of HCN, $\hco$ and HNC}

The normalized $J$=3-2 HCN and $\hco$ luminosities are, just like the
ground-state luminosities, well-correlated with the $J$=1-0 CO
luminosities. The super-unity slope of about 1.12 is identical to the
slope of 1.11 found by Li$\eta$ (2020) for the relation between
HCN(3-2) and FIR luminosities in matched apertures, and less than the
more poorly defined slope of 1.26-1.35 found by Bussman $\etal$ (2008)
in the larger HHST beam.  In Fig.\,\ref{rattrans} the individual
HCN(2-1) (taken from Krips $\etal$ 2008), HCN(3-2), HCN(4-3), and
HNC(3-2) intensities relative to the ground-state are plotted as a
function of the CO(1-0) luminosity. With increasing transition the
average line intensities drop (columns 9-11 in Table\,\ref{rats}, and
Fig.\,\ref{rattrans}). Using that figure to remove the bias due to
unequal sampling, we find that the average intensities of the HCN
excitation ladder are represented by the series 1.00:0.65:0.45:0.25
which is steeper than the comparable CO ladder of 1.00:0.92:0.70:0.57
(Israel 2020). The averages for $\hco$ are practically identical
(Table\,\ref{rats}) columns 13-14). In all $J$ transitions, individual
HCN and $\hco$ line intensities are very similar (Fig.\,\ref{ratspec}.
Because of the paucity of especially JCMT $J$=4-3 $\hco$ observations,
we added to this figure line ratios from the $ALMA$ observations
of compact (ultra)luminous galaxies from Imanishi $\etal$ (2018) and
the APEX observations of normal galaxies from Zhang $\etal$
(2014). The APEX $J$=4-3 and JCMT $J$=3-2 beams are almost identical
facilitating the construction of $J$=4-3/$J$=3-2 ratios.

With each additional step on the ladder, average HCN and $\hco$
intensities decrease but within each step, individual ratios rise with
increasing CO luminosity (Table\,\ref{ratincrease},
Fig\,\ref{rattrans}).  The $J$=1-0, $J$=3-2, and $J$=4-3 intensities
HCN and $\hco$ increase in tandem (leftmost panels in
Fig.\,\ref{ratspec}). In each transition, their ratio is a constant
and unrelated to either CO (rightmost panels in Fig.\,.\ref{ratspec}
or FIR luminosity (Tan $\etal$ 2018).

A similar lack of correlation characterizes the HCN/HNC ratios in the
two transitions.  Again the spread in the $J$=3-2 values is about
double that in the $J$=1-0 values but the small sample reveals no
distinction between AGNs and starburst.  In Figs.\,\ref{gasprop} and
\ref{isorat} we found the $\hco$(1-0)/HCN(1-0) species ratio to be
inversely correlated with both the `dense-gas fraction' HCN(1-0)/CO
and the isotopologue ratio $\co$/$\thirco$ but independent from the CO
luminosity and the `star formation efficiency' FIR/CO. The excitation
of HCN (and $\hco$) is also unrelated to any of these quantities as
illustrated by plots of the $J$=3-2/$J$=1-0 ratios against them.
Fig.\,\ref{32Aratspec} shows that the excitation of HCN is not
meaningfully related to the isotopogical $\co/\thirco$ ratio (left
panel) or to HCN/CO (not shown) and FIR/CO (right panel). Finally,
Fig.\,\ref{32Bratspec} shows that the excitation of HNC is not related
to the luminosities of HCN (left panel) or CO, and that the HCN/HNC
isomer and the HCN(3-2)/HCN(1-0) excitation ratio are likewise
unrelated.

\section{Discussion}

\subsection{Luminosity plots provide little information}

The results in this paper confirm that the linear relations between
the HCN(1-0), CO(1-0), and FIR luminosities of whole galaxies, first
established by Solomon $\etal$ (1992) and Gao $\&$ Solomon (2004a),
also hold for constant aperture measurements of galaxy centers.  They
also show that the surface brightness of CO, HCN, and $\hco$ drops in
similar ways when the area covered by the aperture increases.
Variations in surface brightness and intensity ratios are almost
negligible compared to the variation in galaxy distances.  Distance is
the single factor dominating the luminosities in fixed apertures. This
is the key factor explaining the near-linearity of
luminosity-luminosity relations and it also explains why luminosity
correlations look progressively better as greater ranges in distance
are considered. As a result, the luminosity-luminosity relations
published in the literature, and also presented here, are mostly
trivial and provide very little information on the physical properties
of the galaxies sampled\footnote{see also Kennicutt (1990)}.
Specifically, the observation that linear luminosity relations hold
over a wide range of luminosities does not provide evidence that the
extra-galactic star formation rate is directly proportional to the mass
of dense gas. The scatter in the various luminosity-luminosity plots
carries more information than their linearity.  The non-negligible
dispersion involves factors of five or more and indicates significant
variation among individual galaxies even as a systematical change with
luminosity is absent. Intensity ratios provide a much better means of
evaluating the information conveyed by molecular line emission than
luminosities.

The luminosity plots do not reveal systematical differences between
the various galaxy types either. The only exception is the $\hco$
intensity which is always below that of HCN in AGN galaxies. Over the
observed range $9.0\,\leq$ log $L(FIR)$ (L$_\odot$) $\leq\,12.5$, the
linearity of the various CO, HCN, or $\hco$ luminosity relations does,
however, imply that interstellar medium (ISM) properties do not change
as a function of galaxy luminosity.

\subsection{HCN and HCO$^+$ trace the same gas}

The average $J$=1-0 $\hco$ to HCN intensity ratio is about 0.9 but
individual values can be up to three times higher or lower. The
behavior of $\hco$ and HCN is practically the same in all observed
transitions (Fig.\,\ref{ratspec}) which implies that both molecules
essentially trace the same gas, even at the higher critical densities.
Over a wide range of luminosities, the central regions of other
galaxies are thus much like the inner Milky Way, where Evans $\etal$
(2020) found that HCN(1-0) and $\hco$(1-0) both seem to probe the same
molecular cloud material.  When large areas are sampled
instantaneously, as in extra-galactic measurements, the integrated HCN
or $\hco$ emission from ensembles of molecular gas clouds is easily
dominated by extended regions of low density, down to 10$^2$ $\cc$
(Evans $\etal$ 2020). This may go some way to explain the similar
behavior of the emission from HCN and $\hco$, as well as CO,
notwithstanding the different critical densities sampled
(Table\,\ref{crit}).  From a practical point of view, in most of the
analysis HCN and $\hco$ are interchangeable molecular species.

AGNs do have systematically higher HCN/$\hco$ ratios than star-burst
galaxies (Kohno $\etal$ 2001, 2008; Imanishi $\etal$ 2007; Krips
$\etal$ 2008, Grac\'ia-Carpio $\etal$ 2008, and various later papers).
As HCN/CO ratios also appeared to be higher in AGN-dominated galaxes,
these authors concluded to an HCN over-luminosity reflecting an HCN
overabundance. Figs.\,\ref{gaocomp} and \ref{gasprop} confirm that
ground-state HCN intensities exceed those of $\hco$ in all AGN
galaxies (except NGC~4258, Li $\etal$ 2019) unlike star-burst
galaxies. The situation is complicated by overlooked AGNs, embedded in
ULIRGs (Imanishi $\etal$ 2007, Imanishi 2009, Li $\etal$
2021). High-resolution observations such as provided by ALMA may bring
to light significant small-scale variation that is smoothed out in the
larger-beam observations considered here. AGNs such as those in
e.g. NGC~7469 (Izumi $\etal$ 2020) and NGC~1068 (Butterworth $\etal$
2022) become manifest by the change in line behavior in their
immediate ($\sim$100 pc) surroundings.  Fig.\,\ref{ratspec} includes
the HCN/$\hco$ ratios of distant ULIRGS (12.0$\leq$log
$L$(FIR)$\leq12.3$) in the $J$=3-2 and $J$=4-3 transitions observed
with ALMA (Imanishi $\etal$ 2018). The distinction between AGN and
starburst galaxies is less obvious in the higher transitions. In these
galaxies, however, the compact nuclear regions (diameter $\leq$ 500
pc) have again higher HCN/HCO$^+$ ratios than the surrounding extended
regions.

Attempts to explain high HCN/HCO$^+$ ratios by analyzing the few
observed optically thick molecular lines in terms of star formation or
(X-ray) chemistry with radiative transfer models failed to produce
conclusive results (cf. Costagliola $\etal$ 2011, Izumi $\etal$ 2016,
Imanishi $\etal$ 2016, Privon $\etal$ 2020).  We expect such modelling
to be successful only when molecular species are selected specifically
capable of diagnostically distinguishing different excitation
mechanisms and, as in the case of CO (Israel 2020), only when multiple
transitions, including optically thin lines in addition to optically
thick lines, are considered.

Although the AGNs in our large sample have systematically higher
HCN(1-0)/HCO$^{+}$(1-0) ratios, they do not have systematically higher
ratios HCN(1-0)/CO(1-0) (Fig.\,\ref{gaocomp}), and Aladro $\etal$
(2015) found that their HCN(1-0)/CS(3-2) ratios are not systematically
higher, either. This suggests that in AGNs, $\hco$ is suppressed
rather than HCN enhanced. This suggestion is shared, on different
grounds, by Imanishi $\etal$ (2022). Papadopoulos (2007) has argued
that environmental conditions differently affecting the chemistries of
ions and neutral molecules naturally lead to $\hco$ suppression. For
instance, in the turbulent molecular clouds that dominate in galaxy
centers (cf. Israel 2020), free electrons from the ionized outer
layers are transported inwards, effectively suppressing the $\hco$ ion
in addition to exciting HCN. Better understanding requires a
greater variety of diagnostic line measurements to constrain
models. In the absence of independent evidence whether $\hco$ or HCN
is primarily affected by AGN environments, any conclusion obtained
from modelling is bound to remain speculative.

\subsection{Modest optical depths suggest large translucent fraction}

Decreases in optical depth $\tau$ and in isotopic abundance ratio
[$\thirco$]/[$\co$] both express themselves through increasing
isotopologue intensity ratios $I(\co)/I(\thirco$). The observed
isotopologue ratios increase with CO luminosity (Fig.\,\ref{isorat},
left panel) as well as FIR luminosity. This may be due to (a).
significant carbon monoxide photo-dissociation, (b). selective $\co$
nucleosynthesis, or (c). both. (a). Photo-dissociation of CO in the
strong radiation fields of luminous galaxies would diminish the
optical depths of both $\co$ ($\tau_{12CO}$) and $\thirco$
($\tau_{13CO}$) and cause the observed line ratio to approach the
isotopic abundance ratio, which itself may increase with luminosity
(Visser $\etal$ 2009).  Because luminous galaxies are actively forming
stars, low optical depths $\tau_{12CO}$ signify low abundances
[$\co$]/[$\h2$] rather than low overall ISM optical depths. (b)
Whereas selective photo-dissociation of $\thirco$ appears to be
unimportant (Visser $\etal$ 2009, Romano $\etal$ 2017), selective
nucleosynthesis of $\co$ may occur in star-forming galaxies rendering
$\co$ overabundant (Sage $\etal$ 1991; Wilson 1999; Romano
2017). Thus, $\tau_{12CO}$ would increase with respect to
$\tau_{13CO}$. The increased aperture filling fraction of optically
thick emission raises observed $\co$ intensity relative to $\thirco$
intensity.

The observed HCN-to-CO or FIR-to-CO ratios do not correlate with the
$I(\co)/I(\thirco$) isotopologue ratio. This argues against (b).
because enhanced CO intensities would depress these ratios, and
supports (a) where $\co$ intensities do not fall with decreasing
optical depths as long as $\tau_{12CO}\,\geq\,1$.  Inferred optical
depths decrease from $\tau_{13CO}\,\approx\,0.08-0.13$ for the
low-luminosity galaxies to $\tau_{13CO}\,\approx\,0.03-0.07$ for the
luminous galaxies. For isotopic abundance ratios of 40, typical
  for nearby galaxy centers (cf. Viti $\etal$ 2020) these correspond
to $\tau_{12CO}\,\approx\,3-5$ (low luminosity) and
.$\tau_{12CO}\,\approx\,1-3$ (high luminosity). For typical
luminous galaxy isotopic abundances 100 (cf. Viti $\etal$ 2020), the
corresponding $\tau_{12CO}\,\approx\,3-8$ is similar to that of
low-luminosity galaxies. It appears that the ground-state lines of our
sample are just optically thick with optical depths of a few.  The
modest CO optical depths represent averages over relatively large
surface areas and suggest that the ensemble of molecular clouds
sampled contains a significant fraction of translucent clouds in
addition to more dense clouds. In the environments sampled, low CO
optical depths do not imply equally low molecular hydrogen column
densities $N(\h2)$ (cf. Israel 2020). Optical depths of the very weak
$J$=1-0 H$^{13}$CN, H$\thirco^{+}$, and HN$^{13}$C lines have been
determined for a small number of bright galaxies (Nguyen-Q-Rieu
$\etal$ 1992, Costagliola $\etal$ 2011, Jiang $\etal$ 2011, Aladro
$\etal$ 2015, Jim\'enez-Donaire $\etal$ 2017, Li $\etal$ 2019,
2020b). Optical depths $\tau_{H13CN}$ are typically 0.02-0.15 with
higher values 0.25-0.35 for some LIRGs such as NGC~1614 and
UGC~05101. Optical depths $\tau_{H13CO}$ are in the range 0.02-0.10,
not very different from either $\tau_{H13CN}$ or $\tau_{13CO}$. These
low optical depths likewise suggest that a non-negligible fraction of
the emission in these lines comes from translucent gas. As already
referred to in the previous section, HCN(1-0) emission from Milky Way
molecular clouds is easily detected also in extended translucent
regions with densities $N(\h2)\sim500\,\cc$ (Pety $\etal$ 2017, Evans
$\etal$ 2020). At low $\thirco$ optical depths, the HCN/$\hco$
intensity ratio drops to values below unity (Fig.\,\ref{isorat}, right
panel). In that case, a larger fraction of the molecular gas in
luminous galaxies is at reduced (column) densities better traced by
$\hco$ than by HCN.

\subsection{The dense molecular gas is lukewarm}

The model calculations by Schilke $\etal$ (1992) illustrate how the
HCN/HNC ratio increases with rising molecular gas kinetic temperature
and density, presumably caused by temperature-dependent destruction of
HNC in neutral-neutral reactions. Despite the degeneracy between
temperature and density, the analysis of observations in Orion allowed
Hacar $\etal$ (2020) to derive an empirical relation between the
HCN/HNC ratio and the independently determined kinematic gas
temperature. The ratios $I$(HCN)/$I$(HNC) = 1-5 observed by us are in
the well-established part of the relation in their Figure 3, where
$T_{kin}\,\approx\,10\,\times\,I$(HCN)/$I$(HNC). Thus, the kinetic
temperatures of the HCN gas in the sample galaxies are $T_{kin}$ = 10
- 50 K, with a mean of 20 K. These are beam-averaged temperatures: a
temperature of 20 K might, for instance, also represent a mixture of
25 per cent gas at 50 K and and 75 per cent gas at 10 K etc. The
absence of ratios HCN/HNC$\geq$5, however, rules out substantial
contributions by shocked or high-temperature gas (cf. Schilke $\etal$
1992; Hacar $\etal$ 2020).

The analysis by Hacar $\etal$ (2020) also reveals a strong dependence
of individual HCN and HNC intensities on gas temperature,
$I(HNC)\,\propto\,T_{\rm k}$ and $I(HCN)\,\propto\,T_{\rm k}^2$.  At
any HCN intensity the corresponding molecular gas column density
varies by up to an order of magnitude depending on the actual
temperature. The authors suggest that these emissivity effects render
extra-galactic HCN measurements unreliable as tracers of dense
molecular gas content.  For the observed HNC/HCN ratios, the PDR
models by Meijerink $\etal$ (2007) require molecular gas column
densities $N(H_2)\,\lessapprox\,10^{22}$ cm$^{-3}$.

\subsection{Most of the lines are sub-thermally excited}

In the models by Schilke $\etal$ (1992), relatively modest typical gas
densities $n_H=n(HI)+2n(H_2)\sim10^4$ cm$^{-3}$ accompany low HCN/HNC
ratios.  The single-phase LVG models explored by Grac\'ia-Carpio
$\etal$ (2008) to explain their $J$=1-0, $J$=3-2 HCN and $\hco$
measurements also yield densities $n(\h2)\approx6\times10^4$ cm$^{-3}$
as well as abundances [HCN]/[$\hco$]$\approx$6.  We use the Meijerink
$\etal$ (2007) grid of models of gas excitation by UV photons (PDR)
and by X-ray photons (XDR) to further investigate the gas density.
The observed HCN/$\hco$ ratios fit the PDR excitation for densities
$n_H\,\approx\,(5-20)\times10^4$cm$^{-3}$ and UV irradiation
$G\,=\,(5-50)\times10^3\,$G$_0$\footnote{$G_0\,=\,1.6\times10^{-3}$
  erg cm$^{-2}$ s$^{-1}$, Habing 1968}. At the lowest CO luminosities
($L$(CO)=10$^7-10^8$ K km/s pc$^2$), the implied densities are at the
lower end with low implied irradiation $G\,\leq10^2\,$G$_0$.  The
irradiation requirements are much less if other heating mechanisms
such as XDR or mechanical heating (Meijerink $\etal$ 2006; Kazandjian
$\etal$ 2012, 2015) apply. Their work shows that XDR excitation is
unlikely, as also concluded by P\'erez-Beapuits $\etal$ (2007) who
suggested similar densities for a small sample of Seyfert galaxies.

Densities $n_H\,\approx\,5-20\times10^4$ are above the critical
density for $\hco$(1-0) but just below the critical density for
HCN(1-0) and HNC(1-0). They are well below the critical densities for
all other transitions of these molecules. All these lines are thus
sub-thermally excited.  The HCN/CO model ratios from Meierink $\etal$
(2007) suggest that roughly between 5 and 20 per cent of the observed
CO intensity originates in the dense gas that radiates in HCN and
$\hco$ whereas the remaining 80-95 per cent then comes, as found
before, from less dense gas with typical densities
$n_H\,\approx\,10^3$ cm$^{-3}$\footnote{These are emission fractions,
  not mass fractions which may be different depending on the
  excitation of the lines.}  The isotopologue ratio $\co$/$\thirco$ is
correlated with the dust temperature derived from the far-infrared
continuum (Costagliola $\etal$ 2011, and references therein) but not
with the temperature of the dense gas derived from the HCN/HNC isomer
ratio (Fig.\,\ref{hncrat}). This confirms that most of the CO-emitting
gas must be quite distinct from the HCN-emitting gas.

\subsection{HCN/CO and FIR/CO do not trace dense gas ratio or star  formation efficiency}
The ground-state HCN/CO and HCN/FIR intensity ratios are largely
unrelated to either distance or luminosity.

The intensity ratios FIR/HCN (or FIR/CO) and HCN/CO have been widely
portrayed as proxies for star formation efficiency\footnote{FIR/HCN
  and FIR/CO do not have the dimension of an efficiency
  factor. Instead, they are the inverse of the time in which the
  present amount of (dense) gas will be depleted at the present rate
  of star formation -- assuming that far-infrared and molecular line
  intensities indeed measure what they are purported to do.}
respectively the fraction of molecular gas that resides in dense gas
(see e.g. Gao $\&$ Solomon 2004b, more recently Tan $\etal$ 2018;
Jim\'enez-Donaire $\etal$ 2019). This assumes that FIR continuum
luminosities are directly proportional to star formation rates, and
that velocity-integrated CO and HCN line intensities are directly
proportional to the column densities of all molecular gas and dense
molecular gas, respectively.

Unfortunately, that is not exactly true for CO where factors such as
abundance, dynamics, and excitation all contribute to significant
individual as well as systematic variations in the relation between
line intensity and underlying gas column density (cf. Pety $\etal$
2017, Israel 2020). Doubts have been expressed to various degrees also
on the use of HCN and $\hco$ as reliable mass tracers by several
authors (e.g. Papadopoulos $\etal$ 2007, 2012; Krips $\etal$ 2008,
Costagliola $\etal$ 2011, Privon $\etal$ 2015, Mills $\&$ Battersby
(2017), Pety $\etal$ 2017, Graci\'a-Carpio $\etal$ 2020, Hacar $\etal$
2020), prompted by considerations of excitation, chemistry, and
radiative transfer. Recently, Li $\etal$ (2020a) specifically
emphasized the large uncertainties in dense gas mass estimated from a
single line transition.

Single dish measurements of galaxies typically cover large areas
sampling a wide range of different ISM conditions.  Studies simulating
such measurements with molecular line surveys of Milky Way areas
(e.g. Stephens $\etal$ 2016, Shimajiri $\etal$ 2017, Evans $\etal$
2020) show that extended sub-thermal emission from marginally dense gas
outshines the emission from high-density filaments and clumps. In
nearby galaxy centers (e.g. Circinus galaxy, NGC~4945, M~51, NGC~253,
NGC~1808), high $J$=1-0 HCN/CO ratios occur that are directly
ascribable to environmental conditions other than mass (Curran $\etal$
2001, Matsushita $\etal$ 2015, Walter $\etal$ 2017, Salak $\etal$
2018). Molecular line intensities are a function of gas properties
(chemical abundance, density, opacity) and environmental conditions
(irradiation, turbulence, shocks) that dominate chemistry and excitation
mechanism.  These do not necessarily average out and molecular gas
masses are not simple, linear functions of measured line luminosities.

The observational results in this paper emphasize the inconsistency of
HCN/CO ratios as an indicator of the dense gas fraction. The variation
from galaxy to galaxy is the same at any given intensity or luminosity
ratio. No ratio of HCN or $\hco$ to CO or FIR intensity is
systematically correlated with the luminosity of a molecular line, with 
the FIR continuum luminosity, with the observed isotopological ratio
$I(^{12}$)CO/$I(^{13}$), or with distance. Nor is there any systematic
difference in HCN/CO ratio between normal galaxies, starburst
galaxies and AGNs. In the large number of central regions sampled, the
near-constancy of the various line ratio averages over a wide range of
luminosities is well-established.  If, contrary to our assertion, the
HCN and CO luminosities were to reliably measure the respective masses
of very dense and modestly dense gas, there would still be no relation
between the 'dense gas fraction' HCN/CO and such quantities as total
molecular gas mass or CO optical depth. The purported dense gas
fraction would also be wholly unrelated to surface brightness or
surface area covered.

Much the same can be said for the interpretation of the ratios FIR/CO,
FIR/HCN, and FIR/$\hco$ as proxies for star formation efficiencies
(more correctly: inverse dense-gas-depletion times).  The overwhelming
lack of correlation between tracers of different gas density and
either the FIR luminosity (also noted by Privon $\etal$ 2015 and Li
$\etal$ 2021), the FIR-to-HCN ratio, or the FIR-to-CO ratio is counter
to what is expected in the case of proxies for star formation rates
and efficiencies.  It would mean that the star formation process is
insensitive to the relative amounts of dense ($\hco$) and very dense
gas (HCN).  The star formation efficiency would also be independent of
surface area covered or galaxy type, and there would be a surprising
lack of any connection between the rate and the efficiency of star
formation. Specifically, it would imply the same efficiency of star
formation in local low-luminosity galaxies and in very luminous
galaxies (LIRGs and ULIRGs) where star formation rates are usually
considered to be one or two magnitudes higher. The overall lack of
significant correlations between the relevant data strongly implies
that the ground-state HCN-to-CO, FIR-to-CO, or FIR-to-HCN intensity
ratios are not meaningful as direct proxies for physical quantities
such as the fraction of dense gas in the ISM or the efficiency of star
formation in galaxy centers. Instead, especially Figs.\,\ref{gaocomp}
and \ref{rattrans} suggest that the line intensities of both CO and
HCN primarily reflect the excitation of the molecular gas.

\subsection{The ISM in galaxy centers.}

The ISM in galaxy centers is poor in neutral atomic hydrogen ($\hi$)
and almost all gas is molecular hydrogen ($\h2$). Analysis of the CO
and $\thirco$ emission in various transitions reveals coexisting
molecular gas phases distinguished by density and temperature
(cf. Israel 2020). If we assume PDR-excitation for the galaxies in this
paper, the emission from HCN and $\hco$ originates in molecular gas
at average densities not exceeding $n_H$ = $5\times10^4$ -
$2\times10^5$ $\cc$, low average temperatures of $T_k$ = $10-50$ K,
and very moderate optical depths $\tau_{CO}$ = 1 - 5. The gas traced
by HCN and $\hco$ is itself only a small fraction ($\leq20\%$) of the
molecular gas traced by CO with substantially lower densities of about
$n_H$ = $10^3$ $\cc$, the precise fraction depending on the actual
line ratios and the assumed mode of excitation.

Loenen $\etal$ (2008) used HCN, HNC, and $\hco$ line ratios to
investigate the excitation mechanism. In addition to the PDR and XDR
models from Meijerink $\etal$ (2007), they also considered models with
mechanical heating of cloud volumes in addition to PDR surface
heating. They found that most of the line ratios from the Baan $\etal$
(2008) sample require PDRs dominated by mechanical heating. The bottom
panels in Fig.\,\ref{hncrat} are the counterparts to Figs. 1 and 2 by
Loenen $\etal$ (2008). They exhibit overall similarity but
significantly tighter distributions, reflecting the greater accuracy
of our more homogeneous and sensitive database. The combined relative
intensities of the HCN, $\hco$ and HNC lines conclusively rule out XDR
dominating the excitation in all galaxies as well as pure PDR
excitation in most galaxies (cf. Meijerink $\etal$ 2007, Loenen
$\etal$ 2008). The molecular gas excitation in our sample is almost
certainly dominated by mechanical heating. As a consequence, the
density and especially the irradiation requirements are relaxed to
lower values. Following Kazandjian $\etal$ (2015), average densities
are $n_H\,=\,10^4-10^5\,\cc$ and the implied average irradiation drops
from $G\,=3\times10^4$ G$_0$ to $G=3\times10^2$ G$_o$ for mechanical
heating fractions $\alpha$ increasing from 0.1 to 0.5. The decrease by
two orders of magnitude reflects the greater efficiency of volume over
surface heating.

The relatively low density, low optical depth, and low irradiation of
the molecular gas in galaxy centers are characteristic of an ISM not
actively engaged in star formation. There is no independent evidence
directly and unambiguously linking molecular line intensities such as
$J$=1-0 CO, HCN, or $\hco$ to gas column density or mass.  In this
paper, we also observed an overall lack of significant correlations
between line ratios and properties that could be interpreted as
related to large-scale physical processes. Taken together, this forces
the conclusion that, in spite of much effort over the last few
decades, single-dish molecular line observations allow little more
than assumption-driven speculation and are of little use in the
quantitative determination of star formation in galaxy centers.

The significant dispersion of measured line intensities and ratios
around the average values, typically a factor of three either way,
implies different ISM characteristics for different galaxies,
including varying gas phase combinations. Individual ISM descriptions
surpassing the average treatment given in this paper can be derived
from detailed multi-transition, multiple-species radiative transfer
modelling. Preliminary examples have already been published, such as
M~82 (Loenen $\etal$ 2010), NGC~253 (Rosenberg $\etal$ 2014a), and
Arp~299 (Rosenberg $\etal$, 2014b). The results discussed in this
paper show that only thus we may hope to identify the actual ISM
physics that underlies the observed patterns or lack thereof. Such a
more detailed treatment of the galaxies in the present sample is
deferred to a subsequent paper.

\section{Conclusions}

\begin{enumerate}
\item This paper presents new $IRAM$ and $JCMT$ observations of 46
  bright galaxies in the $J$=1-0, $J$=3-2, and $J$4-3 transitions of
  HCN, $\hco$ and HNC. These are complemented by similar observations
  as well as $\co$(1-0) and $\thirco$(1-0) from published IRAM and
  JCMT surveys. The resulting extensive database covers 130 galaxies
  in HCN(1-0) and $\hco$(1-0) and 94 galaxies in HNC(1-0). In
  addition, it includes 12 galaxies in HCN(2-1), about 50 galaxies in
  HCN(3-2) and $\hco$(3-2), 25 galaxies in HCN(4-3), and 18 galaxies
  in HNC(3-2) and $\hco$(4-3).
\item The observed intensities were normalized to a common resolution
  of $22"$ in order to produce meaningful line intensity ratios. The
  analysis systematically explores luminosity-luminosity relations as
  well as relations between line ratios, and relations between line
  ratios and luminosities.
\item As expected from previous work, the $J$=1-0 HCN, $\hco$, and HNC
  luminosities are all linearly related to CO(1-0) and far-infrared
  (FIR) luminosities.  We also find that this is true for the
  luminosities in the higher transitions of HCN(3-2), $\hco$(3-2),
  HCN(3-2), HCN(4-3), and $\hco$(4-3). Very little can be concluded
  from this, however, as the luminosity-luminostity relations are
  essentially trivial because the luminosities are dominated by the
  variation in distance and not by the physics of the galaxies
  sampled.
\item Individual galaxy luminosities and line ratios show significant
  dispersion around the mean in all comparisons, implying significant
  differences in molecular gas properties between individual galaxies.
  The dispersion is uncorrelated with luminosity or line ratio and
  more likely originates in the detailed ISM physics than in systematic
  large-scale galaxy properties. Analysis of the ISM in individual
  galaxies is deferred to a later paper.
\item The average normalized HCN and $\hco$ transitions ladders,
  $J=$(n+1$\rightarrow$n)/$J$=(1-0), as well as the isotopological ratio
  $J$=1-0 $\co$/$\thirco$ are positively correlated with CO and FIR
  luminosity. No other line ratio shows such a clear correlation with
  luminosity.
\item HCN and $\hco$ have almost equal intensities and behave very
  similarly across the entire sample. These two molecules trace the
  the same gas, notwithstanding a significant difference in critical
  density. In AGN-dominated galaxies, ground-state HCN intensities
  always exceed those of $\hco$. Suppression of $\hco$ intensity is
  more likely than the alternative of HCN-enhancement in starburst
  galaxies.
\item Our radiative transfer models show that only $5-20\%$ of
  the observed CO emission originates in the HCN/$\hco$-emitting
  molecular gas. The HCN and $\hco$ emission represents a mixture of
  dense gas clouds and an undetermined but significant amount of
  translucent molecular gas. Except for $\hco$(1-0), all observed
  lines are sub-thermally excited.
\item The observed CO, HCN, and $\hco$ emission is not simply related
  to molecular gas column density or mass. These lines reflect the
  excitation of the gas, they are affected by the gas opacity and
  abundance, but they are not reliable mass tracers.
\item The HCN/CO line intensity ratio cannot be used as a proxy for
  the dense gas fraction, and the FIR/HCN and FIR/CO intensity ratios
  are also meaningless as proxies for star formation efficiencies
  or even molecular gas depletion times. Because the molecular lines
  do not reliably trace mass, comprehensive understanding of star
  formation requires a more appropriate determination of gas mass.
\item The molecular line emission from galaxy centers rules out a
  dominant heating contribution by X-rays (XDRs) but is fully
  consistent with UV-photon heating (PDRs) enhanced by a significant
  mechanical heating contribution due to turbulence or shocks.
\item The densest molecular gas in the galaxy centers sampled by
  ground-state HCN and $\hco$ lines has relatively low average kinetic
  temperatures $T_{\rm kin}\,=\,10-50$ K, relatively low average
  densities $n_{\rm H}\,=\,10^4-10^5\,\cc$, and relatively low optical
  depths of only a few. Most of the gas sampled by CO has densities
  $n_{\rm H}\,\leq=\,10^3\,\cc$. If the mechanical heating fraction is
  $50\%$, the energy input required is only $G\,\sim\,300$ G$_{0}$.
\end{enumerate}

\section*{Acknowledgements}

Most of the JCMT observations were obtained in service mode. I am
especially indebted to the $JCMT$ operators and observers, who helped
to collect the large data base described in this paper.

\end{document}